\documentclass{aa}
\usepackage{natbib}
\usepackage{graphicx}
\usepackage{txfonts}
\usepackage{booktabs}
\usepackage[breaklinks,colorlinks=true,citecolor=blue,linkcolor=magenta,urlcolor=blue]{hyperref}

\usepackage{caption}
\DeclareCaptionLabelFormat{continued}{#1~#2 (continued)}

\newcommand{\teff}{$T_{\rm eff}$}
\newcommand{\lsiv}{\object{LS\,IV$-$14$\degr$116}}
\newcommand{\feige}{\object{Feige\,46}}

\begin{document}

\title{Heavy-metal enrichment of intermediate He-sdOB stars:\\
the pulsators \feige\ and \lsiv\ revisited}

\author{M. Dorsch\inst{1,2}\and M. Latour\inst{3}\and U. Heber\inst{2}\and A. Irrgang\inst{2}\and S. Charpinet\inst{4}\and C. S. Jeffery\inst{5}}  

\institute{
Institut für Physik und Astronomie, Universität Potsdam, Haus 28, Karl-Liebknecht-Str. 24/25, 14476 Potsdam-Golm, Germany\\
\email{matti.dorsch@fau.de} 
\and
Dr. Karl Remeis-Observatory \& ECAP, Friedrich-Alexander University Erlangen-N\"{u}rnberg,
Sternwartstr. 7, 96049 Bamberg, Germany
\and
Institute for Astrophysics, Georg-August-University,
Friedrich-Hund-Platz 1, 37077 G\"{o}ttingen, Germany
\and 
Institut de Recherche en Astrophysique et Plan\'etologie, CNRS, Universit\'e de Toulouse, CNES, 14 Avenue Edouard Belin, 31400 Toulouse, France
\and
Armagh Observatory and Planetarium, College Hill, Armagh BT61 9DG, Northern Ireland
}

\date{Received ; accepted }

\abstract
{Hot subdwarf stars of spectral types O and B represent a poorly understood phase in the evolution of low-mass stars, in particular of close compact binaries. A variety of phenomena are observed, which make them important tools for several astronomical disciplines. For instance, the richness of oscillations of many subdwarfs are important for asteroseismology. Furthermore, hot subdwarfs are among the most chemically peculiar stars known. 
Two intermediate He-rich hot subdwarf stars, \lsiv\ and \feige, are particularly interesting, because they show extreme enrichments of heavy elements such as Ge, Sr, Y, and Zr, which are strikingly similar in both stars.
In addition, both stars show light oscillations at periods incompatible with standard pulsation theory and form the class of V366\,Aqr variables. 
We investigated whether the similar chemical compositions extend to more complete abundance patterns in both stars and validate the pulsations in \feige\ using its recent TESS light curve. High-resolution optical and near-ultraviolet spectroscopy are combined with non-local thermodynamical-equilibrium model atmospheres and synthetic spectra calculated with \textsc{Tlusty} and \textsc{Synspec} to consistently determine detailed metal abundance patterns in both stars.
Many previously unidentified lines were identified for the first time with transitions originating from \ion{Ga}{iii}, \ion{Ge}{iii-iv}, \ion{Se}{iii}, \ion{Kr}{iii}, \ion{Sr}{ii-iii}, \ion{Y}{iii}, \ion{Zr}{iii-iv}, and \ion{Sn}{iv}, most of which have not yet been observed in any star.
The abundance patterns of 19 metals in both stars are almost identical, light metals being only slightly more abundant in \feige,\ while Zr, Sn, and Pb are slightly less enhanced compared to \lsiv.
Both abundance patterns are distinctively different from those of normal He-poor hot subdwarfs of a similar temperature.
The extreme enrichment in heavy metals of more than 4\,dex compared to the Sun is likely the result of strong atmospheric diffusion processes that operate similarly in both stars while their similar patterns of C, N, O, and Ne abundances might provide clues to their as yet unclear evolutionary history.
Finally, we find that the periods of the pulsation modes in \feige\ are stable to better than $\dot{P} \lesssim 10^{-8}$ s/s. This is not compatible with $\dot P$ predicted for pulsations driven by the $\epsilon$-mechanism and excited by helium-shell flashes in a star that is evolving, for example, onto the  extended horizontal branch.
}


 \authorrunning{Dorsch et al.}
 \titlerunning{\feige\ and \lsiv\ revisited}

\maketitle

\section{Introduction}

Most hot subdwarf stars are compact, core He-burning objects of spectral types O (sdO) and B (sdB) with a thin hydrogen-rich envelope \citep[see][for reviews]{heb09,heb16}.
The bulk of sdB stars form the hot end of the horizontal branch (HB), the so-called extended horizontal branch (EHB).
Due to their lack of an extended hydrogen envelope, hot subdwarf stars are not able to sustain a hydrogen-burning shell \citep{Sweigart1987}.
They are thought to have a He-burning lifetime of about 100\,Myr, after which point they directly evolve towards the white dwarf (WD) cooling sequence \citep{dor93}.

Despite their lack of a thick hydrogen envelope, the atmospheres of most sdBs are dominated by hydrogen as a result of atomic diffusion, that is, the balance between radiative levitation and gravitational settling, damped by turbulence and mass loss \citep{mich11, hu11}.
In contrast, many sdO stars are extremely He-enhanced and show almost no hydrogen in their atmospheres \citep{Stroeer2007, Nemeth2012, Fontaine2014}.
Helium-rich sdO stars are thought to be the result of either a delayed He-flash at the top of the red giant branch \citep[RGB,][]{bert08} or the merging of two low-mass stars: for example, two He-WDs \citep{zhang12}.
Unlike the He-poor sdB stars, these He-sdOs do not seem to be influenced by diffusion processes \citep[due to convection caused by the ionisation of He\,\textsc{ii};][]{Groth1985}.
Two questions arise: will most He-sdOs evolve to become He-poor sdBs, or do they represent a distinct population? 
And at which point in the stellar evolution does atmospheric diffusion become important? 
Both \feige\ and \lsiv\ are part of the small population of intermediately He-rich sdOB (iHe-sdOB) stars that is of special interest when trying to address these questions \citep{2012ASPC..452...41J}.
They share many physical properties, which make them a unique pair not only among the iHe-sdOBs.

\begin{figure}
\begin{centering}
\includegraphics[width=1\columnwidth]{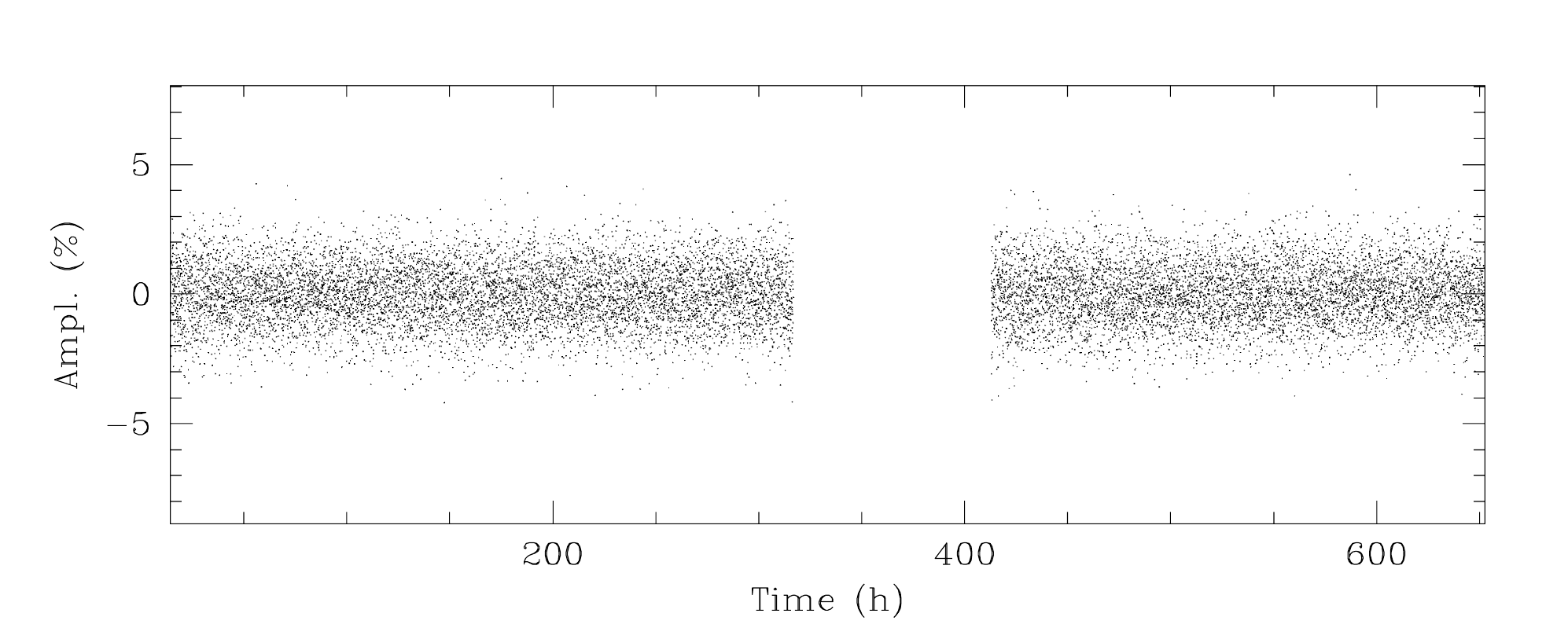}

\includegraphics[width=1\columnwidth]{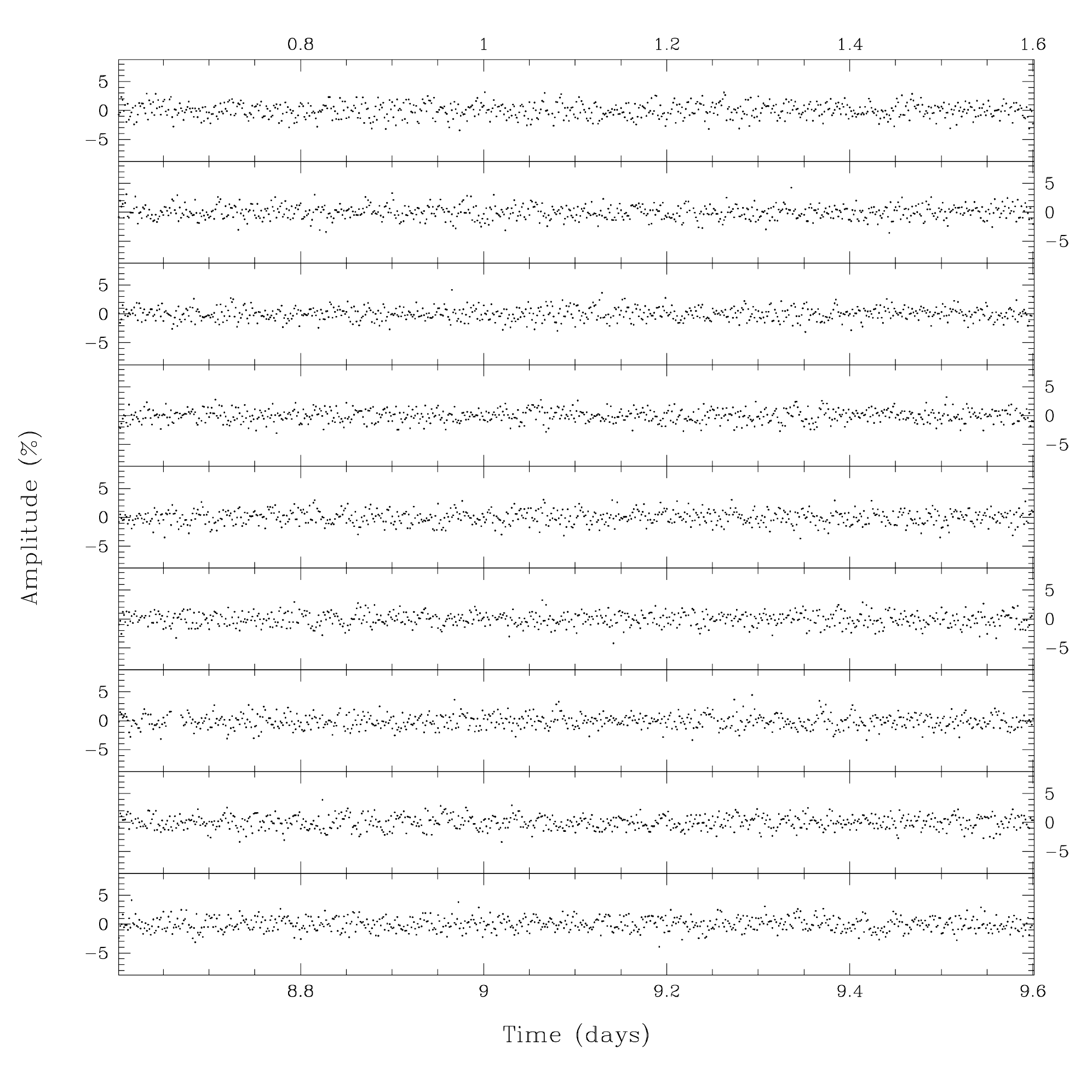}

\includegraphics[width=1\columnwidth]{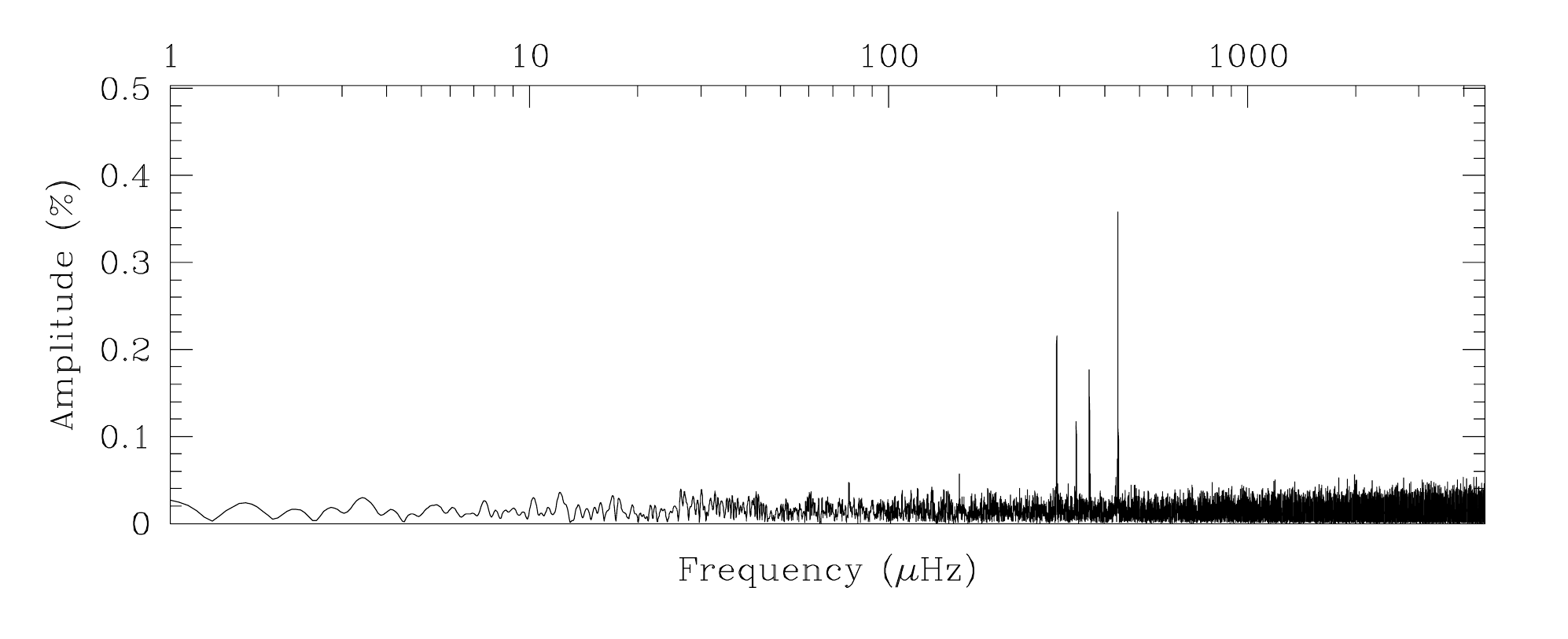}
\par\end{centering}
\caption{Photometry obtained for Feige\,46 (TIC\,371813244) with TESS. \emph{\emph{Top
panel}:} light curve from Sector 22 (amplitude is in percent of the
mean brightness of the star) spanning 26.59\,d (638.16\,h) sampled every
120\,s. Gaps in this time series are caused by the mid-sector interruption
during data download and measurements removed from the light curve
because of a non-optimal quality warning. \emph{\emph{Middle panel}:} close-up
view of the light curve covering the first nine days, where modulations
are visible. \emph{\emph{Bottom panel}:} Lomb-Scargle Periodogram of the
light curve up to the Nyquist frequency limit ($\sim4167$\,$\mu$Hz).
Significant periodic signal is clearly detected in the $250-500$\,
$\mu$Hz range. \label{fig:TESS-F46-LSP}}
\end{figure}

Kinematic analyses of \lsiv\ \citep{ran15} and \feige\ \citep{Latour2019a} have shown that both stars are likely to be members of the Galactic halo unlike most of the helium-rich hot subdwarfs \citep{2017MNRAS.467...68M}.
Both stars show light variations attributed to pulsations.
Since its light variations were discovered by \cite{ahm05}, \lsiv\ remained the sole member of its class of pulsating stars, now termed V366 Aqr variables, until \cite{Latour2019a} identified similar pulsations in \feige. \cite{ahm05} identified two periods of 1950\,s and 2900\,s in the light variations of \lsiv.
These pulsations were confirmed in follow-up observations by \cite{jeff11} and \cite{green11}, who identified four additional periods up to 5084\,s. Pulsational light variations in sdB stars are well established. 
Both pressure (p-mode) and gravity (g-mode) oscillations have been observed in hot subdwarf stars -- the former have periods of a few minutes (short periods), whereas the periods of the latter range from 30 minutes to a few hours (long periods; for recent compilations, see \citealt{2017MNRAS.466.5020H} and \citealt{2018OAst...27..157R}).

The pulsations observed in He-poor sdB stars are thought to be driven by an opacity ($\kappa$-) mechanism that is related to an iron and nickel opacity bump in the thin stellar envelope.
This mechanism can produce both short-period oscillations \citep{cha96,cha97} at the temperature of \lsiv\ and \feige\ ($\sim$35\,000\,K), and long-period oscillations \citep{green03,jeff06} at lower temperatures.
The detection of long periods in \lsiv\ is remarkable, because the $\kappa$-mechanism predicts that short-period pulsations should be excited at the high effective temperature and surface gravity of \lsiv, which, however, are not observed.
How the observed long-period pulsations are excited in \lsiv\ remains an open question.
\cite{bat18} and \cite{bert11, Miller2020} showed that gravity modes stochastically excited by He-flash driven convection are able to produce long-period pulsation similar to that observed in \lsiv.
This would place \lsiv\ in an evolutionary state immediately following one of the first He-core flashes, subsequent to either a late hot He-flash or the merging of two He-WDs.
Alternatively, \cite{saio19} showed that the pulsation of \lsiv\ could also be explained by carbon and oxygen opacity bumps, but would require very substantial C/O enrichment at temperatures around $10^6$ K.

Another striking peculiarity of \lsiv\ and  \feige\ is their chemical composition characterised by extreme overabundances of heavy metals.
\cite{nas11} found \lsiv\ to be enriched in strontium, yttrium, and zirconium, to the order of 10\,000 times the solar values.
A very similar abundance pattern was found in \feige\ by \cite{Latour2019b}.
Whether or not this atmospheric enrichment in heavy metals extends to the envelope, where it could influence the driving of pulsations via the $\kappa$-mechanism is not known.
Other recently discovered heavy-metal subdwarfs include the lead-rich iHe-sdOBs [CW83]\,0825+15 \citep{jeff17}, EC\,22536-4304 \citep{jeff19}, PG\,1559+048, and FBS\,1749+373 \citep{Naslim2020}.
This extreme enrichment compared to solar values is thought to be the result of strong atmospheric diffusion processes.
While the population of known heavy-metal subdwarfs continues to grow, it remains too small to relate the observed differences in enrichment to specific ranges in their atmospheric parameters.
In addition, theoretical diffusion calculations for iHe-sdOB stars are still lacking.

In this investigation, we focus on the determination and comparison of the detailed abundance patterns of \lsiv\ and \feige. 
We recently obtained high-resolution spectra for \feige\ at the ESO VLT, while archival spectra were retrieved for \lsiv.
A coarse inspection of the spectra showed that they were strikingly similar.
The same metal lines are detected in both stars at very similar strengths, indicating that the abundances are similar as well.
It is therefore tempting to study both stars jointly.

Before addressing the main aim of the study, we start with a short account of the recent TESS light curve of \feige\ in Sect.~\ref{sect:tess}.
Photometric measurements, \textit{Gaia} astrometry, and the spectroscopic surface gravity and effective temperature are combined to derive the mass, radius, and luminosity of each star in Sect.~\ref{sect:sed}.
In Sect.~\ref{sect:obs}, we give an overview of the available spectra. 
Our spectral analysis is described in Sect.~\ref{sect:analysis}.
We summarise our results in Sect.~\ref{sect:conclusions}.

\begin{table*}
\caption{List of modes detected in the TESS time series of Feige\,46.\label{tab:Frequencies}}
\centering{}%
\begin{tabular}{lllllr}
\hline 
\hline 
\noalign{\vskip5bp}
Frequency & Freq. change$^{a}$ & Period & Period change & Amplitude$^{b}$ & S/N\tabularnewline
($\mu$Hz) & ($\mu$Hz) & (s) & (s) & (\%) & \tabularnewline[5bp]
\hline 
\noalign{\vskip5bp}
$435.948\pm0.130^{c}$ & $+0.156\pm0.130$ & $2293.85\pm0.68^{c}$ & $-0.82\pm0.68$ & $0.133\pm0.014$ & $9.4$\tabularnewline
$435.573\pm0.050^{c}$ & $+0.045\pm0.050$ & $2295.83\pm0.26^{c}$ & $-0.23\pm0.26$ & $0.344\pm0.014$ & $24.1$\tabularnewline
$363.606\pm0.035$ &   & $2750.23\pm0.26$ &   & $0.097\pm0.014$ & $6.8$\tabularnewline
$362.625\pm0.019$ & $+0.009\pm0.019$ & $2757.67\pm0.14$ & $-0.07\pm0.14$ & $0.182\pm0.014$ & $12.8$\tabularnewline
$333.358\pm0.029$ & $-0.066\pm0.029$ & $2999.78\pm0.26$ & $+0.60\pm0.27$ & $0.117\pm0.014$ & $8.2$\tabularnewline
$294.017\pm0.016$ & $-0.039\pm0.016$ & $3401.17\pm0.18$ & $+0.46\pm0.19$ & $0.217\pm0.014$ & $15.3$\tabularnewline
\hline 
\end{tabular}
\tablefoot{
$^{a}$Relative to the measurement in \cite{Latour2019a}.
$^{b}$Given amplitude values are uncorrected for contamination (see text).
$^{c}$Formal fitting errors were increased by a factor of 5 to loosely account for the 
poorly resolved peaks.
}
\end{table*}

\section{The TESS light curve of Feige\,46}\label{sect:tess}

\feige\ was observed with TESS in Sector 22, from February 19 to
March 17, 2020. The light curve covers a time baseline of $26.59$
days sampled nearly continuously every 120 seconds, except for a four-day
interruption mid-run, which is typical of TESS data. 
It is therefore shorter 
than the light curve used in \cite{Latour2019a}, 
which was taken between February 26 and May 25, 2018 
(for a 87.76-day time baseline) with the Mont4K CCD camera at the 
1.55m Kuiper telescope of Steward Observatory on Mt Bigelow, resulting 
in a lower frequency resolution of $0.44$\,$\mu$Hz compared to $0.13$\,$\mu$Hz. 
However, the duty cycle is vastly improved with TESS and 
daily frequency aliases in Fourier transforms are no longer present.
Due to the large pixels of TESS ($\sim$21 arcsec), 
the light curve is likely affected by a slightly fainter visual companion
located 13.3 arcsec northwest of Feige\,46 ($G_{{\rm RP}}=13.95$ compared 
to $13.55$ for Feige\,46). 
According to the TESS contamination indicator 
(crowdsap), only $61.4\%$ of the collected light is attributed 
to Feige\,46. 
Assuming the contaminating star is not variable, this 
blend only affects the measured amplitudes of Feige\,46 brightness 
variations, which have to be scaled up by a factor of $1.63$. 
The 
pulsation frequencies remain unaffected. 

\begin{figure}
\begin{centering}
\includegraphics[width=1\columnwidth]{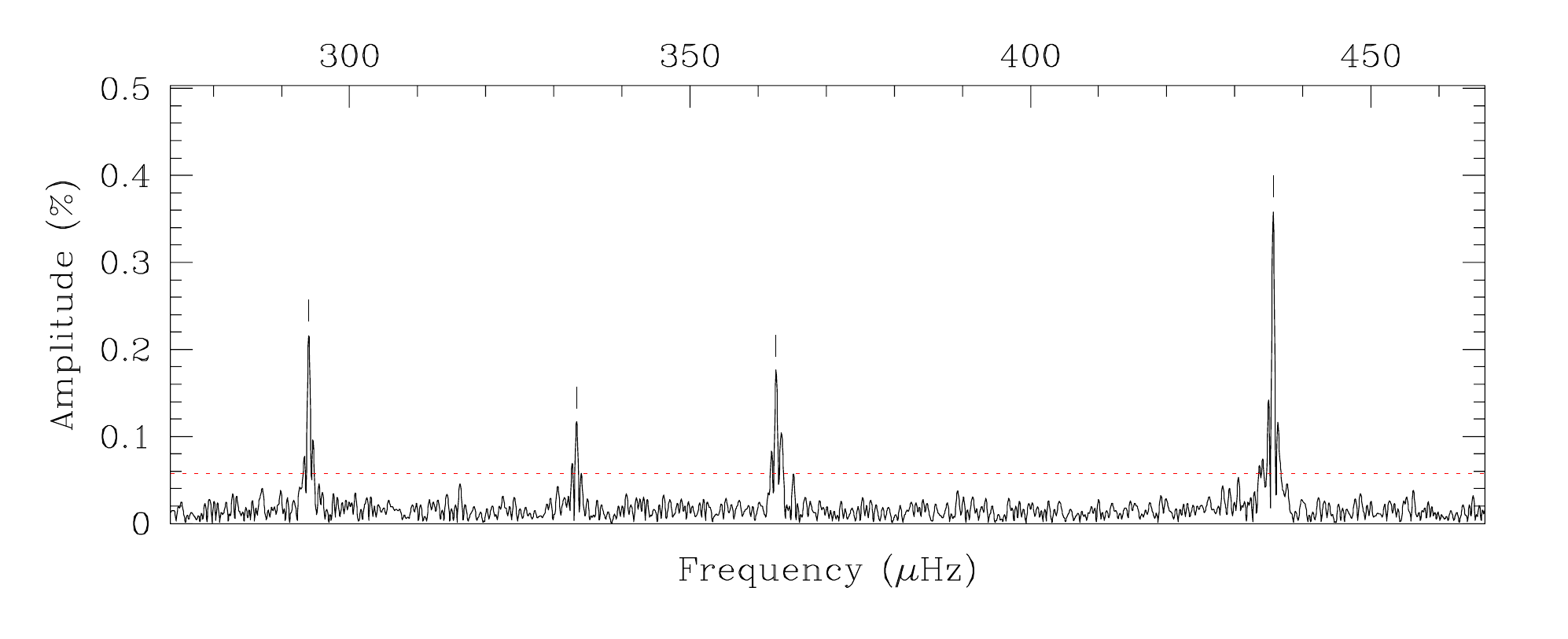}
\par\end{centering}
\begin{centering}
\includegraphics[width=1\columnwidth]{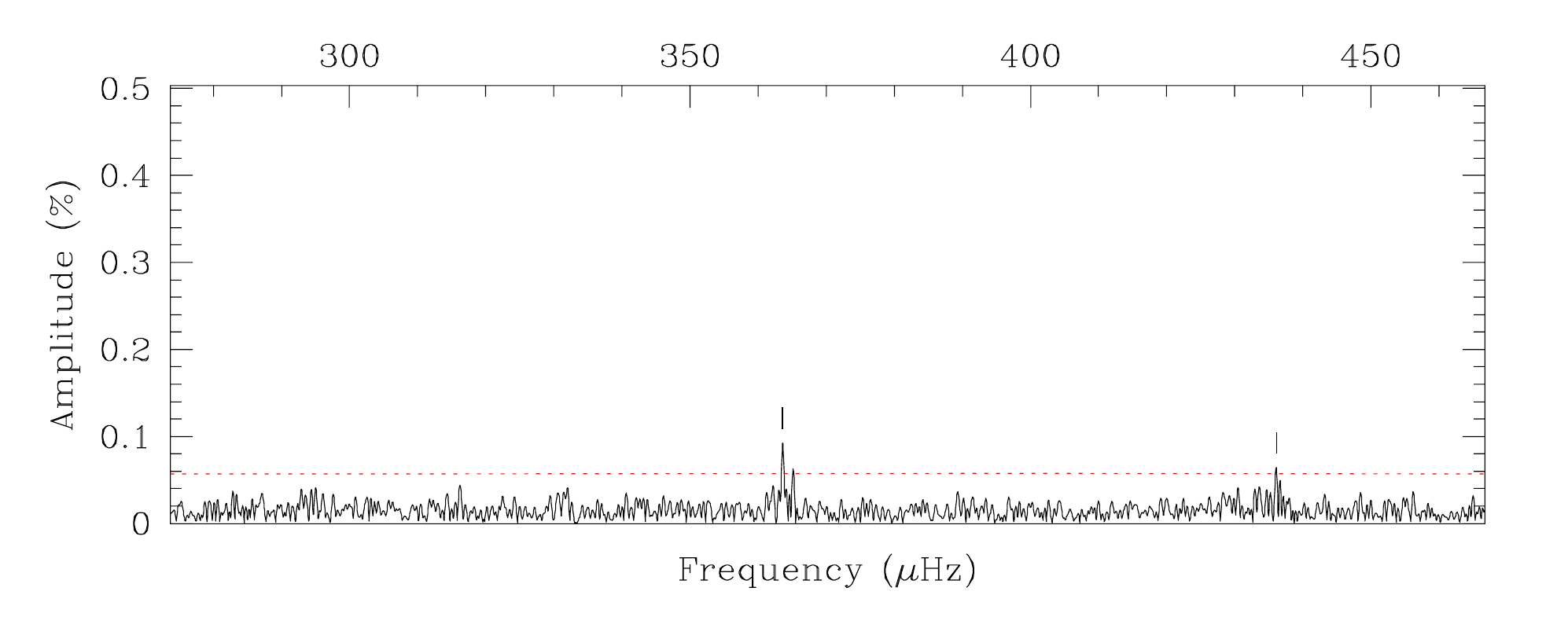}
\par\end{centering}
\begin{centering}
\includegraphics[width=1\columnwidth]{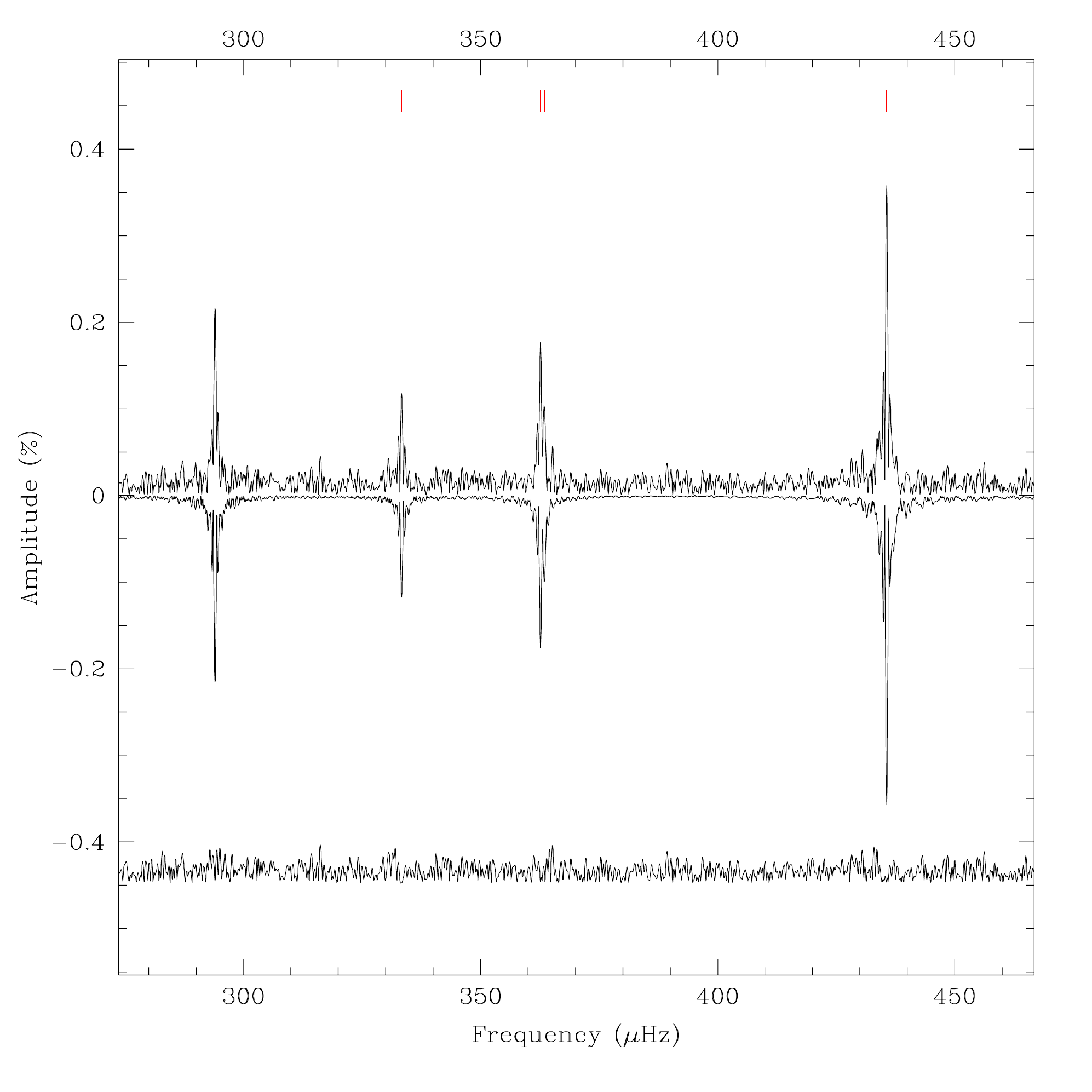}
\par\end{centering}
\caption{Pre-whitening sequence of detected pulsation modes. \emph{\emph{Top panel:}}
Lomb-Scargle Periodogram of the TESS time series in the relevant $270-420$
$\mu$Hz frequency range. The horizontal dotted line indicates four times
the median noise level and corresponds to the chosen detection limit.
\emph{\emph{Middle panel:}} residual periodogram after pre-whitening the four dominant
peaks identified in the top panel. Two additional low-amplitude peaks
are identified. \emph{\emph{Bottom panel:}} from top to bottom are the original, reconstructed
(from the frequencies, amplitudes, and phases of the six fitted sine
waves; plotted upside down) and residual periodograms after completing
the pre-whitening process. \label{fig:TESS-F46-Prewhitening}}
\end{figure}

Figure \ref{fig:TESS-F46-LSP} illustrates the TESS observations obtained
for Feige\,46. A close-up view of the light curve (middle panel) suggests
the presence of periodic light modulations, which become clearly apparent
in the Lomb-Scargle periodogram (LSP; bottom panel), which covers the
entire frequency range (in log scale) accessible to these data (i.~e.~up to the Nyquist frequency limit corresponding to the 120\,s sampling).
Significant peaks are found in the $250-500$ $\mu$Hz frequency
range, where the pulsations were indeed expected, while nothing above a
4-$\sigma$ detection threshold emerges elsewhere in the spectrum. We
extracted the periodic modulations in Feige\,46 by following a similar
approach to \cite{Latour2019a}, using a standard Fourier analysis and pre-whitening techniques (see, e.~g.~\citealt{Billeres2000}). 
This was accomplished efficiently with our dedicated
time-series analysis software, FELIX (\citealt{charpinet2010}; \citealt{Zong2016}). 
The entire pre-whitening procedure is illustrated in
Fig. \ref{fig:TESS-F46-Prewhitening}, and the extracted mode parameters
are listed in Table \ref{tab:Frequencies}.

Most peaks, except the largest amplitude one, are easily reproduced
by fitting a pure sinusoidal component to the time series, leaving
no residual behind in Fourier space. For the largest peak around 453\,$\mu$Hz, however, a single frequency fit leaves a significant residual,
and we find its structure to be better reproduced when a blend of
two close, poorly resolved frequency components is instead assumed. 
We favour this solution considering that this peak was clearly resolved 
in two independent components, interpreted as a possible rotational multiplet, 
in \cite{Latour2019a}. 
Due to this blend, the frequencies of these two components are less accurately 
measured in the TESS data (see Table \ref{tab:Frequencies}). 
Overall, we detect five of the six
periods found by \cite{Latour2019a}. Their peak with the
lowest amplitude, at 2586\,s, is not visible in the TESS run. However,
we find a new period at 2750\,s that is close to the already known
period at 2758\,s. These modes are separated by $\sim$1\,$\mu$Hz and
might be part of a rotation multiplet, but better data are required
to reliably detect and identify rotational splitting in this star.

Since the TESS light curve and that of \cite{Latour2019a}
were obtained, on average, 698 days apart, it is possible to constrain
a potential period decay, as predicted by \cite{bat18}.
These authors propose that pulsators like Feige\,46 are fast evolving
stars experiencing helium sub-flashes before reaching the extreme
horizontal branch. The pulsations would be driven by the $\epsilon$-mechanism associated with these sub-flashes. \cite{bat18} predicted period changes in the $10^{-5}-10^{-7}$ s/s range for late hot flasher
models, with the fastest rates corresponding to pulsations driven
by the first He-flash and the slowest rates corresponding to subsequent
flashes (see their Table 3). The longest periods ($\sim$2000\,s) are
only excited during the first one or two He-flashes, and would therefore
be associated with the fastest period changes. In this context, the
periods observed in Feige\,46, which are all longer than 2000\,s, would
be expected to change rapidly, at a rate close to $\sim$10$^{-5}$\,s/s.
We do find period differences between the two observing runs (see Table
\ref{tab:Frequencies}), but these are difficult to associate with
certainty to secular variations, because of frequency resolution limitations
and the likely presence of poorly resolved rotational splittings.
The important finding, however, is that these period variations are,
at most, of the order of one second (to be conservative), which would
correspond to a rate of period change of $\sim$10$^{-8}$\,s/s or less. 
This is orders of magnitude slower than the rates predicted by \cite{bat18}. 
In other words, if the periods were to change at $10^{-5}$\,s/s, this effect would by far dominate the period differences between
the two epochs of observation, but this is clearly not observed in
Feige\,46. \\
It would be interesting to check for period decay in \lsiv\ as well. We are aware of three photometric observation runs, performed in 2004 \citep{ahm05}, 2005 \citep{jeff11}, and 2010 \citep{green11}. Comparing the periods found by \cite{ahm05} and \cite{jeff11} with those stated in \cite{green11} suggests an upper limit on the rate of period change of $\sim$10$^{-6}$\,s/s or less, which is less than what was predicted by \cite{bat18}. A future consistent analysis of all data sets might improve on this upper limit, especially since \lsiv\ is scheduled to be observed with the CHEOPS satellite.

\section{Parallax, spectral energy distribution, and stellar parameters}\label{sect:sed}

The \textit{Gaia} mission recently provided parallaxes for a large number of hot subdwarf stars. This allows atmospheric parameters  to be converted to the fundamental stellar parameters: mass, radius, and luminosity, without relying on predictions from evolutionary models. 
The parallax measurements for \lsiv\ and \feige\ are of excellent quality, with uncertainties of less than 5\%. 
In addition, photometry is required to derive the angular diameters ($\theta$) of the stars. 
\begin{figure*}
\sidecaption
\includegraphics[width=11.7cm]{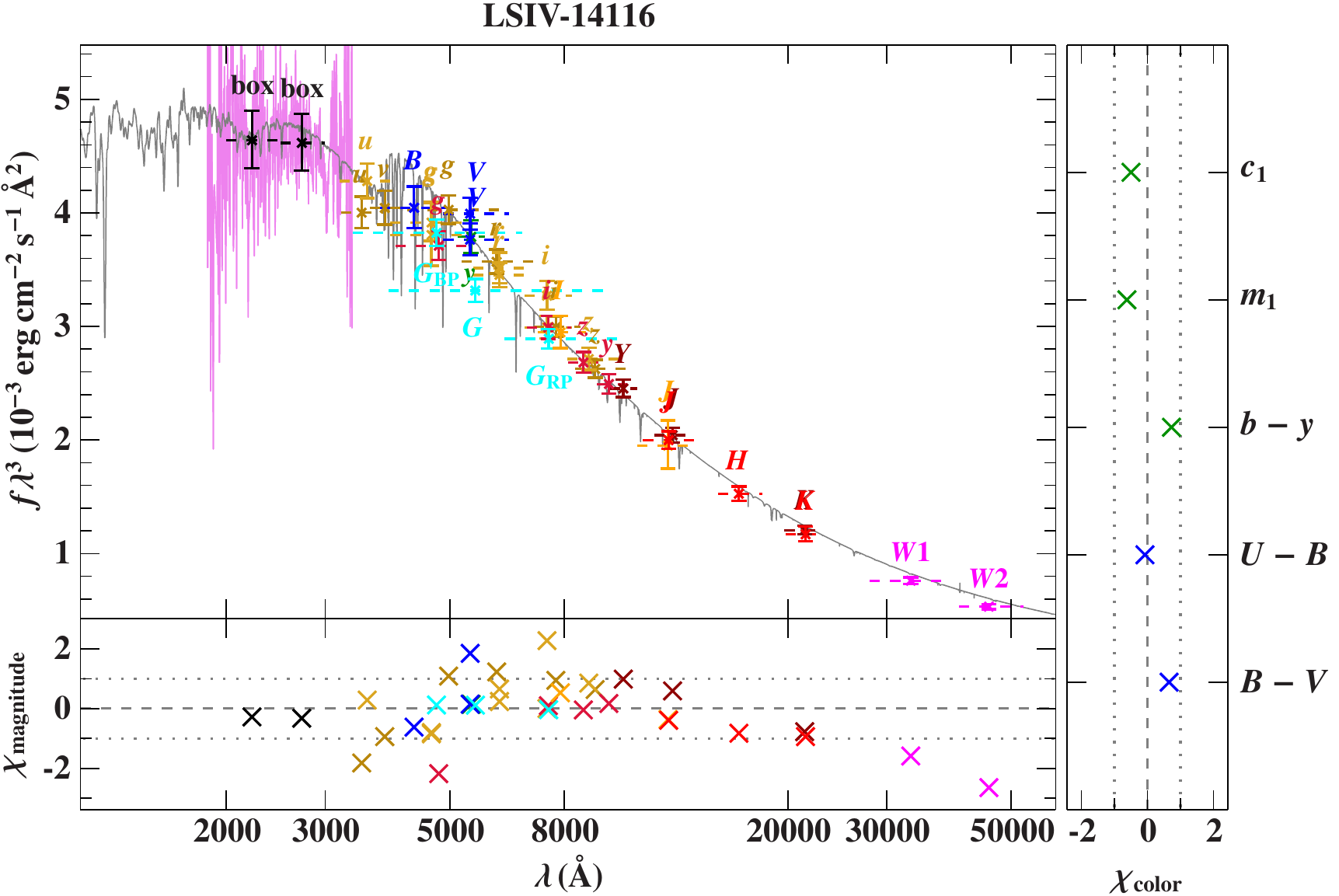}
\caption{Comparison of smoothed final synthetic spectrum of \lsiv\ (grey line) with photometric data.
The two black data points labelled `box' are binned fluxes from an IUE spectrum \citep[LWP10814LL, magenta line,][]{Wamsteker2000_INES}. 
Filter-averaged fluxes are shown as coloured data points that were converted from observed magnitudes (the dashed horizontal lines indicate filter widths). 
The residual panels at the bottom and on the right sides, respectively, show the differences between synthetic and observed magnitudes/colours.
The following colour codes are used to identify the photometric systems: SDSS \cite[yellow,][]{Henden_2015,SDSS_2015}, SkyMapper \cite[dark yellow,][]{Wolf_2018}, Pan-STARRS1 \cite[red,][]{2016arXiv161205560C}, Johnson-Cousins \cite[blue,][]{Henden_2015,O'Donoghue_2013}, Str\"omgren \cite[green,][]{Hauck_1998}, \textit{Gaia}  \cite[cyan,][]{Gaia2018_VizieR}, VISTA \cite[dark red,][]{McMahon_2013}, DENIS \cite[orange,][]{vizier:B/denis}, 2MASS \cite[bright red,][]{Cutri2003_2MASS}, and WISE \cite[magenta,][]{Schlafly_2019}.}
\label{fig:sed_fit}
\end{figure*}
We combined apparent magnitudes from the ultraviolet to the infrared to construct the observed spectral energy distribution (SED) of \lsiv\ (see Fig.~\ref{fig:sed_fit}). 
Our final synthetic spectrum of \lsiv\ was then  scaled to fit this SED using $\chi^2$ minimisation based on the method described by \cite{2018OAst...27...35H}.
Interstellar reddening is considered after \cite{Fitzpatrick2019}, assuming an extinction parameter $R\,(55)=3.02$.
Fit parameters are the angular diameter $\theta$ and $E$(44$-$55), which is the monochromatic analogon of the colour excess $E$($B$--$V$).
To derive the stellar radius $R$, the \textit{Gaia} parallax $\varpi$ is combined with the resulting angular diameter $\theta = 2R\varpi$. 
The stellar mass $M$ is then derived using the spectroscopic surface gravity $g = GM/R^2$, where $G$ is the gravitational constant ($\log g = 5.85$ for \lsiv).
The stellar luminosity $L$ is based on the spectroscopic effective temperature ($T_\mathrm{eff} = 35500$\,K).
We repeated the SED fit for \feige using $\log g = 5.93$ and $T_\mathrm{eff} = 36100$\,K (see Fig.~\ref{fig:sed_feige}).
The atmospheric parameters used are the same as those used for the spectroscopic analysis and are  described in Sect.~\ref{sect:methods}.
For both stars, we assume systematic errors of 0.1\,dex in $\log g$ and 1000\,K in \teff.
The results of this analysis are listed in Table \ref{tab:SED}.
The derived stellar mass for \lsiv\ ($0.38\pm 0.10$\,$M_\odot$) is somewhat less than the value obtained for \feige\ ($0.53\pm 0.14$\,$M_\odot$). Given the uncertainties both masses are consistent with the canonical mass suggested by evolution models \citep[$\sim$0.46\,$M_\odot$,][]{dor93,2003MNRAS.341..669H}. 

\begin{table}
\setstretch{1.2}
\caption{Parallax and parameters derived from the SED fitting.
The atmospheric parameters \teff\ and $\log g$ are derived from spectroscopy and discussed in Sect.~\ref{sect:methods}.}
\label{tab:SED}
\vspace{-8pt}
\begin{center}
\begin{tabular}{lcc}
\toprule
\toprule
 & \lsiv  & \feige\\ 
\midrule
$\varpi$\,(mas)    &  $2.38  \pm 0.09$  &  $1.86  \pm 0.07$  \\
$d$\,(pc)               & $420 \pm 15$  & $538 \pm 19$          \\
$\theta\,(10^{-11}\,\mathrm{rad})$      & $1.310 \pm 0.007$     & $1.093 \pm 0.008$              \\
$E$(44$-$55)            & $0.033  \pm 0.005$    & $0.011 \pm 0.005$     \\
\teff\,(K)    &  $35500  \pm 1000$  &  $36100 \pm 1000$  \\
$\log g$    &  $5.85  \pm 0.10$  &  $5.93  \pm 0.10$  \\
$R/R_\odot$             & $0.122  \pm 0.005$    & $0.130\pm 0.006$      \\
$M/M_\odot$             & $0.38\pm 0.10$        & $0.53 \pm 0.14$\\
$L/L_\odot$     & $21\pm 3$ & $26\pm4$  \\
\bottomrule
\end{tabular}
\end{center}
\end{table}

\section{Spectroscopic observations}\label{sect:obs}
\begin{table}
\setstretch{1.2}
\captionof{table}{UVES spectra used in the present analysis. For \lsiv, only spectra with sufficient S/N for cross-correlation were used. Total exposure times are given per wavelength range and resolution.}
\label{tab:obs}
\vspace{-11pt}
\begin{center}
\resizebox{\columnwidth}{!}{
\begin{tabular}{l c c c c c}
\toprule
\toprule
Star & Range / \AA  & R  &  $n_\mathrm{exp}$ &  $\sum t_\mathrm{exp}$ / s & Run ID \\ 
\midrule
\feige\ &    $3305-4525$   &   40970    &   \phantom{0}4   & \phantom{0}5920 & 0104.D-0206(A) \\
        &    $4620-6645$   &   42310    &   \phantom{0}4   & \phantom{0}5920 & \\
\lsiv\  &    $3290-4525$   &   40970    &   12             & \phantom{0}3600 & 087.D-0950(A) \\
        &    $4788-6835$   &   42310    &   15             & \phantom{0}4500 & \\
        &    $3290-4525$   &   49620    &   18             & \phantom{0}3600 & 095.D-0733(A) \\
        &    $4788-6835$   &   51690    &   18             & \phantom{0}3600 & \\
        &    $3290-4525$   &   58640    &   64             & 12800           & \\
        &    $4788-6835$   &   66320    &   71             & 14200           & \\
\bottomrule 
\end{tabular}
}
\end{center}
\end{table}
We obtained four VLT/UVES spectra of \feige\ in February 2020 with a total exposure time of 5920\,s (ID 0104.D-0206(A)).
These spectra have a resolution of $R$ $\approx$ 41000 and cover the spectral range from 3305\,\AA\ to 6645\,\AA,\ with gaps at 4525 -- 4620 \AA\ and 5599 -- 5678 \AA.
The individual spectra were stacked after cross-correlation to obtain a single spectrum with an increased  signal-to-noise ratio (S/N) of about 80.
The radial velocity obtained, $v_\mathrm{rad}=89$ km\,s$^{-1}$, is fully consistent with the value found by \cite{dri87}: $90\pm 4$ km\,s$^{-1}$.
For the spectral analysis, the observed spectrum was shifted to the stellar rest frame.
We refer the reader to \cite{Latour2019b} for the description and analysis of older spectra of \feige, including ultraviolet (UV) observations.

\lsiv\ has been observed extensively with the UVES spectrograph. 
A total of 788 spectra are available in the ESO archive (corresponding to 394 exposures).
Spectra were taken as part of two programmes: on 7 September, 2011 (ID 087.D-0950(A)) and between 23 and 27 August, 2015 (ID 095.D-0733(A)).
These programs used time-resolved spectroscopy in order to relate the observed photometric variability to radial velocity variations \citep{jeff15,Martin2017}.
We combined spectra from both runs to create a high-S/N spectrum that is suitable for a detailed abundance analysis.
For each resolution, spectra with the highest S/N (typically 16 to 25) were cross-correlated and stacked.
These stacked spectra were convolved to the lowest common resolutions ($R=40970$ for the blue range, and $R=42310$ for the red range) and were then co-added.
We then shifted the spectrum to the stellar rest frame, correcting for the high radial velocity of about $v_\mathrm{rad}=-154$ km\,s$^{-1}$.
The final spectrum has a mean effective S/N of about 200, which is limited by small-scale artefacts.
Details of the UVES spectra used in the present analysis are given in Table \ref{tab:obs}.

\cite{ran15} carried out spectropolarimetry of \lsiv\ with VLT/FORS2 to search for a magnetic field.
While no polarisation could be detected, their observations produced a flux spectrum of excellent quality (spectral resolution $\Delta \lambda \approx 1.8$\,\AA, S/N $\approx$ 700).
In contrast to the UVES spectra, this long-slit spectrum is not affected by the normalisation issues that frequently occur in the reduction procedure of Echelle spectra.
The FORS2 spectrum is therefore useful for determining atmospheric parameters based on broad hydrogen and helium lines.

\begin{table}
\setstretch{1.2}
\captionof{table}{Sources of oscillator strengths for detected lines of heavy metals in \feige\ and \lsiv.
}
\label{tab:fsource}
\vspace{-11pt}
\begin{center}
\begin{tabular}{lrr}
\toprule
\toprule
 Ion     & $N_\mathrm{ident}$  & Reference      \\ 
\midrule
Ga\,\textsc{iii}        & 9 & \cite{oreilly98}                          \\
Ge\,\textsc{iii}        & 3 & \cite{nas11}              \\
Ge\,\textsc{iv}         & 6 & \cite{oreilly98}          \\
Kr\,\textsc{iii}        & 17 & \cite{Raineri1998}                               \\
Sr\,\textsc{ii}         & 2 & \cite{Fernandez2020}                              \\
                        & 3 & Kurucz/Linelists \\
Sr\,\textsc{iii}        & 35 & Kurucz/Atoms\\
Y\,\textsc{iii}         & 2 & \cite{nas11}                              \\
        & 3 & \cite{Fernandez2020}                              \\
Zr\,\textsc{iii}        & 2 & Kurucz/Linelists \\
Zr\,\textsc{iv}         & 16 &\cite{Rauch2017}                          \\
Sn\,\textsc{iv}         & 2 & \cite{Kaur2020}                           \\
Pb\,\textsc{iv}         & 1 & \cite{safronova04}                                \\
\bottomrule
\end{tabular}
\end{center}
\end{table}

\section{Spectroscopic analysis}
\label{sect:analysis}
The excellent UVES spectra enable a detailed abundance analysis, as well as a consistent comparison of abundances between \lsiv\ and \feige, which is described in the following section.

\subsection{Methods}\label{sect:methods}

To minimise systematic errors, we analysed the spectra of both \feige\ and \lsiv\ using the same fitting method and the same type of model atmospheres, following the procedure described in \cite{Latour2019b} and \cite{dorsch2019}.
This analysis is based on model atmospheres and synthetic spectra computed using the hydrostatic, homogeneous, plane-parallel, non-local thermodynamic equilibrium (NLTE) codes \textsc{Tlusty} and \textsc{Synspec} \citep{hubeny88, Lanz2003, hub11}.
We used the most recent public versions as described in \cite{hubeny17a,hubeny17b,hubeny17c}.

Our line list is based on atomic data provided by R.~Kurucz.\footnote{\url{http://kurucz.harvard.edu/linelists/gfnew/gfall08oct17.dat}; see also \cite{Kurucz2018}.\label{note1}}
We extended this line list to include lines from additional heavy ions. 
The atomic data previously collected are described in \cite{dorsch2019} and \cite{Latour2019b}.
This list was further extended to model the rich spectrum of \feige.
The main sources for detected lines of heavy ions are listed in Table \ref{tab:fsource}.
Heavy elements (here $Z>30$) in ionisation stages \textsc{i-iii} are included in LTE using the treatment of \cite{Proffitt2001}, who added ionisation energies and partition functions from R.~Kurucz's ATLAS9 code \citep{Kurucz1993} to \textsc{Synspec}.
Partition functions for higher ionisation stages are calculated as described in \cite{Latour2019b}.

As in our previous analysis of \feige,  all model atmospheres were calculated using the atmospheric parameters derived by \cite{Latour2019a} (\teff\ = 36\,100 K, $\log g$ = 5.93, and a helium abundance of $\log \epsilon_\mathrm{He} / \epsilon_\mathrm{H} = -0.32$).
Atmospheric parameters for \lsiv\ were derived by \cite{ran15} based on a high S/N FORS2 spectrum (\teff\ = 35\,150 K, $\log g$ = 5.88, $\log \epsilon_\mathrm{He} / \epsilon_\mathrm{H} = -0.62$).
We used a grid of line-blanketed NLTE models to re-fit the same FORS2 spectrum, and we obtained \teff\ = 35\,500 K, $\log g$ = 5.85, $\log \epsilon_\mathrm{He} / \epsilon_\mathrm{H} = -0.60$, which is fully compatible with the results of \cite{ran15}.
The model grid used for this fit includes H, He, C, N, O, Ne, Mg, Al, Si, and Fe in NLTE with abundances appropriate for \lsiv.

Using the atmospheric parameters reported above for each star, we then constructed series of models, varying the abundance of one element at a time.
These models also include nickel in NLTE.
Based on these grids, we determined metal abundances using the $\chi^2$-fitting program SPAS developed by \cite{hirsch09}.

Both \feige\ and \lsiv\ show slightly broadened lines that are best reproduced at a projected rotational velocity of $v_\mathrm{rot} \sin i = 9$\,km\,s$^{-1}$. 
This broadening might not be caused solely by rotation, but instead likely results from unresolved (high-order) pulsations.
Indeed, \cite{jeff15} found that the principal pulsation mode in \lsiv\ (1950\,s) leads to radial velocity variations with a semi-amplitude of about 5.5\,km\,s$^{-1}$.
They also came to the conclusion that other pulsation periods lead to additional unresolved motion.
Similar variability could be present in \feige, which would explain the observed broadening given that the UVES exposure times (1480\,s) cover a significant fraction of the shortest period observed in \feige\ (2295\,s).
However, the exposure times of the UVES spectra of \lsiv\ were much shorter (200\,s or 300\,s).
The remaining broadening (despite cross-correlating individual exposures before co-adding) may be explained by a combination of uncertainties in the cross-correlation, high-order pulsations, unresolved motion due to multiple periods, and actual rotation.
\cite{jeff15} also find evidence for differential pulsation: line strength and pulsation amplitude might be correlated.
Therefore, correlating single spectra using specific strong lines would not perfectly mitigate the broadening in the stacked spectrum for weak lines.
However, differential pulsation was not confirmed in the radial velocity study of \cite{Martin2017}.
Additional broadening may be caused by microturbulence ($v_\mathrm{tb}$). 
However, as shown by \cite{Latour2019b}, a microturbulence of 5\,km$^{-1}$ is too high to simultaneously reproduce UV and optical lines in \feige.
We therefore adopted $v_\mathrm{tb}=2$\,km\,s$^{-1}$ for both stars, which leads to negligible broadening.

\begin{figure*}

        \includegraphics[width=1\textwidth]{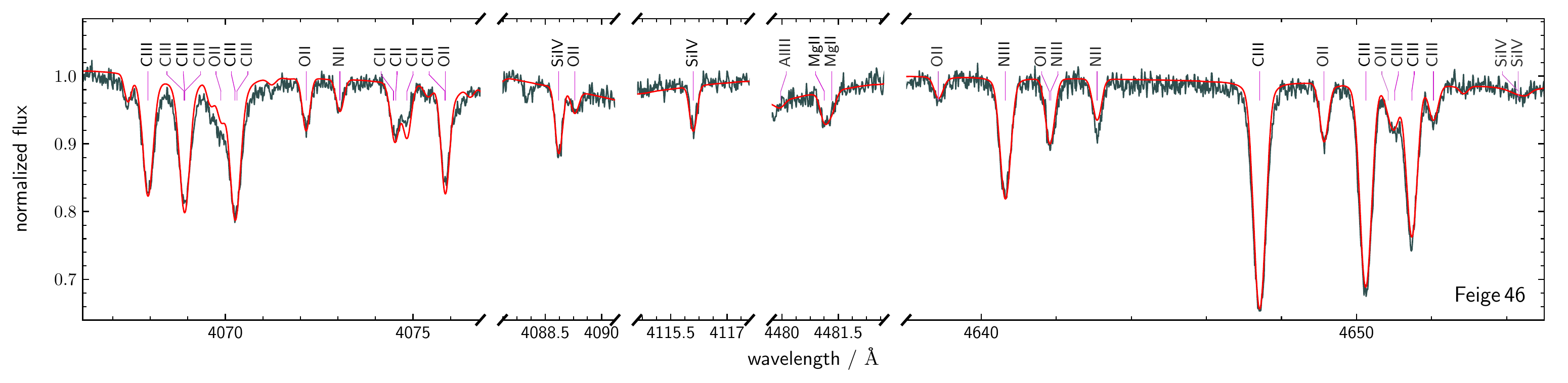}\vspace{-15pt}
        \includegraphics[width=1\textwidth]{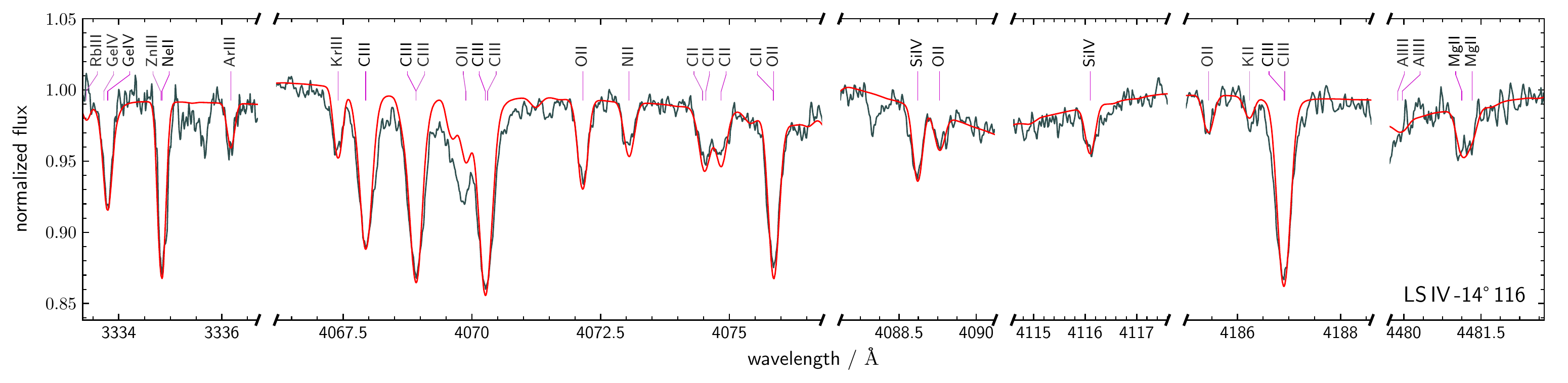}\vspace{-15pt}
        \includegraphics[width=1\textwidth]{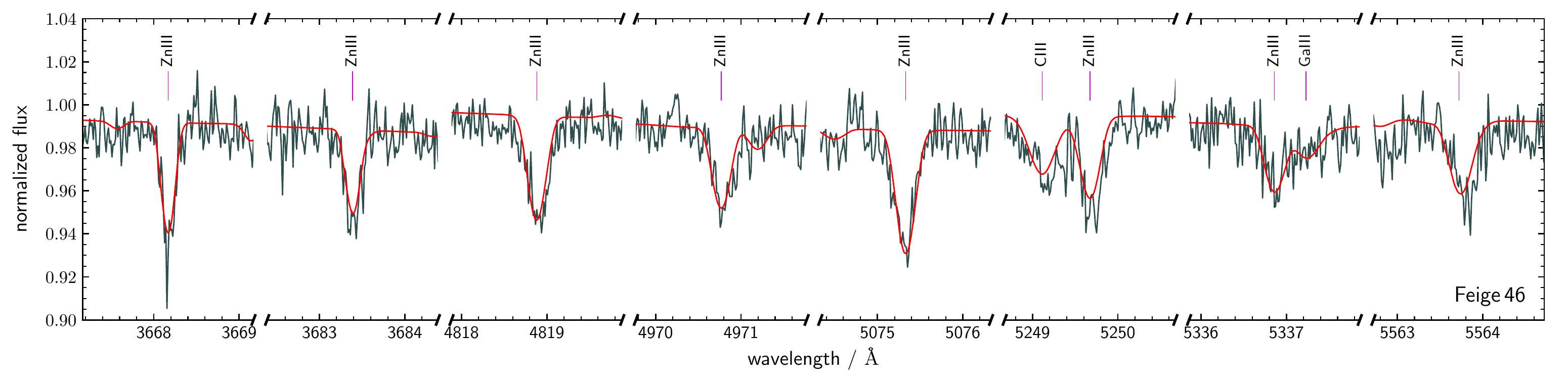}\vspace{-15pt} 
        \includegraphics[width=1\textwidth]{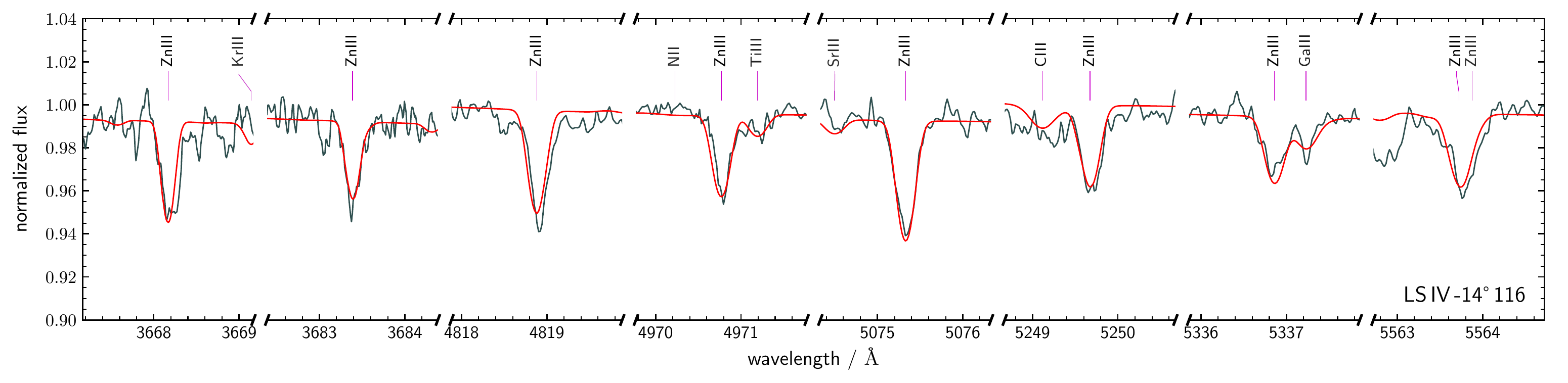}\vspace{-15pt}
        \includegraphics[width=1\textwidth]{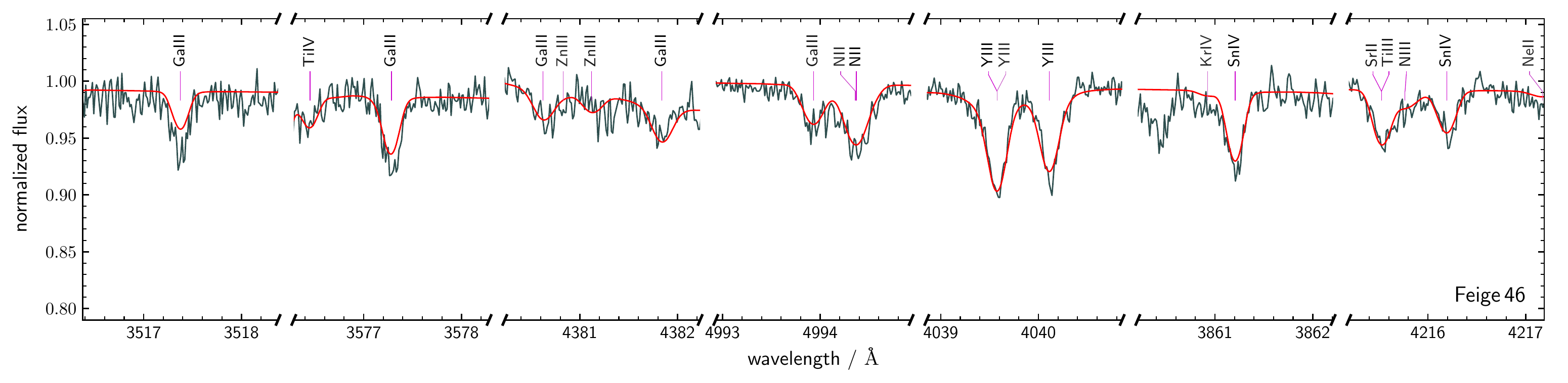}\vspace{-15pt}
        \includegraphics[width=1\textwidth]{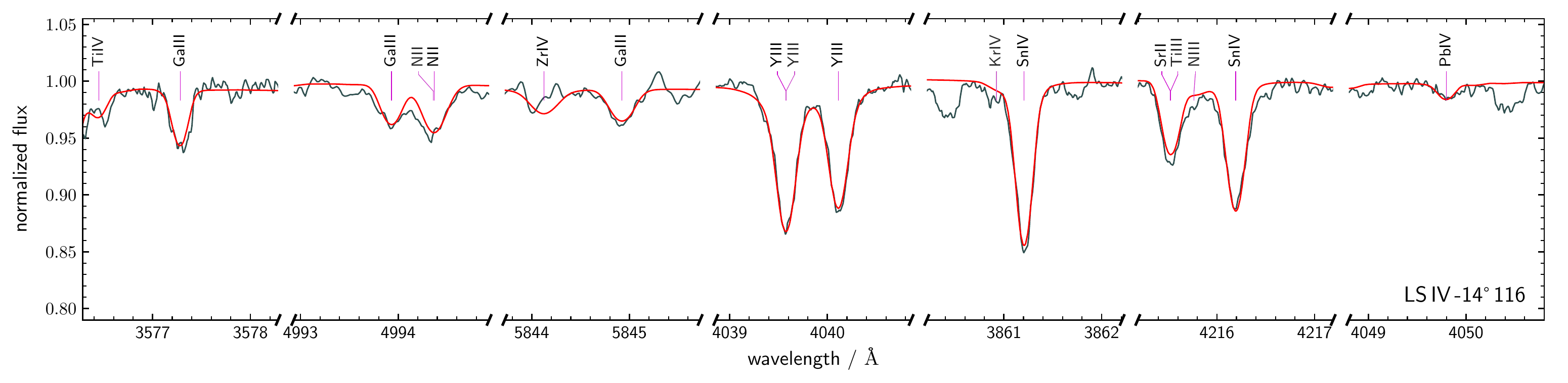}\vspace{-6pt}
        \caption{Representative regions in the UVES spectra of \feige\ and \lsiv. The best fit models are shown in red.}
        \label{lines_1}
\end{figure*}

\begin{figure*}
        \includegraphics[width=1\textwidth]{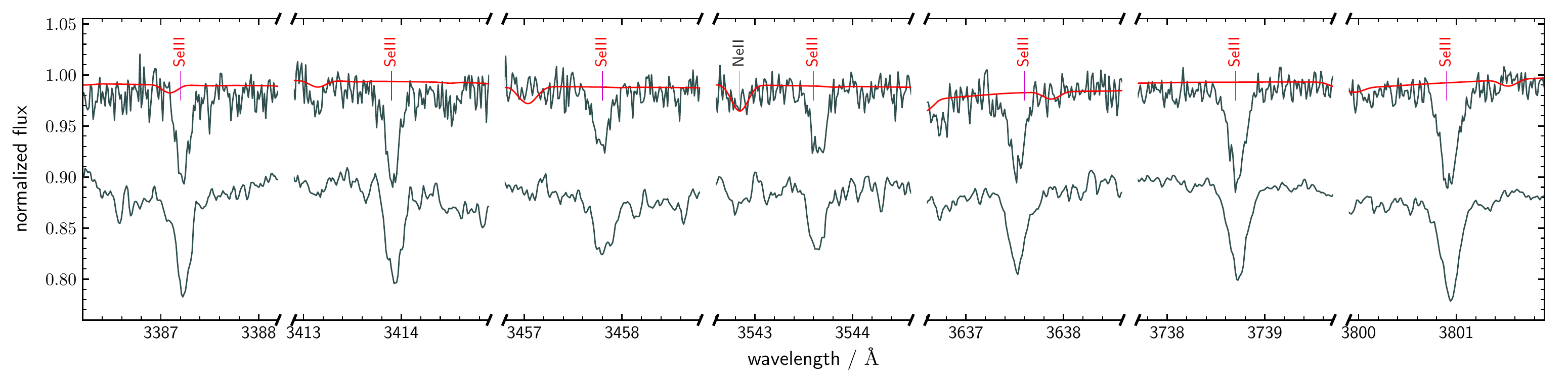}\vspace{-15pt}
        \includegraphics[width=1\textwidth]{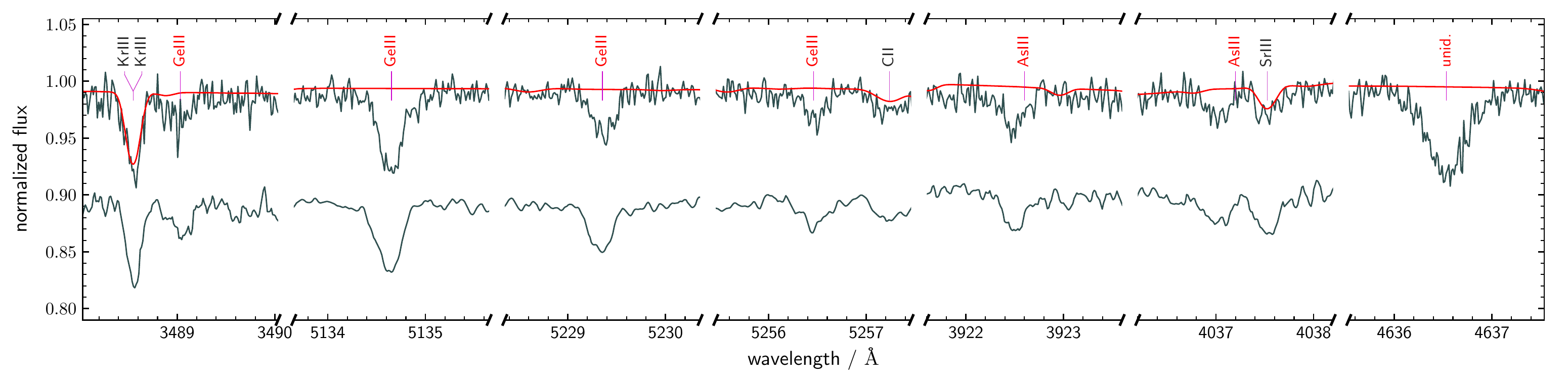}\vspace{-6pt}
        \caption{Additional regions in the UVES spectrum of \feige\ showing newly identified lines that are lacking oscillator strengths and the strongest unidentified line observed. 
        The UVES spectrum of \lsiv\ is shown for comparison, offset by $-0.1$.}
        \label{lines_unid}
\end{figure*}

\begin{table}
\centering
\caption{Updated line positions. Observed positions are accurate to about 0.02\,\AA\ depending on the specific line strengths. }
\label{table_lshift}
\setstretch{1.1}
\begin{tabular}{l c c c}
\toprule
\toprule
Ion  &  $\lambda_\mathrm{lit}$ / \AA & $\lambda_\mathrm{obs}$ / \AA & $\Delta \lambda$ / \AA \\
\midrule
\ion{Zn}{iii}   &   5075.243   &   5075.330 & $+0.087$ \\
\ion{Zn}{iii}   &   5157.431   &   5157.580 & $+0.149$ \\[2pt]
\ion{Ge}{iii}   &   4178.960   &   4179.078 & $+0.118$ \\[2pt]
\ion{Ge}{iv}    &   3320.410   &   3320.530 & $+0.120$ \\
\ion{Ge}{iv}    &   3333.640   &   3333.785 & $+0.145$ \\
\ion{Ge}{iv}    &   3554.190   &   3554.257 & $+0.067$ \\
\ion{Ge}{iv}    &   3676.650   &   3676.735 & $+0.085$ \\
\ion{Ge}{iv}    &   4979.190   &   4979.987 & $+0.797$ \\
\ion{Ge}{iv}    &   5072.900   &   5073.330 & $+0.430$ \\[2pt]
\ion{Kr}{iii}   &   3311.540   &   3311.490 & $-0.050$ \\
\ion{Kr}{iii}   &   3474.750   &   3474.650 & $-0.100$ \\[2pt]
\ion{Sr}{iii}   &   3976.706   &   3976.033 & $-0.673$ \\
\ion{Sr}{iii}   &   3991.587   &   3992.272 & $+0.685$ \\[2pt]
\ion{Y}{iii}    &   4039.602   &   4039.576 & $-0.026$ \\[2pt]
\ion{Zr}{iv}    &   5462.333   &   5462.380 & $+0.047$ \\
\ion{Zr}{iv}    &   5779.843   &   5779.880 & $+0.037$ \\[2pt]
\ion{Sn}{iv}    &   3862.051   &   3861.207 & $-0.844$ \\
\ion{Sn}{iv}    &   4217.184   &   4216.192 & $-0.992$ \\
\bottomrule
\end{tabular}
\end{table}

\begin{figure*}
        \includegraphics[width=1\textwidth]{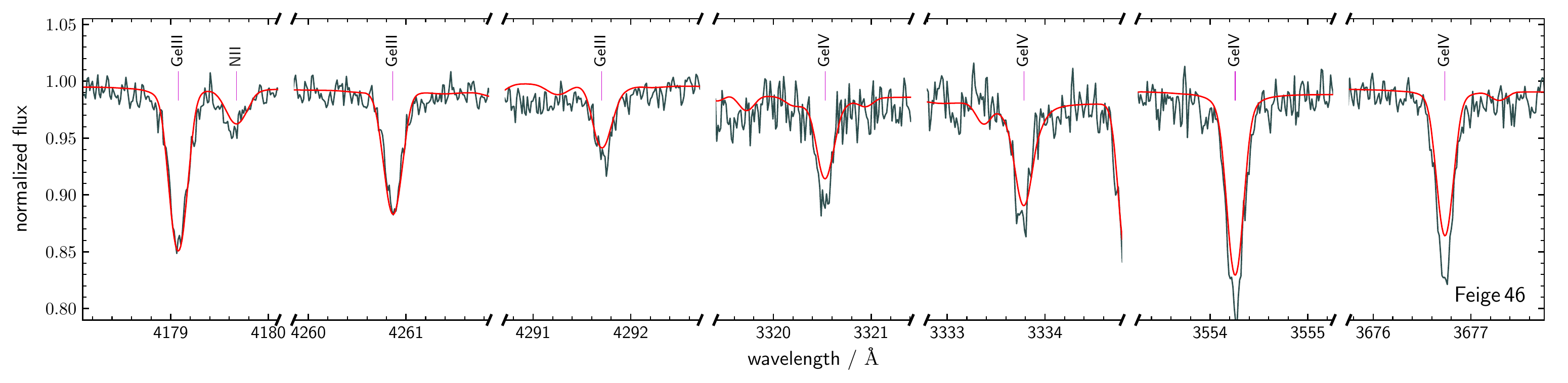}\vspace{-15pt}
        \includegraphics[width=1\textwidth]{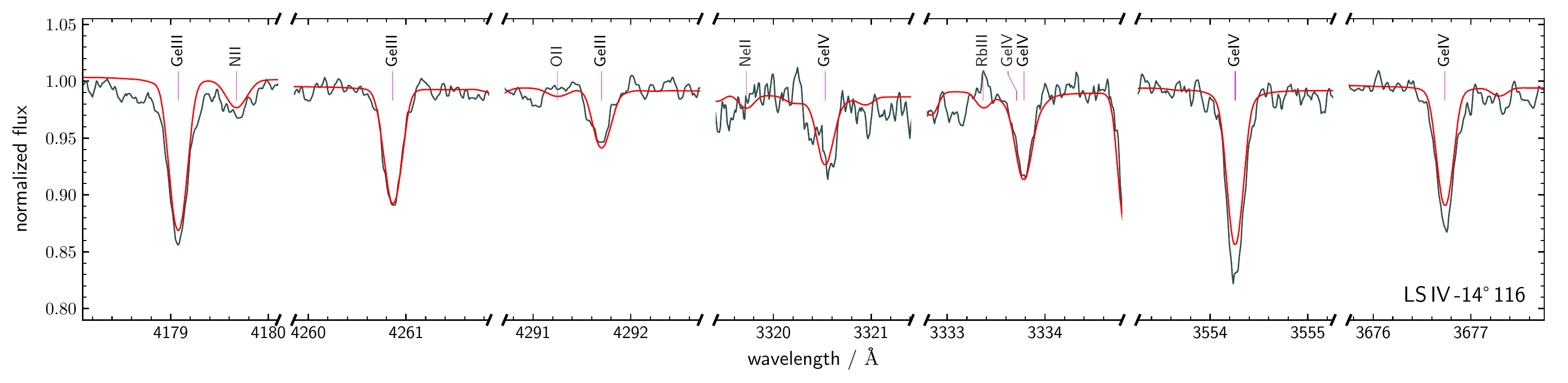}\vspace{-15pt}
        \includegraphics[width=1\textwidth]{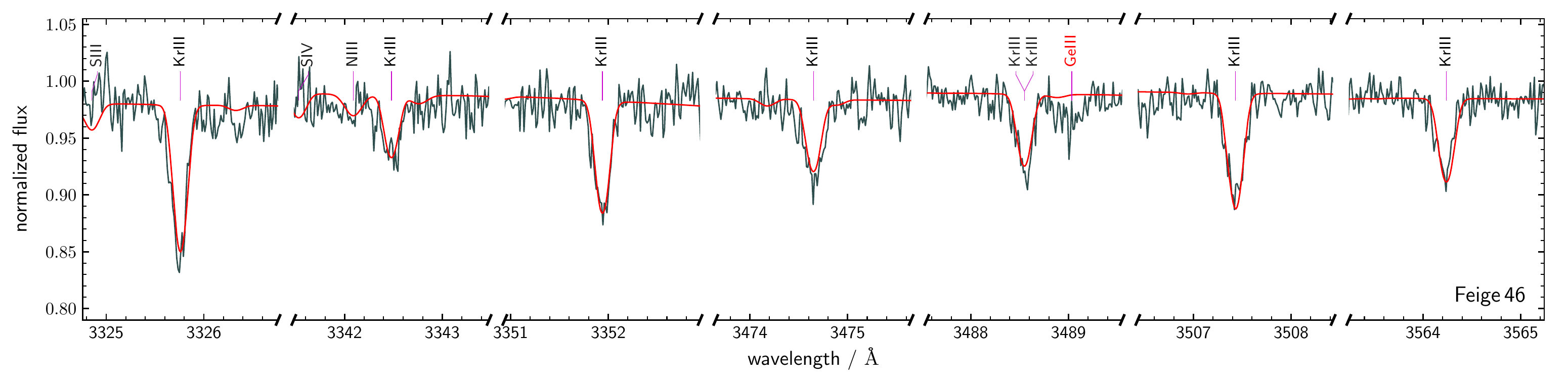}\vspace{-15pt}
        \includegraphics[width=1\textwidth]{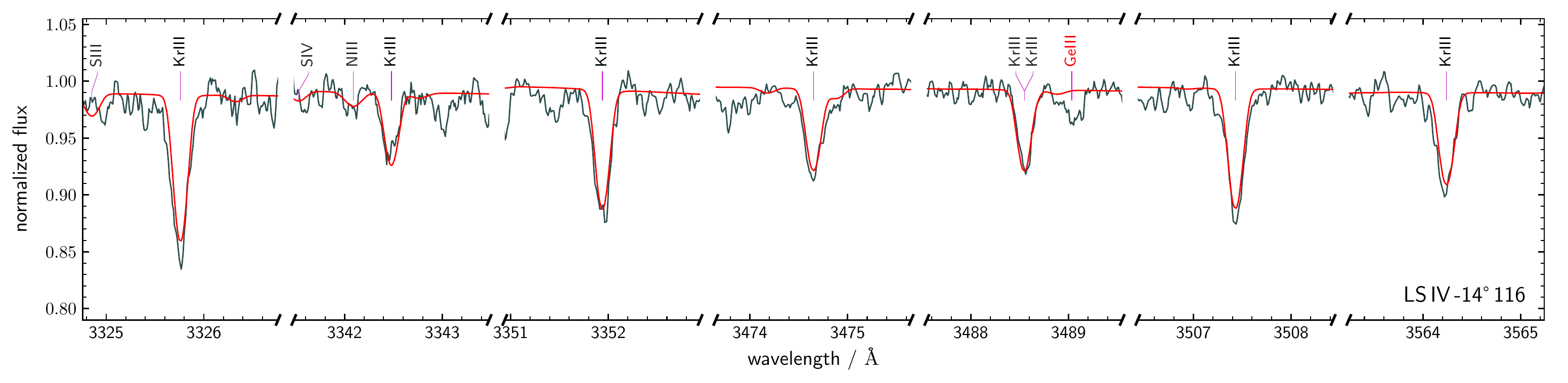}\vspace{-15pt}
        \includegraphics[width=1\textwidth]{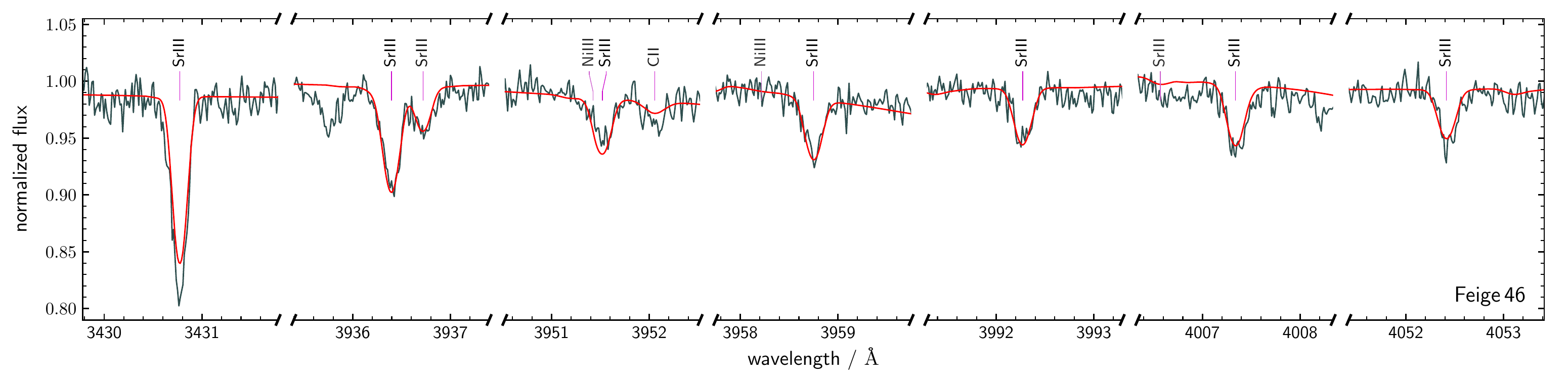}\vspace{-15pt}
        \includegraphics[width=1\textwidth]{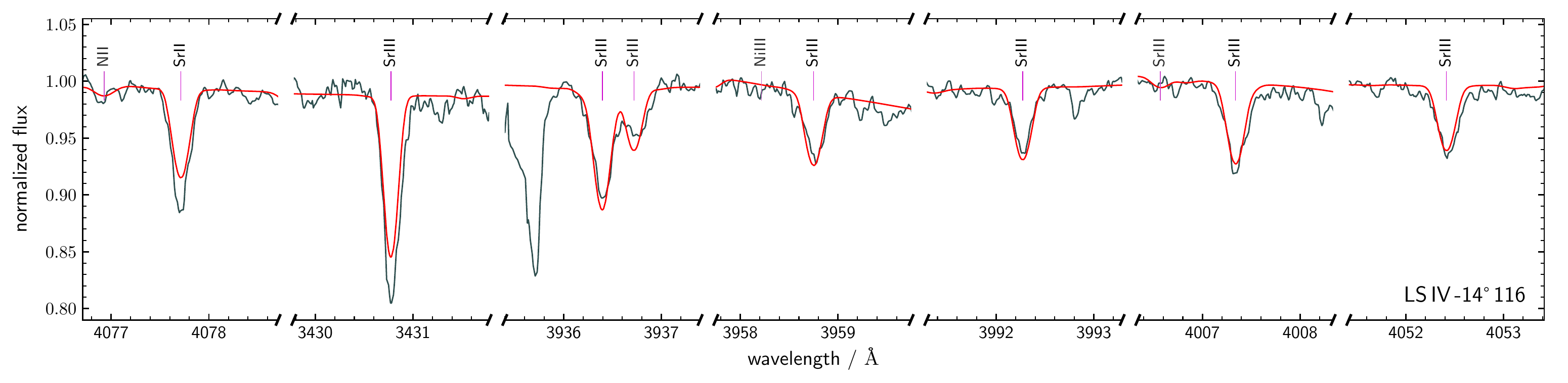}\vspace{-6pt}
        \caption{Strongest lines identified in the UVES spectra of \feige\ and \lsiv\ for elements Ge, Kr, and Sr.} 
        \label{lines_2}
\end{figure*}

\begin{figure*}

        \includegraphics[width=1\textwidth]{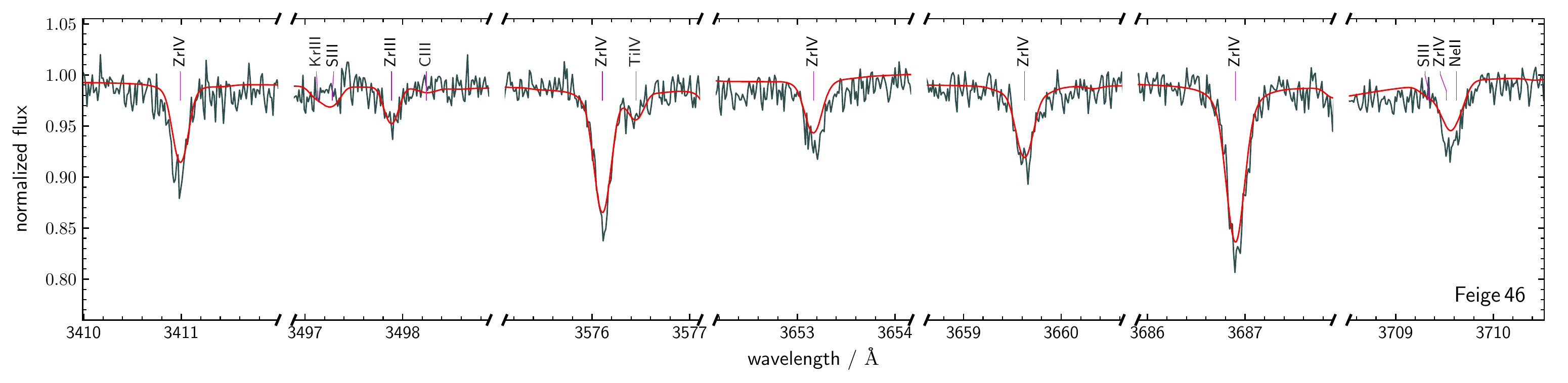}\vspace{-15pt}
        \includegraphics[width=1\textwidth]{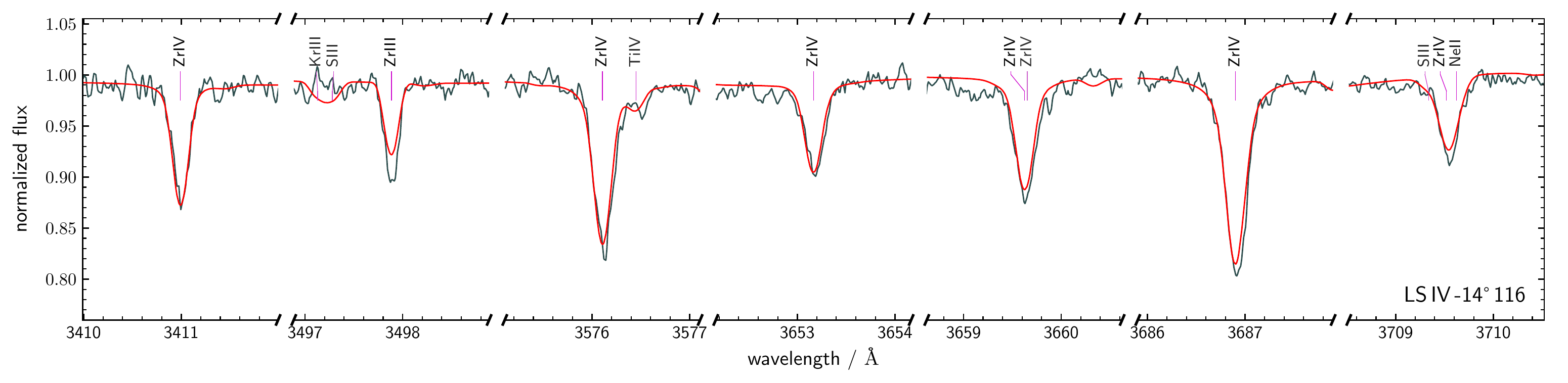}\vspace{-15pt}
        \includegraphics[width=1\textwidth]{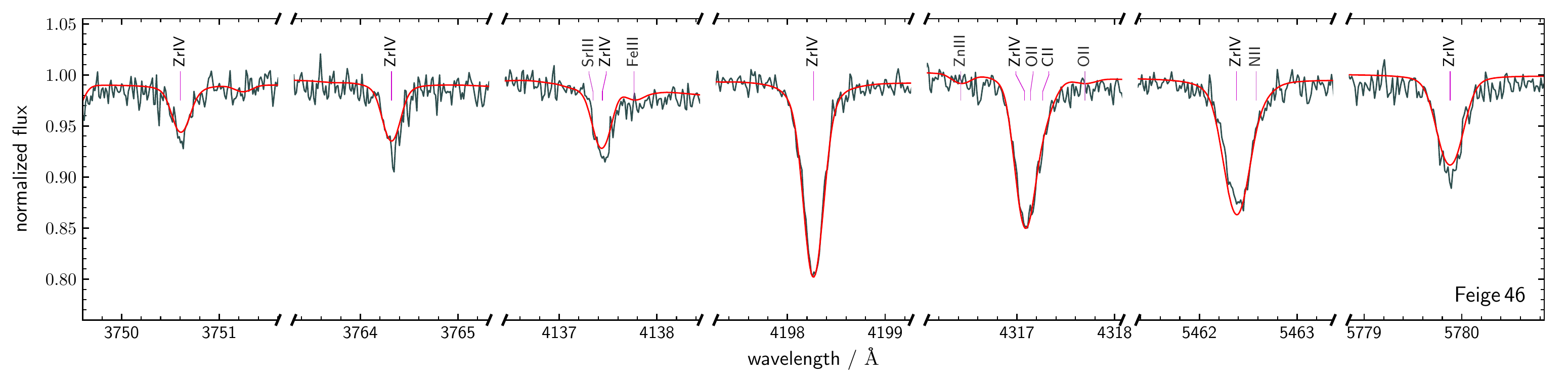}\vspace{-15pt}
        \includegraphics[width=1\textwidth]{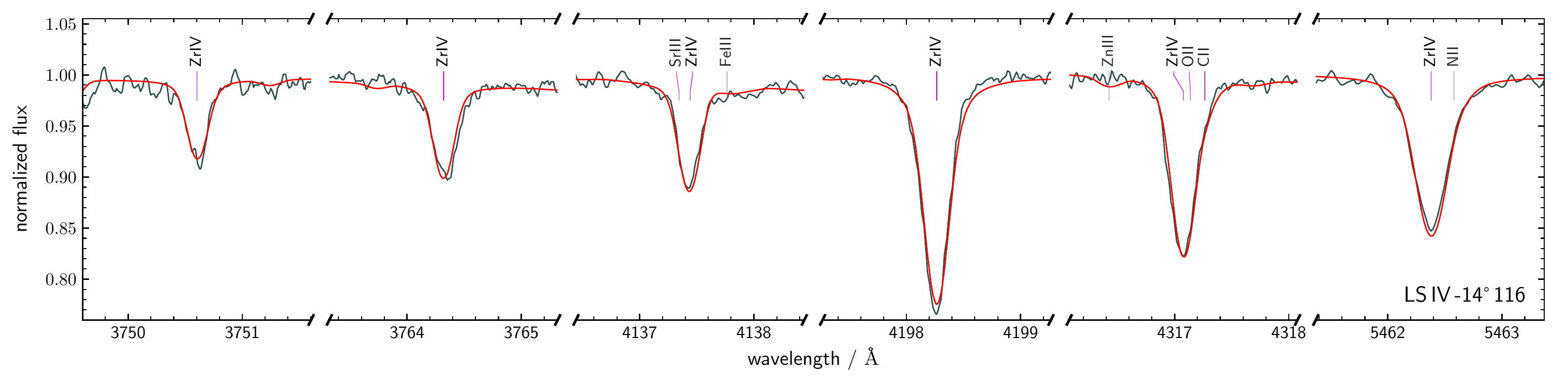}\vspace{-6pt}
        \caption{\ion{Zr}{iv} lines and one \ion{Zr}{iii} line identified in UVES spectra of \feige\ and \lsiv\ at the best fit abundances.}
        \label{lines_zr}
\end{figure*}

\subsection{Individual abundances}

In the following section, we present, in detail, the result of our abundance analysis for each element.
A summary of the abundances derived for \feige\ and \lsiv\ is given in Tables \ref{table_abund_feige} and \ref{table_abund_lsiv}, respectively.
Abundances stated in the text are always relative to solar values.
The full UVES spectra along with the final models for both stars are shown in Appendix \ref{sect:fullspec}.

\subsubsection{Light elements and the iron group: Carbon to zinc}

Examples of the strongest lines from light elements for both stars along with the final models are shown in the top panels of Fig.~\ref{lines_1}. 
The following paragraphs summarise the derivation of abundances for light metals and the iron group.

\paragraph{Carbon, nitrogen, and oxygen:} plenty of carbon, nitrogen, and oxygen lines are available to determine abundances, including the lines shown in Fig.~\ref{lines_1}. 
Both stars have a carbon abundance close to the solar number fraction that is slightly enhanced for \feige\ (+0.25 dex) and somewhat depleted for \lsiv\ ($-0.19$ dex). Nitrogen is overabundant in both stars by 0.46 dex and 0.28 dex, respectively, while oxygen is significantly underabundant by $-1.03$ dex and $-1.23$ dex. On average, the CNO content of \lsiv\ is lower than that of \feige\ by about 0.2\,dex.    
Although the general fit for carbon lines is good, there is some discrepancy between the strongest \ion{C}{ii} and \ion{C}{iii} lines.
We attribute this mostly to NLTE effects that are not perfectly modelled. 
For instance, the \ion{C}{ii} 4267.3\,\AA\ doublet is too strong in our synthetic spectra, while the C\,\textsc{iii} triplet 4152.5, 4156.5, 4162.9\,\AA\ is slightly too weak (see Fig.~\ref{fig:fspec_feige} and \ref{fig:fspec_lsiv}).
\ion{C}{ii}\,5661.9\,\AA\ is predicted to be in emission although no line is observed at this position in the UVES spectrum of \lsiv.
Some nitrogen lines display similar behaviour: \ion{N}{ii} 4630.5, 4643.1, 4803.3, 5005.2, 5179.5, and 5710.8\,\AA\ are too weak in our models and were not considered for determining the nitrogen abundance.
These lines also appeared in emission in the synthetic spectra of the iHe-sdO HD\,127493 \citep{dorsch2019}, who used the same model atoms.
Resolving these issues is a complex task because
almost all optical lines of \ion{C}{ii-iii} and \ion{N}{ii} originate from high-lying levels.
The population of these levels is very sensitive to the photo-ionisation (radiative bound-free) cross-sections used. 
The development of new \textsc{Tlusty} model atoms would be required for at least C\,\textsc{ii-iii} and N\,\textsc{ii,} which is an elaborate process and beyond of the scope of the present investigation.
For the time being, the best fit to lines of C, N, and O can be considered satisfactory.
The derived abundances of C, N, and O for \feige\ do not differ significantly from the values given by \cite{Latour2019b} (see Fig. \ref{comp:latour2019}). 

\vspace{-12pt}
\paragraph{Neon:} the slightly sub-solar  neon abundance for both stars is based on several Ne\,\textsc{ii} lines in the blue range, for example \ion{Ne}{ii} 3334.8, 3664.1, and 3694.2\,\AA.

\vspace{-12pt}
\paragraph{Magnesium:} the Mg\,\textsc{ii} 4481\,\AA\ doublet is observed in both stars and best reproduced at abundances of $-0.79$ dex for \feige,\ and $-1.07$ dex for \lsiv.  

\vspace{-12pt}
\paragraph{Aluminium:} the strongest predicted aluminium lines, \ion{Al}{iii} 4479.9, 4512.6, and 5696.6\,\AA, are not detected in \feige\ or \lsiv. 
The upper limit derived from these lines is slightly sub-solar.

\vspace{-12pt}
\paragraph{Silicon:} sub-solar silicon abundances are based mainly on the \ion{Si}{iv} 4088.9, 4116.1\,\AA\ doublet.

\vspace{-12pt}
\paragraph{Phosphorus:}
the only phosphorus line observed in \feige, \ion{P}{iii} 4222.2\,\AA, is very weak but present in \lsiv\ as well. The derived abundance based on this line is solar for \feige\ and slightly sub-solar for \lsiv.

\vspace{-12pt}
\paragraph{Sulphur:} 
no sulphur lines are detected in either star. The upper limit derived for \feige\ is consistent with the value found by \cite{Latour2019b} from the UV spectrum.

\vspace{-12pt}
\paragraph{Argon:} the argon abundance for \feige\ is based on the weak \ion{Ar}{iii} 3311.6 and 3503.6\,\AA\ lines.
The same lines could not be used for \lsiv, where the Ar abundance (about solar) is instead based on \ion{Ar}{iii} 3336.1 and 3511.2\,\AA.
Significant uncertainty ($\sim$0.2\,dex) is introduced by the continuum placement since all Ar lines are very weak.

\vspace{-12pt}
\paragraph{Calcium:} the upper limits derived for calcium are based on the non-detection of the \ion{Ca}{ii} 3933.7\,\AA\ resonance line, which is well separated from interstellar lines in both stars.
These upper limits indicate severe underabundances (by about 0.7\,dex) for both stars, which is consistent with the non-detection of the \ion{Ca}{iii} 4233.7 and 4240.7\,\AA\ lines that are usually observed in He-poor sdOB stars.

\vspace{-12pt}
\paragraph{Titanium:} weak titanium lines are observed in both stars. We used \ion{Ti}{iii} 3354.7, 4215.5, 4285.6\,\AA\ and \ion{Ti}{iv} 3541.4, 3576.5, 4971.2, and 5398.9\,\AA\ to derive super-solar abundances.

\vspace{-12pt}
\paragraph{Chromium, manganese, iron, and cobalt:} no lines from the iron-peak elements chromium, manganese, iron, and cobalt are observed in UVES spectra of either star.
For completeness, we list the abundances derived from UV lines for \feige\ by \cite{Latour2019b}
in Table \ref{table_abund_feige}.
The absence of high-resolution UV spectra of \lsiv\ means that no information on the abundance of these elements can be obtained for that star, except for iron.
The iron upper limit for \lsiv\ (0.35 times solar) is based on the non-detection of \ion{Fe}{iii} 5243.3 and 5891.9\,\AA, which are too strong in the final model.
\ion{Fe}{iii} 4137.8 and 4164.7\,\AA\ are well reproduced at this abundance.

\vspace{-12pt}
\paragraph{Nickel:}
several weak nickel lines (\ion{Ni}{iii}) could be used to derive abundances for both stars, for example \ion{Ni}{iii} 5332.2, 5436.9, 5481.3 and 5482.3\,\AA. 
The Ni abundance derived from the optical lines for \feige\ is the same as that obtained from the UV lines: overabundant by about 1 dex with respect to solar.

\vspace{-12pt}
\paragraph{Zinc:} the zinc abundances for \feige\ and \lsiv\ (about 300 times solar) are based on 13 and 16 strong lines, respectively (e.~g.~\ion{Zn}{iii} 3683.4, 4818.9, 4970.8, 5075.2, 5249.7, and 5563.7\,\AA; see Fig.~\ref{lines_1}).

\subsubsection{Heavy metals}

\begin{table}
\centering
\caption{Abundance results for \feige\ by number relative to hydrogen ($\log \epsilon$/$\epsilon_{\mathrm{H}}$), by number fraction ($\log \epsilon$), and number fraction relative to solar ($\log \epsilon$/$\epsilon_{\odot}$).  The number of resolved lines used per ionisation stage is given in the last column.
}\label{table_abund_feige}
\vspace{-10pt}
\setstretch{1.1}
\begin{center}
\resizebox{\columnwidth}{!}{
\begin{tabular}{l@{\hspace{2pt}}rrrr}
\toprule
\toprule
 Element  & \multicolumn{1}{c}{$\log \epsilon/\epsilon_{\mathrm{H}}$} & \multicolumn{1}{c}{$\log \epsilon$} & \multicolumn{1}{c}{$\log \epsilon/\epsilon_{\odot}$} & \multicolumn{1}{c}{$N_\mathrm{lines}$}\\
\midrule
H       &   $0.00\pm0.00$ &               $-0.17\pm0.02$ &                                   $-0.13\pm0.02$ \\
He      &  $-0.32\pm0.05$ &               $-0.49\pm0.03$ &                                    $0.62\pm0.04$ \\
\ion{C}{ii-iv}       &  $-3.19\pm0.13$ &               $-3.36\pm0.13$ &                                    $0.25\pm0.14$ & 6/16/1\\
\ion{N}{ii-iii}       &  $-3.57\pm0.08$ &               $-3.74\pm0.08$ &                                    $0.46\pm0.10$ & 23/14\\
\ion{O}{ii-iii}       &  $-4.21\pm0.10$ &               $-4.38\pm0.10$ &                                   $-1.03\pm0.11$ & 12/1 \\
\ion{Ne}{ii}      &  $-4.31\pm0.07$ &               $-4.48\pm0.07$ &                                   $-0.38\pm0.12$ & 18 \\
\ion{Mg}{ii}      &  $-5.05\pm0.02$ &               $-5.22\pm0.02$ &                                   $-0.79\pm0.04$ & 1\\
\ion{Al}{iii}       &  <$-6.16^{+0.40}_{}$ &   <$-6.33^{+0.40}_{}$ &                       <$-0.74^{+0.40}_{}$  &  \\ 
\ion{Si}{iii-iv}      &  $-5.51\pm0.03$ &               $-5.68\pm0.03$ &                                   $-1.15\pm0.04$ & 1/3 \\
\ion{P}{iii}       &  $-6.44\pm0.05$ &               $-6.61\pm0.05$ &                                    $0.02\pm0.06$ & 1\\
S       &  <$-5.60^{+0.30}_{}$ &  <$-5.77^{+0.30}_{}$ &                 <$-0.85^{+0.30}_{}$ \\
\ion{Ar}{iii}      &  $-5.75\pm0.14$ &            $-5.92\pm0.14$ &                             $-0.28\pm0.20$ & 3 \\
Ca      &  <$-6.15^{+0.40}_{}$ &  <$-6.32^{+0.40}_{}$ &                 <$-0.62^{+0.40}_{}$ \\
\ion{Ti}{iii-iv}       &  $-5.51\pm0.12$ &               $-5.68\pm0.12$ &                                    $1.41\pm0.13$ & 3/2\\
$^\ast$Cr      &  $-5.68\pm0.17$ &            $-5.85\pm0.17$ &                              $0.55\pm0.18$ \\
$^\ast$Mn      &  <$-5.69^{+0.40}_{}$ &  <$-5.86^{+0.40}_{}$ &                  <$0.75^{+0.40}_{}$ \\
$^\ast$Fe      &  $-4.64\pm0.14$ &               $-4.81\pm0.14$ &                                   $-0.27\pm0.15$ \\
$^\ast$Co      &  $-5.85\pm0.21$ &               $-6.02\pm0.21$ &                                    $1.03\pm0.23$ \\
\ion{Ni}{iii}      &  $-4.53\pm0.19$ &               $-4.70\pm0.19$ &                                    $1.12\pm0.19$ & 8 \\
\ion{Zn}{iii}      &  $-4.79\pm0.12$ &               $-4.96\pm0.12$ &                                    $2.51\pm0.13$ & 13 \\
\ion{Ga}{iii}      &  $-5.48\pm0.12$ &               $-5.66\pm0.12$ &                                    $3.34\pm0.15$ & 10 \\
\ion{Ge}{iii-iv}      &  $-4.89\pm0.15$ &               $-5.06\pm0.15$ &                                    $3.33\pm0.19$ & 3/3 \\
\ion{Kr}{iii}      &  $-4.90\pm0.07$ &               $-5.07\pm0.07$ &                                    $3.72\pm0.10$ & 11 \\
\ion{Sr}{ii-iii}      &       $-4.51\pm0.09$ &                    $-4.68\pm0.09$ &                                      $4.49\pm0.12$ & 3/19 \\
\ion{Y}{iii}       &       $-5.23\pm0.02$ &                    $-5.40\pm0.02$ &                                      $4.43\pm0.05$ & 2 \\
\ion{Zr}{iii-iv}      &       $-5.00\pm0.08$ &                    $-5.17\pm0.08$ &                                      $4.29\pm0.09$ & 1/12 \\
\ion{Sn}{iv}      &  $-6.26\pm0.06$ &               $-6.43\pm0.06$ &                                    $3.57\pm0.12$ & 2 \\
$^\ast$\ion{Pb}{iv}       &  <$-7.29^{+0.60}_{}$ &  <$-7.46^{+0.60}_{}$ &                  <$2.83^{+0.60}_{}$ \\
\bottomrule
\end{tabular}
}
\end{center}
\vspace{-10pt}
\tablefoot{$^\ast$ results for Cr, Mn, Fe, Co, and Pb are from \cite{Latour2019b} and based on UV data.}
\end{table}

\begin{table}
\centering
\caption{Same as Table \ref{table_abund_feige}, but for \lsiv.}
\label{table_abund_lsiv}
\setstretch{1.1}
\resizebox{\columnwidth}{!}{
\begin{tabular}{l@{\hspace{2pt}}rrrr}
\toprule
\toprule
 Element  & \multicolumn{1}{c}{$\log \epsilon/\epsilon_{\mathrm{H}}$} & \multicolumn{1}{c}{$\log \epsilon$} & \multicolumn{1}{c}{$\log \epsilon/\epsilon_{\odot}$} & \multicolumn{1}{c}{$N_\mathrm{lines}$}\\
\midrule
H       &        $0.00\pm0.00$ &                    $-0.10\pm0.02$ &                                     $-0.06\pm0.02$ \\
He      &       $-0.60\pm0.10$ &                    $-0.70\pm0.08$ &                                      $0.41\pm0.08$ \\
\ion{C}{ii-iii}        &       $-3.70\pm0.12$ &                    $-3.80\pm0.12$ &                                     $-0.19\pm0.13$ & 6/9 \\
\ion{N}{ii-iii}       &       $-3.82\pm0.06$ &                    $-3.92\pm0.06$ &                                      $0.28\pm0.08$ & 20/3 \\
\ion{O}{ii-iii}       &       $-4.48\pm0.10$ &                    $-4.57\pm0.10$ &                                     $-1.23\pm0.11$ & 11/1 \\
\ion{Ne}{ii}      &       $-4.50\pm0.06$ &                    $-4.60\pm0.06$ &                                     $-0.49\pm0.12$ & 13 \\
\ion{Mg}{ii}      &       $-5.40\pm0.02$ &                    $-5.50\pm0.02$ &                                     $-1.07\pm0.05$ & 1 \\
\ion{Al}{iii}     &  <$-6.42^{+0.30}_{}$  &        <$-6.52^{+0.30}_{}$ &                         <$-0.93^{+0.30}_{}$ &  \\ 
\ion{Si}{iii-iv}      &       $-6.03\pm0.07$ &                    $-6.13\pm0.07$ &                                     $-1.60\pm0.07$ & 1/2 \\
\ion{P}{iii}        &       $-6.76\pm0.06$ &                    $-6.85\pm0.05$ &                                     $-0.23\pm0.06$ & 1 \\
S       &  <$-6.00^{+0.30}_{}$ &     <$-6.10^{+0.30}_{}$ &                      <$-1.18^{+0.30}_{}$ \\
\ion{Ar}{iii}      &       $-5.55\pm0.04$ &                    $-5.64\pm0.04$ &                                     $-0.01\pm0.14$ & 2 \\
Ca      &  <$-6.36^{+0.30}_{}$ &     <$-6.46^{+0.30}_{}$ &                      <$-0.76^{+0.30}_{}$ \\
\ion{Ti}{iii-iv}      &       $-5.69\pm0.12$ &                    $-5.79\pm0.12$ &                                      $1.30\pm0.13$ & 3/2 \\
Fe      &  <$-4.90^{+0.30}_{}$ &     <$-5.00^{+0.30}_{}$ &                      <$-0.46^{+0.30}_{}$ \\
\ion{Ni}{iii}       &       $-4.62\pm0.13$ &                    $-4.72\pm0.13$ &                                      $1.10\pm0.14$ & 14 \\
\ion{Zn}{iii}      &       $-4.92\pm0.08$ &                    $-5.02\pm0.08$ &                                      $2.46\pm0.09$ & 15 \\
\ion{Ga}{iii}      &       $-5.62\pm0.06$ &                    $-5.72\pm0.06$ &                                      $3.28\pm0.11$ & 7 \\
\ion{Ge}{iii-iv}      &       $-5.05\pm0.10$ &                    $-5.14\pm0.10$ &                                      $3.24\pm0.14$ & 3/5 \\
\ion{Kr}{iii}      &       $-4.91\pm0.10$ &                    $-5.01\pm0.11$ &                                      $3.77\pm0.12$ & 10 \\
\ion{Sr}{ii-iii}      &       $-4.44\pm0.09$ &                    $-4.54\pm0.09$ &                                      $4.63\pm0.11$ & 4/21 \\
\ion{Y}{iii}       &       $-5.13\pm0.01$ &                    $-5.23\pm0.01$ &                                      $4.60\pm0.05$ & 2 \\
\ion{Zr}{iii-iv}      &            $-4.76\pm0.09$ &                         $-4.85\pm0.09$ &                                      $4.60\pm0.10$ & 1/13 \\
\ion{Sn}{iv}      &            $-5.56\pm0.04$ &                         $-5.65\pm0.04$ &                                      $4.34\pm0.11$ & 2 \\
\ion{Pb}{iv}      &            $-6.75\pm0.40$ &                         $-6.84\pm0.40$ &                                      $3.44\pm0.42$ & 1 \\
\bottomrule
\end{tabular}
}
\end{table}

From a spectroscopic perspective, the prevalence of strong lines of heavy elements (here $Z > 30$) is the most striking feature of \lsiv\ and \feige. Nevertheless, many lines of heavy metals remained either undetected in the previous analyses \citep{nas11,Latour2019b}, owing to the limited S/N and wavelength coverage of the spectra available, or unidentified due to the scarcity of atomic data.
Therefore, we set out to identify these lines that are present both in \lsiv\ and \feige.

Oscillator strengths are available for many ions that are expected to show spectral lines in the programme stars.
However, several of these lines have remained unidentified so far because their rest wavelengths are not known with sufficient precision.
The large wavelength coverage and good S/N of our spectra allowed us to identify lines of such ions from predicted relative intensities by adjusting the theoretical wavelengths to match the position of observed lines.
These empirical wavelengths may also be useful in future atomic structure calculations.

The 102 detected heavy-metal lines with available oscillator strengths are listed in Table \ref{table_idf}.
This includes strong previously unidentified lines noted by \cite{nas11} at 4007\,\AA\ and 4216\,\AA, that now appear to be \ion{Sr}{iii} and \ion{Sn}{iv}.
Lines that required significant shifts to match observed lines are additionally listed in Table \ref{table_lshift}.
The 21 newly identified lines that lack oscillator strengths are listed in Table \ref{table_idpos}; some are shown in Fig.~\ref{lines_unid}.
The 51 remaining unidentified lines are listed in Table \ref{table_unid}.
In the following paragraphs, we briefly describe the analysis for each heavy element detected. 
The strongest modelled lines for each heavy element are shown in Fig.~\ref{lines_1} (Ga, Y, Sn, Pb), Fig.~\ref{lines_2} (Ge, Kr, Sr), and Fig.~\ref{lines_zr} (Zr) for both stars.

\vspace{-12pt}
\paragraph{Gallium:} we identified several \ion{Ga}{iii} lines in the spectra of \feige\ and \lsiv. 
Oscillator strengths for optical \ion{Ga}{iii} lines were derived by \cite{oreilly98}.
In particular, \ion{Ga}{iii} 3517.4, 3577.3, 3806.7, 4380.6, 4381.8, 4993.9, 5337.2, 5358.2, 5844.9, and 5993.9\,\AA\ could be used to derive an abundance of about 2000 times solar for both stars. 
To our knowledge, they have never been observed in any star.

\begin{figure*}
        \centering
        \includegraphics[width=0.95\textwidth]{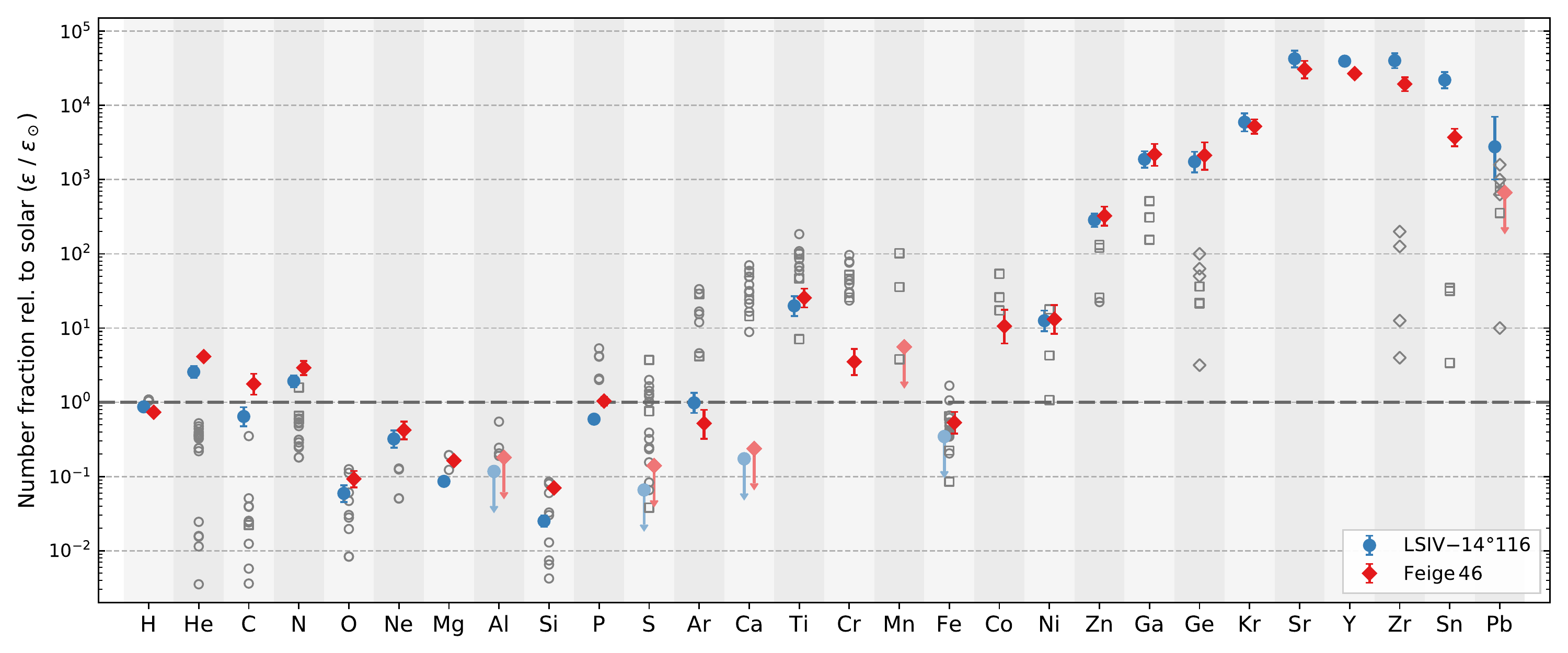}
        \caption{Abundance patterns of \lsiv\ and \feige\ relative to that of the Sun (by number fraction). Only elements with an abundance measurement are shown. Upper limits are marked with an arrow and less saturated colours. For comparison, abundance measurements for He-poor sdOB stars (33000\,K < \teff\ < 36500\,K) are shown as grey open circles \citep[][based on optical data]{Geier2013}, diamonds \citep[][based on far-UV data]{chayer2006}, and squares \citep[][based on UV data]{otoole06}.}
        \label{fig:pattern}
\end{figure*}

\vspace{-12pt}
\paragraph{Germanium:} \cite{nas11} identified and provide oscillator strengths for three \ion{Ge}{iii} lines in the optical spectrum of \lsiv.
Oscillator strengths for optical lines of \ion{Ge}{iv} were provided by \cite{oreilly98}. However, these \ion{Ge}{iv} lines have never been used to derive abundances, and their wavelengths had to be shifted to match the observed ones as listed in Table \ref{table_lshift}.
We used these three \ion{Ge}{iii} lines as well as four \ion{Ge}{iv} lines to derive a germanium abundance of 2000 times solar for both stars.
There is a mismatch between \ion{Ge}{iii} and \ion{Ge}{iv} lines, which systematically appear too weak in our synthetic spectra. 
This may be due to NLTE effects or systematic differences between the atomic data used for \ion{Ge}{iii} and \ion{Ge}{iv}. 

An effective temperature of 35920\,K would be required for \lsiv\ to simultaneously reproduce both ionisation stages. 
However, this temperature is too high by about 400\,K to be able to reproduce the ionisation balance of most other elements.

\vspace{-12pt}
\paragraph{Arsenic:} two weak, unidentified lines at 3922.5\,\AA\ and 4037\,\AA\ are observed close to experimental wavelengths of the As\,\textsc{iii} lines provided by \cite{Lang1928}, as listed in NIST.\footnote{National Institute of Standards and Technology,  \url{https://physics.nist.gov/PhysRefData/ASD/lines_form.html}; see also \cite{NIST_ASD}.}
We are not aware of oscillator strengths for optical \ion{As}{iii} lines, and, therefore, cannot derive the abundance.

\vspace{-12pt}
\paragraph{Selenium:} fifteen previously unidentified lines were identified with \ion{Se}{iii} using the experimental wavelengths provided by \cite{Badami1933} as listed in NIST (see Fig.~\ref{lines_unid}). 
This is the first time these lines have been observed in any star, and they are visible in both \feige\ and \lsiv.
A list of identifications is given in Table \ref{table_idpos}.
Unfortunately, no oscillator strengths are available for optical \ion{Se}{iii} lines.

\vspace{-12pt}
\paragraph{Krypton:} \ion{Kr}{iii} shows many lines in the UVES spectra of \feige\ and \lsiv\ that have never been identified in any star as far as we know.\footnote{Around 2012, N.~Naslim reported the possible presence of krypton lines to one of us (CSJ); this could not be confirmed at the time.} Fortunately, oscillator strengths are provided by \cite{Raineri1998} allowing us to determine the krypton abundance.
Some lines were shifted to match the observed position; they are listed in Table \ref{table_lshift}.
We have used \ion{Kr}{iii} 3325.76, 3342.48, 3351.94, 3474.65, 3488.55, 3564.24, 3641.35, 3690.66, and 4067.40\,\AA\ to derive an abundance of about 5500 times solar for both stars. 
The predicted \ion{Kr}{iii} 3308.22, 3396.72\,\AA\ lines do not match observed lines.
The alternative oscillator strengths for these two lines provided by \cite{Eser2018} are even larger.
These lines might require large shifts or have inaccurate oscillator strengths.

\vspace{-12pt}
\paragraph{Strontium:} 
in total, 35 previously unidentified lines can clearly be attributed to \ion{Sr}{iii}: for example, the strong 3430.8, 3936.4\,\AA\ lines.
To our knowledge, these lines have never before been reported in stellar spectra. 
Wavelengths and oscillator strengths for \ion{Sr}{ii-iii} were provided by R.~Kurucz, allowing us to determine the strontium abundance.
The resonance lines \ion{Sr}{ii} 4077.7, 4215.5\,\AA\ used by \cite{nas11} to derive the strontium abundance in \lsiv\ are also observed in \feige.
To model these lines, we used oscillator strengths from \cite{Fernandez2020}, who recently investigated \ion{Sr}{ii} in detail (along with \ion{Y}{iii} and \ion{Zr}{iv}).
Both stars also show \ion{Sr}{ii} lines at 3380.7, 3464.5, and 4305.4\,\AA. 
Fitting four \ion{Sr}{ii} lines (three for \feige) as well as 21 \ion{Sr}{iii} lines (19 for \feige) results in an abundance of 43\,000 times solar for \lsiv\ and 31\,000 times solar for \feige.

\vspace{-12pt}
\paragraph{Yttrium:} \cite{nas11} identified two strong yttrium lines in the spectrum of \lsiv: \ion{Y}{iii} 4039.602 and 4040.112\,\AA.
Fitting these lines (\ion{Y}{iii} 4039.6\,\AA\ at a slightly revised position) results in abundances of 27\,000 times solar for \feige\ and 40\,000 times solar for \lsiv.
Oscillator strengths for additional \ion{Y}{iii} lines observed at 5102.9, 5238.1, and 5602.2\,\AA\ are provided by \cite{Fernandez2020}.
However, these lines are not consistent with \ion{Y}{iii} 4039.6, 4040.1\,\AA\ and were therefore not considered for the abundance determination.

 \begin{figure*}
        \centering
        \includegraphics[width=0.95\textwidth]{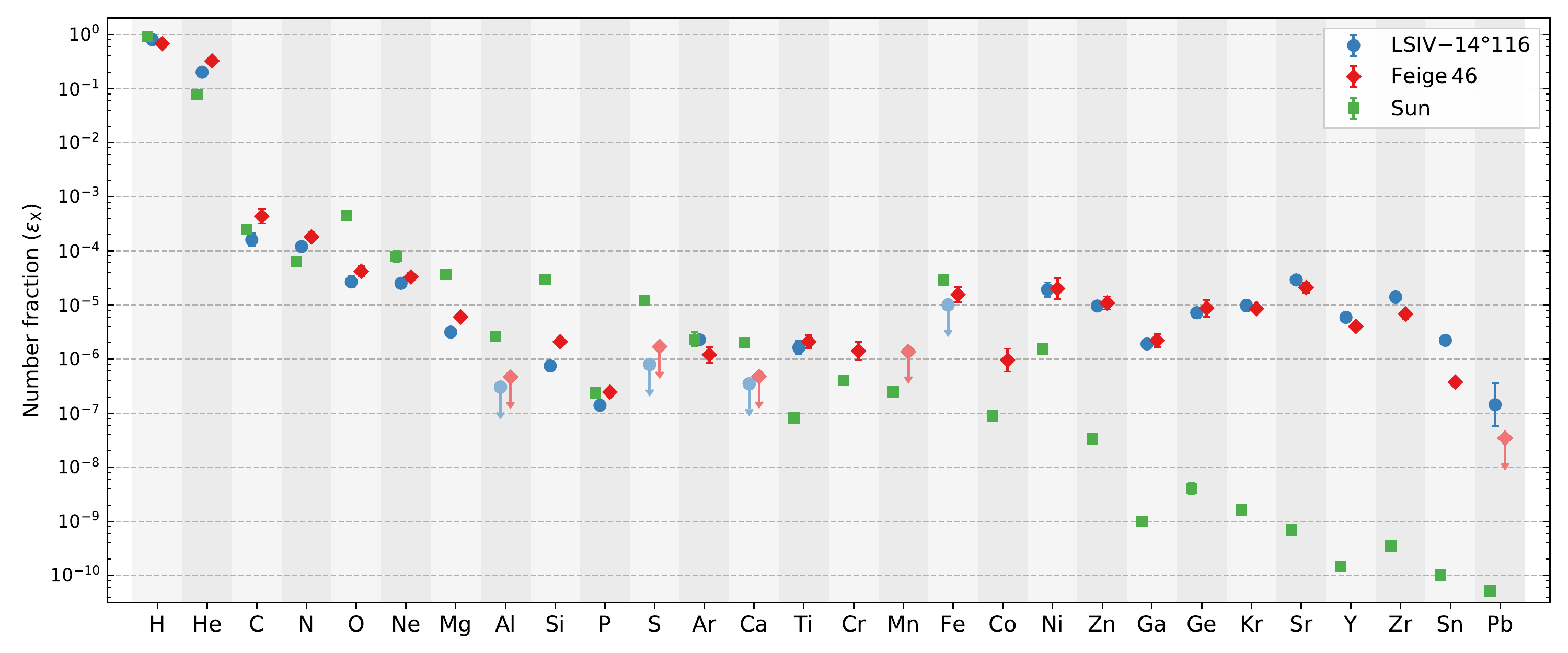}
        \caption{Atmospheric abundances for \lsiv\ and \feige\ by number fraction \cite[with solar values from][]{asplund09}.}
        \label{fig:pattern_nf}
 \end{figure*}

\vspace{-12pt}
\paragraph{Zirconium:} by far the strongest lines from heavy metals in the optical spectrum of both stars originate from zirconium \textsc{iv} transitions (see Fig.~\ref{lines_zr}).
Oscillator strengths for four Zr\,\textsc{iv} lines were provided by \cite{nas11} and for two additional lines by \cite{nas13}.
\cite{Rauch2017} also provide oscillator strengths for a large number of UV and optical Zr\,\textsc{iv} lines, while
\cite{Fernandez2020} have recently computed oscillator strengths for eight Zr\,\textsc{iv} lines that are observed in the UVES spectra of both stars.
We exclusively rely on data from \citet{Rauch2017}, as they provide the most extensive list.
A single strong Zr\,\textsc{iii} line is observed at 3497.9\,\AA\ and is somewhat too weak in our models.
We used this line as well as several Zr\,\textsc{iv} lines to determine the abundance in both stars, including the four Zr\,\textsc{iv} lines used by \cite{nas11}.
The best fit Zr abundance for \lsiv, 40\,000 times solar, is significantly higher than that for \feige\ (20\,000 times solar).
As shown in Fig.~\ref{lines_zr}, Zr\,\textsc{iv} lines are very well reproduced in both stars (with the exception of Zr\,\textsc{iv} 3919.3 and 5462.3\,\AA, which are too strong in our models). 
In addition, we slightly revised the position of two \ion{Zr}{iv} lines: \ion{Zr}{iv} 5462.38 and 5779.88\,\AA.

\vspace{-12pt}
\paragraph{Tin:} 
strong spectral lines of \ion{Sn}{iv} at 3862.1 and 4217.2\,\AA\ are visible in the UVES spectra of \feige\ and \lsiv. These lines have not been previously identified in any star.
To model these lines, we used oscillator strengths provided by \cite{Kaur2020}, but the rest wavelengths had to be adjusted (see Table \ref{table_lshift}).
The abundance of tin derived from the two newly identified lines turns out to be 22\,000 times solar for \lsiv\ and 3700 times solar for \feige, which is consistent with the value derived from UV lines by \cite{Latour2019b}.

\vspace{-12pt}
\paragraph{Lead:} the lead abundance of \lsiv, 2800 times solar, is based on a very weak \ion{Pb}{iv} 4049.8\,\AA\ line.
No other lead lines are detected, thus the abundance has a large uncertainty.
\cite{Latour2019b} determined an upper limit for lead in \feige, 680 times solar.
It is based on the Pb\,\textsc{iv}\,1313\,\AA\ resonance line and is likely close to the actual abundance.
Although this upper limit is consistent with the non-detection of Pb lines in the optical spectrum of \feige, it should be confirmed by UV observations of higher quality than the low S/N IUE spectrum that is currently available.

\vspace{-12pt}
\paragraph{Undetected elements:} 
we searched unsuccessfully for lines of fluorine, sodium, chlorine, potassium, scandium, vanadium, rubidium, and xenon. Details can be found in Appendix \ref{appendix_undetected_lines}.

\subsection{The emerging abundance pattern}

Both stars show almost the same abundance pattern, as illustrated in Fig.~\ref{fig:pattern}.
When compared to solar values, nitrogen is enhanced and oxygen depleted.
Carbon is slightly super-solar in \feige\ and slightly sub-solar in \lsiv.
The light metals C, N, O, Ne, Mg, Si, and P are all slightly more abundant in \feige,\ but otherwise follow the same pattern as in \lsiv.
The abundances of elements from argon to krypton (when known) are almost identical, and calcium is depleted in both stars.
Heavy elements are enriched to very high values, from zinc at about 300 times solar, to zirconium well above 20\,000 times solar.
While being highly enriched when compared to solar values, the concentration of Sr, Y, Zr, Sn, and Pb in the line-forming region of \feige\ is progressively less extreme when compared to that of \lsiv. 
This enrichment is nevertheless notable when compared to He-poor sdOB stars, which have been observed to be enhanced in Zn, Ga, Ge, Zr, and Sn to about 10 to 200 times the solar value \citep{otoole06,chayer2006,bla08}.
The extreme enrichment in heavy metals and the abundances of lighter metals are different to those observed in He-poor sdOB stars.
In particular, the argon and calcium abundances in \lsiv\ and \feige\ are significantly lower than the super-solar values \cite{Geier2013} obtained for He-poor sdOBs of similar temperatures.

Such strong deficiency  when compared to He-poor sdOB stars (2\,dex for calcuium) cannot be explained by a lower initial metallicity that might be expected for \lsiv\ and \feige\ due to their halo kinematics. 
It is worth mentioning that this calcium deficiency is not observed in lead-rich iHe-sdOB stars such as [CW83]\,0825+15 \citep{jeff17}, FBS\,1749+373, and PG\,1559+048 \citep{Naslim2020}. These stars show calcium abundances in line with those observed in He-poor sdOB stars.
In contrast to this, the carbon and nitrogen abundances in \lsiv\ and \feige\ are higher than in the average He-poor sdOB star.
Such enrichment could be caused by an excess of material processed by H-burning (CNO-cycle) and He-burning (3$\alpha$) in the atmospheres of \lsiv\ and \feige\ when compared to He-poor sdOB stars, as predicted for both hot flasher \citep{bert08} and merger scenarios \citep{zhang12}. This is consistent with the positive correlation between the helium and carbon abundances of sdOB stars in the globular cluster $\omega$\,Cen, as found by \cite{Latour2014}.
In addition, the abundances of C, N, and O might still be affected by diffusion processes to some degree (in both He-poor and iHe-sdOB stars).

\section{Discussion and conclusions}\label{sect:conclusions}

We performed a detailed spectral analysis of \feige\ and \lsiv.
This consistent analysis of both stars enables an accurate and direct comparison of their abundance patterns, which would be hampered by the use of different analysis methods.

The abundance patterns of both stars, as well as their differences, can likely be explained with atmospheric diffusion processes.
In terms of diffusion, it is convenient to consider the abundance pattern as number fraction without the comparison to solar values (Fig.~\ref{fig:pattern_nf}).
It is easy to recognise that, overall, the abundance of light metals from carbon to phosphor drops by three orders of magnitude from $\log \epsilon \approx-3.5$ to about $-$6.5\,dex.
Unlike in the Sun, the abundances of heavier elements (except calcium) do not continue to drop further, but follow a more constant pattern.
A comparison of detailed diffusion calculations with these observed abundance patterns is required to resolve the question of whether diffusion alone is enough to produce such high enrichment of heavy metals.
In addition, atmospheric models that consider atmospheric stratification are required to determine whether the observed heavy-metal enrichment can be explained by thin layers of enriched material in the line-forming region. 

Thanks to the excellent quality and wavelength coverage, we were able to identify many previously unidentified lines in the UVES spectra of \feige\ and \lsiv\ with transitions of heavy ions.
Strong lines with available oscillator strengths originate from ions such as \ion{Ge}{iv}, \ion{Kr}{iii}, \ion{Sr}{iii}, \ion{Zr}{iii}, and \ion{Sn}{iv}. Their identification will enable the determination of abundances in future analyses of other heavy-metal stars, even with spectra of lower quality.
Atomic data are still lacking for some heavy elements and ionisation stages \textsc{iii-iv}, including several newly identified lines of \ion{Ge}{iii}, \ion{Se}{iii}, and \ion{Y}{iii}.
We also provide observed wavelengths for these lines that may be useful in future atomic structure calculations.
About 50, mostly weak, lines detected in the spectra of \lsiv\ and \feige\ remain unidentified and could belong to elements not yet identified in either star.

We also analysed the TESS light curve of \feige\ and detected five of the six modes found by \cite{Latour2019a}.
The period stability of these five pulsation modes ($\dot P \lesssim 10^{-8}$ s/s) is not compatible with the stronger period decay predicted by \cite{bat18} for a star quickly evolving through a series of late helium flashes.
This questions the idea that \feige\ may be a pre-EHB object with pulsations driven by the $\epsilon$-mechanism generated by late helium flashes.

Stellar parameters (mass, radius, and luminosity) were derived from the high-quality \textit{Gaia} parallax by combining it with the atmospheric parameters and the spectral energy distribution.
The results for both stars are limited by the uncertainty of the surface gravity, but consistent with the canonical subdwarf mass predicted by hot flasher models  \citep[0.46M$_\odot$,][]{dor93,2003MNRAS.341..669H}.

The similarity of \lsiv\ and \feige\ in terms of atmospheric parameters, abundances, pulsation, and kinematics remains puzzling.
A larger sample of intermediately He-rich sdOB stars with detailed observed abundance patterns is required to draw conclusions regarding the causal relation between these features.
Such a sample would also be required in order to answer the questions: 
\begin{itemize}
\item What makes the heavy-metal stars different from the normal sdOB stars? Other chemically peculiar stars such as
helium-rich main-sequence B stars and Ap stars have strong magnetic fields, and so it has been suggested that the heavy-metal stars are magnetic too, but no magnetic field has been detected in LS IV-14 116 \citep[down to 300\,G,][]{ran15}.\vspace{2pt}
\item Are most iHe-sdOB stars an intermediate stage in the evolution of He-sdOs towards the He-poor sdBs? \vspace{2pt}
\item At which point in their evolution will atmospheric diffusion become important?
\end{itemize}

Fortunately, recent surveys such as the LAMOST survey \citep[e.g.][]{Lei_2020} and the SALT/HRS survey \citep[e.g.][]{2017OAst...26..202J} are discovering many new He-rich subdwarf stars.
Future analyses of a larger sample of stars that share the atmospheric parameters of \lsiv\ and \feige\ (intermediate He-enrichment and \teff\ around 35\,000\,K), but also of their possible progenitors, the extreme He-sdOs, might give important clues towards the evolution of He-rich subdwarf stars. 

\begin{acknowledgements}
We thank Simon Kreuzer for the development of the the photometry query tool and Ingrid Pelisoli for very helpful comments on TESS. 
We thank the referee, Suzanna Randall, for useful suggestions that improved this paper.
M.~L.~acknowledges funding from the Deutsche Forschungsgemeinschaft (grant DR 281/35-1).
S.~C.~acknowledges financial support from the Centre National d’Études Spatiales 
(CNES, France) and from the Agence Nationale de la Recherche (ANR, France) 
under grant ANR-17-CE31-0018.
Based on observations collected at the European Southern Observatory under ESO programmes 0104.D-0206(A), 087.D-0950(A), and 095.D-0733(A).
This paper includes data collected by the TESS mission, which are publicly available from the Mikulski Archive for Space Telescopes (MAST).
Funding for the TESS mission is provided by NASA's Science Mission directorate.
Based on INES data from the IUE satellite.
This work has made use of data from the European Space Agency (ESA) mission {\it Gaia} (\url{https://www.cosmos.esa.int/gaia}), processed by the {\it Gaia} Data Processing and Analysis Consortium (DPAC, \url{https://www.cosmos.esa.int/web/gaia/dpac/consortium}). Funding for the DPAC has been provided by national institutions, in particular the institutions participating in the {\it Gaia} Multilateral Agreement.
Based on observations obtained as part of the VISTA Hemisphere Survey, ESO Progam, 179.A-2010 (PI: McMahon).
The TOSS service (\url{http://dc.g-vo.org/TOSS}) used for this paper was constructed as part of the activities of the German Astrophysical Virtual Observatory.
We acknowledge the use of the Atomic Line List (\url{http://www.pa.uky.edu/~peter/newpage/}).
This research has made use of NASA's Astrophysics Data System.
\end{acknowledgements}

\bibliographystyle{aa}
\bibliography{lsiv_f46_uves.bib}

\begin{appendix}

\section{Additional elements investigated}\label{appendix_undetected_lines}

\begin{table}
\caption{Upper limits for elements that could not be detected.}
\label{tab:abu_failed}
\centering
\setstretch{1.1}
\resizebox{\columnwidth}{!}{
\begin{tabular}{lrrrr}
\toprule
\toprule
  & \multicolumn{2}{c}{$\log \epsilon$} & \multicolumn{2}{c}{$\log \epsilon / \epsilon_\odot$}\\[0pt]
         \hspace{-10pt} & \small{\feige} & \small{\lsiv}  & \small{\feige} & \small{\lsiv} \\
\midrule
F  &   <$-5.79^{+0.50}_{}$ &        <$-5.99^{+0.50}_{}$ &                        <$1.69^{+0.53}_{}$ &                          <$1.49^{+0.53}_{}$ \\
Na &                              &        <$-5.65^{+0.50}_{}$ &                                                  &                          <$0.15^{+0.50}_{}$ \\
Cl &   <$-6.22^{+0.40}_{}$ &        <$-6.41^{+0.40}_{}$ &                        <$0.32^{+0.45}_{}$ &                          <$0.13^{+0.45}_{}$ \\
\bottomrule
\end{tabular}
}
\end{table}

In the following, we describe additional elements that could not be detected in \feige\ and \lsiv.
Upper limits are listed in Table~\ref{tab:abu_failed}.
As in Tables \ref{table_abund_feige} and \ref{table_abund_lsiv}, upper limits are given as best fit values, while their uncertainties represent values that can clearly be excluded.

\ion{F}{ii} 3505.6, 3847.1, 3850.0, and 4246.2\,\AA\ exclude abundances higher than about 100 times solar in \lsiv.
The very weak photospheric resonance lines \ion{Na}{i} 5889.94 and 5889.96\,\AA\ are well separated from the interstellar lines, but unfortunately blended with the stronger \ion{C}{ii}\,5889.78\,\AA.
These lines, as well as \ion{Na}{ii} 3533.1\,\AA,\ seem to exclude abundances higher than about five times solar in \lsiv,\ while no sensible upper limit can be derived for \feige.
Chlorine abundances higher than six times solar for \feige\ and about four times solar for \lsiv\ can be excluded based on the non-detection of \ion{Cl}{iii} 3530.0, 3601.9\,\AA. 

The upper limits derived for the following elements are either too high to be of use, or uncertain because of poorly known line positions.
We therefore refrain from stating even an upper limit.

\ion{K}{ii} 4186.2\,\AA\ seems to fit  a weak line in \lsiv\ at an abundance of about 30 times solar. 
However, this abundance seems to be excluded by \ion{K}{iii} 3322.4, 3358.4, and 3364.3\,\AA, which suggest an upper limit of about ten times solar. 
The upper limit derived from the non-detection of very weak predicted \ion{Sc}{iii} and \ion{V}{iii} lines are next to meaningless for both stars. 
\cite{Zhang2014} provided atomic data for \ion{Rb}{iii}. 
However, because these lines have never been observed, their positions are likely not accurate.
They do not match observed lines in \feige\ or \lsiv.
The same is true for optical \ion{Xe}{iv} lines as predicted by \cite{Rauch2017}.

\section{Comparison with abundances from \cite{Latour2019b}}
 \begin{figure}
        \includegraphics[width=1\columnwidth]{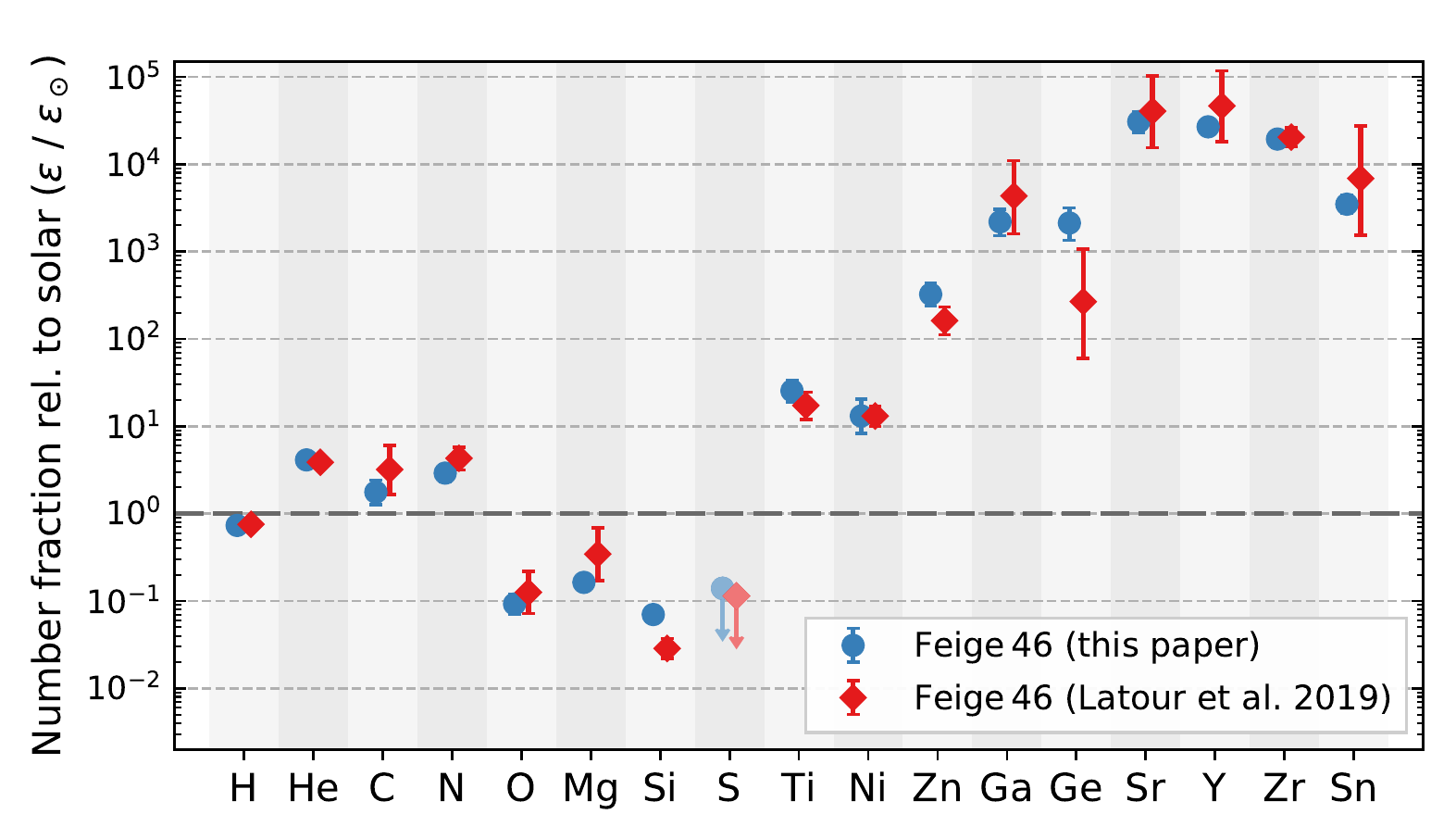}
        \caption{Comparison with results from \cite{Latour2019b}.}
        \label{comp:latour2019}
 \end{figure}
Our results for \feige\ are in good agreement with the abundances derived by \cite{Latour2019b}.
The excellent agreement for Ti, Ni, Zn, Ga, Sr, and Sn is remarkable because these abundances were previously solely based on UV data.
The slight disagreement for Si and Ge may be explained with the low S/N of the CASPEC and IUE spectra used by \cite{Latour2019b}.

\section{Observed wavelengths}

Table \ref{table_idf} lists all detected lines from heavy elements for which oscillator strengths are available. Since observed wavelengths are useful in the calculation of oscillator strengths, we also list in Table \ref{table_idpos} observed positions for lines that could be identified but do not have oscillator strengths available.

Several, mostly weak, lines are visible in the UVES spectra of both \feige\ and \lsiv,\ but they remain unidentified. They are listed in Table \ref{table_unid}. These lines are visible in both stars and therefore likely to be real. They might be due to transitions in heavy ions.

\begin{table}
\centering
\caption{Identified lines of heavy metals ($Z>30$) in the UVES spectra of \lsiv\ and \feige\ for which oscillator strengths are available. Equivalent widths are given for \lsiv.}
\label{table_idf}
\setstretch{1.1}
\begin{minipage}{0.49\columnwidth}
\resizebox{\columnwidth}{!}{
\begin{tabular}{l c c}
\toprule
\toprule
Ion & $\lambda_\mathrm{obs}$ / \AA & EW / m\AA \\
\midrule
\ion{Zr}{iv} & 3297.858 & 20.0 \\
\ion{Sr}{iii} & 3302.730 & \phantom{0}5.2 \\
\ion{Kr}{iii} & 3311.490 & 13.0 \\
\ion{Ge}{iv} & 3320.530 & 20.6 \\
\ion{Kr}{iii} & 3325.752 & 29.9 \\
\ion{Ge}{iv} & 3333.785 & 15.4 \\
\ion{Kr}{iii} & 3342.461 & 10.2 \\
\ion{Kr}{iii} & 3351.938 & 22.4 \\
\ion{Kr}{iii} & 3374.961 & \phantom{0}5.7 \\
\ion{Sr}{ii} & 3380.702 & \phantom{0}7.4 \\
\ion{Zr}{iv} & 3410.999 & 23.3 \\
\ion{Sr}{iii} & 3430.775 & 36.5 \\
\ion{Sr}{iii} & 3444.874 & \phantom{0}8.4 \\
\ion{Sr}{ii} & 3464.480 & 12.6 \\
\ion{Kr}{iii} & 3474.650 & 16.3 \\
\ion{Kr}{iii} & 3488.558 & 12.6 \\
\ion{Zr}{iii} & 3497.889 & 17.4 \\
\ion{Kr}{iii} & 3507.435 & 22.8 \\
\ion{Ga}{iii} & 3517.392 & \phantom{0}9.9 \\
\ion{Kr}{iii} & 3549.408 & \phantom{0}3.9 \\
\ion{Ge}{iv} & 3554.257 & 35.0 \\
\ion{Sr}{iii} & 3559.674 & \phantom{0}8.0 \\
\ion{Kr}{iii} & 3564.223 & 17.8 \\
\ion{Kr}{iv} & 3567.647 & \phantom{0}3.7 \\
\ion{Zr}{iv} & 3576.123 & 39.0 \\
\ion{Ga}{iii} & 3577.291 & 12.0 \\
\ion{Kr}{iii} & 3641.332 & \phantom{0}6.5 \\
\ion{Sr}{iii} & 3650.734 & \phantom{0}9.6 \\
\ion{Zr}{iv} & 3653.182 & 19.8 \\
\ion{Zr}{iv} & 3659.634 & 30.1 \\
\ion{Ge}{iv} & 3676.735 & 26.4 \\
\ion{Zr}{iv} & 3686.914 & 49.6 \\
\ion{Sr}{iii} & 3688.299 & \phantom{0}6.1 \\
\ion{Kr}{iii} & 3690.652 & \phantom{0}5.0 \\
\ion{Zr}{iv} & 3709.552 & 18.0 \\
\ion{Zr}{iv} & 3750.608 & 15.9 \\
\ion{Zr}{iv} & 3764.335 & 25.6 \\
\ion{Kr}{iii} & 3792.666 & \phantom{0}4.1 \\
\ion{Sr}{iii} & 3821.965 & \phantom{0}7.3 \\
\ion{Zr}{iii} & 3829.240 & \phantom{0}1.9 \\
\ion{Sr}{iii} & 3855.913 & \phantom{0}5.6 \\
\ion{Sn}{iv} & 3861.207 & 30.0 \\
\ion{Kr}{iii} & 3868.793 & \phantom{0}3.8 \\
\ion{Sr}{iii} & 3874.278 & 11.6 \\
\ion{Sr}{iii} & 3874.755 & \phantom{0}2.8 \\
\ion{Zr}{iv} & 3919.332 & 11.9 \\
\ion{Sr}{iii} & 3936.403 & 25.9 \\
\ion{Sr}{iii} & 3936.740 & 14.8 \\
\ion{Sr}{iii} & 3951.546 & 12.9 \\
\ion{Sr}{iii} & 3958.762 & 12.8 \\
\ion{Sr}{iii} & 3976.033 & \phantom{0}8.2 \\
[3pt]
\bottomrule
\end{tabular}
}
\end{minipage}
\begin{minipage}{0.49\columnwidth}
\resizebox{\columnwidth}{!}{
\begin{tabular}{l c c}
\toprule
\toprule
Ion & $\lambda_\mathrm{obs}$ / \AA & EW / m\AA \\
\midrule
\ion{Sr}{iii} & 3992.272 & 11.7 \\
\ion{Sr}{iii} & 4007.348 & 16.8 \\
\ion{Sr}{iii} & 4037.534 & \phantom{0}8.3 \\
\ion{Y}{iii} & 4039.576 & 32.8 \\
\ion{Y}{iii} & 4040.115 & 26.0 \\
\ion{Sr}{iii} & 4052.432 & 13.8 \\
\ion{Kr}{iii} & 4067.382 & 10.5 \\
\ion{Sr}{ii} & 4077.711 & 23.4 \\
\ion{Sr}{iii} & 4094.047 & \phantom{0}7.1 \\
\ion{Zr}{iv} & 4137.430 & 23.4 \\
\ion{Kr}{iii} & 4154.452 & \phantom{0}4.2 \\
\ion{Ge}{iii} & 4179.078 & 30.3 \\
\ion{Zr}{iv} & 4198.255 & 65.2 \\
\ion{Sr}{ii} & 4215.531 & 19.6 \\
\ion{Sn}{iv} & 4216.192 & 27.1 \\
\ion{Kr}{iii} & 4226.580 & \phantom{0}6.4 \\
\ion{Ge}{iii} & 4260.865 & 24.2 \\
\ion{Ge}{iii} & 4291.700 & 11.9 \\
\ion{Sr}{ii} & 4305.406 & \phantom{0}5.7 \\
\ion{Zr}{iv} & 4317.073 & 49.8 \\
\ion{Ga}{iii} & 4380.662 & \phantom{0}9.9 \\
\ion{Ga}{iii} & 4381.793 & 11.9 \\[3pt]
\ion{Ge}{iv} & 4979.987 & \phantom{0}7.9 \\
\ion{Ga}{iii} & 4993.940 & 11.7 \\
\ion{Sr}{iii} & 5022.702 & \phantom{0}7.7 \\
\ion{Sr}{iii} & 5071.126 & \phantom{0}4.3 \\
\ion{Ge}{iv} & 5073.330 & 13.2 \\
\ion{Sr}{iii} & 5074.551 & \phantom{0}1.4 \\
\ion{Y}{iii}  & 5102.901 & \phantom{0}8.4 \\
\ion{Sr}{iii} & 5158.291 & \phantom{0}3.5 \\
\ion{Y}{iii}  & 5238.110 & 22.4 \\
\ion{Sr}{iii} & 5257.763 & 13.0 \\
\ion{Sr}{iii} & 5262.211 & \phantom{0}8.2 \\
\ion{Y}{iii}  & 5263.580 & \phantom{0}2.5 \\
\ion{Sr}{iii} & 5288.360 & \phantom{0}7.5 \\
\ion{Ga}{iii} & 5337.238 & \phantom{0}2.9 \\
\ion{Ga}{iii} & 5358.205 & \phantom{0}5.5 \\
\ion{Sr}{iii} & 5391.037 & 11.8 \\
\ion{Sr}{iii} & 5405.448 & \phantom{0}5.1 \\
\ion{Sr}{iii} & 5417.570 & \phantom{0}3.9 \\
\ion{Sr}{iii} & 5421.061 & \phantom{0}7.3 \\
\ion{Sr}{iii} & 5443.479 & 16.9 \\
\ion{Zr}{iv} & 5462.380 & 52.8 \\
\ion{Sr}{iii} & 5463.942 & \phantom{0}9.0 \\
\ion{Y}{iii}  & 5602.151  & \phantom{0}5.8 \\
\ion{Sr}{iii} & 5664.628 & \phantom{0}4.5 \\
\ion{Sr}{iii} & 5689.761 & \phantom{0}5.4 \\
\ion{Zr}{iv} & 5779.880 & 26.4 \\
\ion{Ga}{iii} & 5844.912 & \phantom{0}9.3 \\
\ion{Ga}{iii} & 5993.887 & \phantom{0}6.1 \\
\ion{Zr}{iv} & 6443.235 & 13.5 \\
\bottomrule
\end{tabular}
}
\vfill
\end{minipage}
\end{table}

\begin{table}
\centering
\caption{Observed wavelengths for newly identified lines that lack oscillator strengths in the spectra of \lsiv\ and \feige. Equivalent widths are given for \lsiv.}
\label{table_idpos}
\setstretch{1.1}
\begin{tabular}{l l l c c}
\toprule
\toprule
Ion & $\lambda_\mathrm{lit}$ / \AA  & $\lambda_\mathrm{obs}$ / \AA & $\Delta \lambda$ / \AA & EW / m\AA \\
\midrule
\ion{Ge}{iii} & 3489.034 & 3489.055 & $+0.021$ & \phantom{0}6.3 \\
\ion{Ge}{iii} & 5134.652 & 5134.626 & $-0.026$ & 18.2 \\
\ion{Ge}{iii} & 5229.354 & 5229.336 & $-0.018$ & 12.3 \\
\ion{Ge}{iii} & 5256.459 & 5256.466 & $+0.007$ & \phantom{0}8.8 \\[2pt]
\ion{As}{iii} & 3922.6 & 3922.499 & $-0.101$ & \phantom{0}8.2 \\
\ion{As}{iii} & 4037.2 & 4037.015 & $-0.185$ & \phantom{0}9.3 \\[2pt]
\ion{Se}{iii} & 3387.2 & 3387.232 & $+0.032$ & 16.4 \\
\ion{Se}{iii} & 3413.9 & 3413.931 & $+0.031$ & 17.2 \\
\ion{Se}{iii} & 3428.4 & 3428.398 & $-0.002$ & \phantom{0}9.4 \\
\ion{Se}{iii} & 3457.8 & 3457.817 & $+0.017$ & 17.4 \\
\ion{Se}{iii} & 3543.6 & 3543.638 & $+0.038$ & 12.6 \\
\ion{Se}{iii} & 3570.2 & 3570.191 & $-0.009$ & 10.0 \\
\ion{Se}{iii} & 3637.6 & 3637.526 & $-0.074$ & 15.9 \\
\ion{Se}{iii} & 3711.7 & 3711.683 & $-0.017$ & 12.7 \\
\ion{Se}{iii} & 3738.7 & 3738.727 & $+0.027$ & 20.6 \\
\ion{Se}{iii} & 3743.0 & 3742.921 & $-0.079$ & \phantom{0}6.7 \\
\ion{Se}{iii} & 3800.9 & 3800.938 & $+0.038$ & 21.8 \\
\ion{Se}{iii} & 4046.7 & 4046.733 & $+0.033$ & \phantom{0}6.5 \\
\ion{Se}{iii} & 4083.2 & 4083.164 & $-0.036$ & \phantom{0}8.4 \\
\ion{Se}{iii} & 4169.1 & 4169.070 & $-0.030$ & 15.8 \\
\ion{Se}{iii} & 4637.9 & 4637.896 & $-0.004$ & \phantom{0}5.7 \\[2pt]
\bottomrule
\end{tabular}
\end{table}

\begin{table}
\centering
\caption{Remaining unidentified lines in the spectra of \lsiv\ and \feige. Estimated equivalent widths are given for \lsiv. The detection limit is about 1.5\,m\AA.}
\label{table_unid}
\setstretch{1.1}
\begin{tabular}{l r l}
\toprule
\toprule
$\lambda_\mathrm{obs}$ / \AA & EW / m\AA & Comment\\
\midrule
3330.784 & 14.6 & \ion{Kr}{iii}? \\
3439.421 & 13.3 & \ion{Kr}{iii}, \ion{Rb}{iii}? \\
3457.789 & 11.3 & \ion{Sr}{iii}? \\
3492.674 & 6.2 & \ion{Kr}{iii}, \ion{Rb}{iii}? \\
3530.783 & 7.9 & \ion{Zn}{iii}? \\
3570.183 & 4.5 & \ion{Ne}{ii}? \\
3647.659 & 4.4  \\
3649.103 & 4.2  \\
3853.263 & 7.1  \\
3857.237 & 4.8  \\
3860.431 & 8.0  \\
3863.822 & 7.8  \\
3870.852 & 6.0 & \ion{Ni}{iii}? \\
3873.239 & 3.0  \\
3901.527 & 5.3  \\
3912.595 & 4.5 & \ion{Sr}{iv}? \\
3915.091 & 5.3 \\
3931.572 & 5.2 \\
3931.572 & 5.2 \\
3935.767 & 9.5 \\
4013.975 & 4.1 \\
4037.023 & 6.5 \\
4050.439 & 4.8 \\
4058.837 & 4.4 \\
4059.791 & 2.6 \\
4088.011 & 4.7 \\
4148.989 & 13.0 \\
4181.054 & 7.5 & \ion{Kr}{iv}? \\
4184.853 & 5.1 \\
4210.418 & 1.8 \\
4211.177 & 1.8 \\
4479.618 & 14.0 \\
4636.534 & 35.0 & not covered for \lsiv \\
         &      & broad \\
4814.473 & 14.0 & \\
4820.085 & 9.2 & \\
4879.165 & 5.3 \\
4972.468 & 8.1 \\
5102.885 & 16.5 & weaker in \feige, \ion{Sn}{iii}? \\
5106.656 & 6.6 \\
5114.154 & 12.9 & very broad \\
5135.913 & 13.9 & broad \\
5167.776 & 6.2 \\
5207.452 & 4.1 \\
5208.282 & 3.5 \\
5210.208 & 2.8 & stronger in \feige \\
5221.968 & 9.4 & weaker in \feige \\
5232.749 & 5.9 \\
5234.305 & 5.7 & not det. in \feige \\
5241.974 & 6.2 \\
5562.854 & 8.3 & \ion{Zn}{iii}? \\
6756.452 & 16.3 & not covered for \feige \\
\bottomrule
\end{tabular}
\end{table}

\section{Full spectral comparisons}\label{sect:fullspec}

This section presents the full spectral comparison between the co-added UVES spectra of \lsiv\ and \feige\ and our final synthetic spectra.
The observations have been shifted to laboratory wavelengths and normalised to match the continuum levels of our synthetic spectra.
The synthetic spectra are convolved with a variable Gaussian kernel for a resolution of $R=40970$.
Rotational broadening was considered at $v _\mathrm{rot} \sin i = 9$\,km\,s$^{-1}$ using the auxiliary program, \textsc{rotin3,} delivered with \textsc{Synspec} and a limb-darkening coefficient of 0.3 (which is more appropriate for compact stars than the default of 0.6).
The strongest photospheric metal lines are labelled in black, while identified lines for which no oscillator strengths are available are labelled in red.

\captionsetup[ContinuedFloat]{labelformat=continued}
\begin{figure*}
\centering
\includegraphics[width=24.2cm,angle=90,page=2]{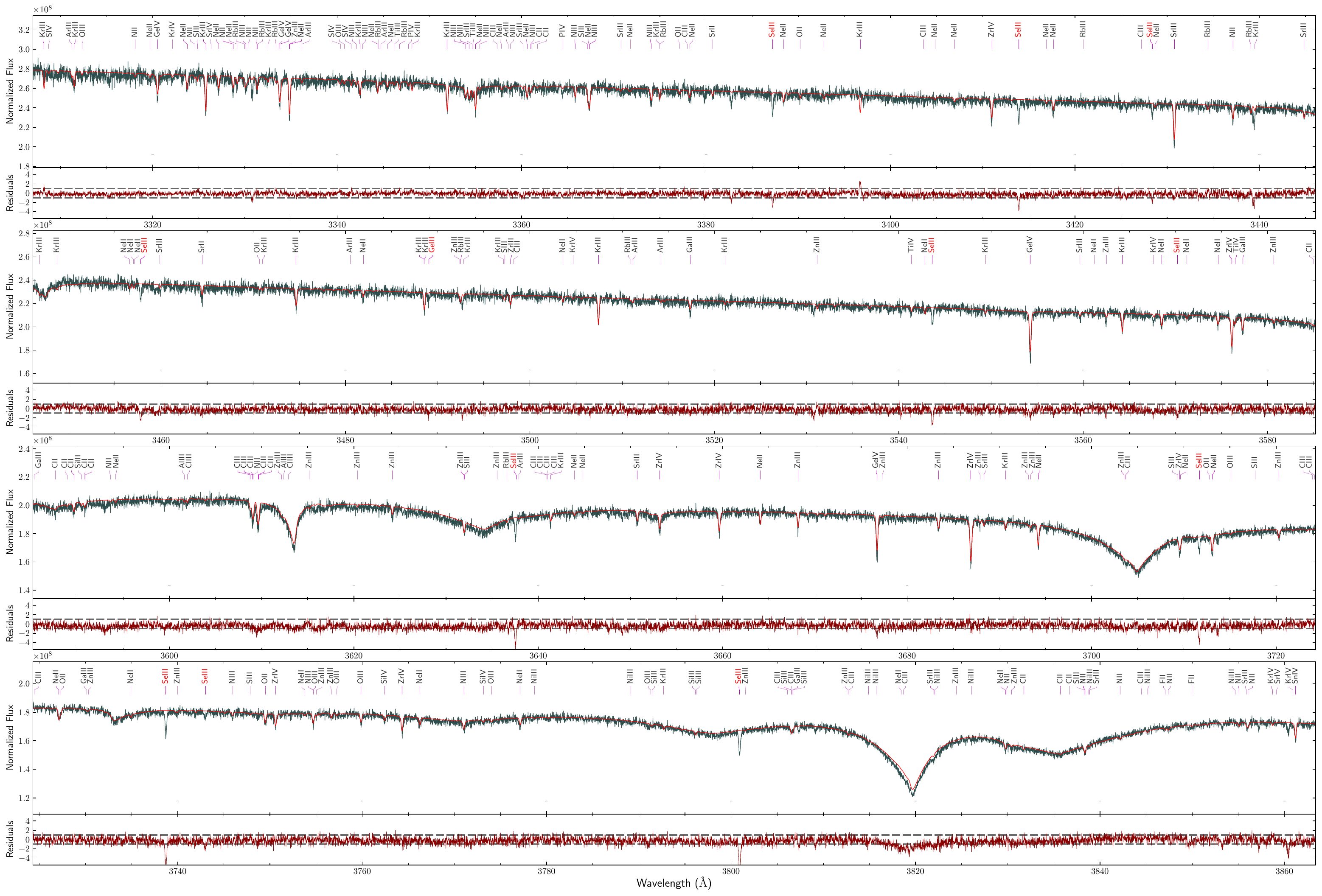}
\caption{UVES spectrum of \feige\ (grey) and final model (red).}
\end{figure*}
\begin{figure*}
\ContinuedFloat
\centering
\includegraphics[width=24.2cm,angle=90,page=3]{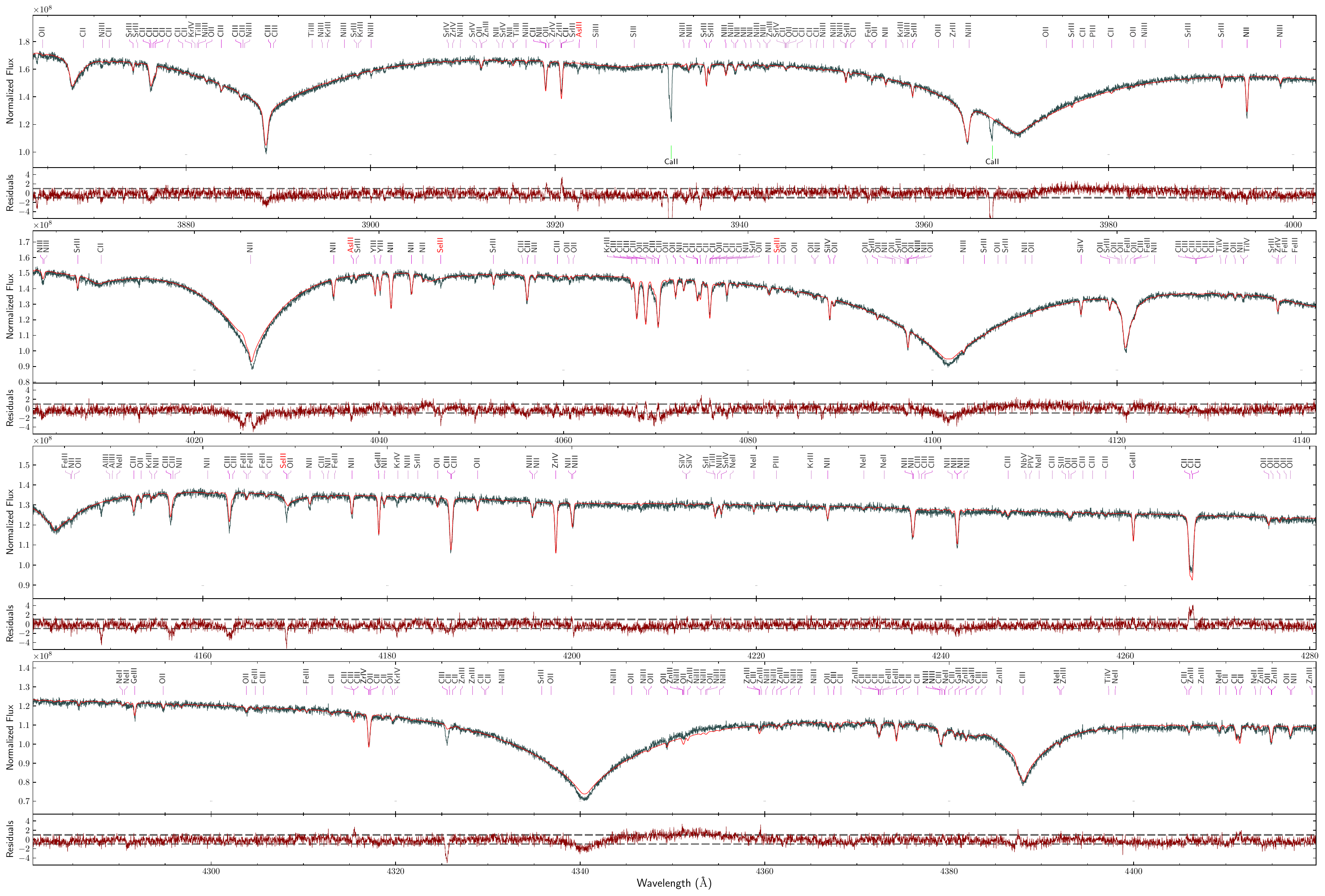}
\caption{UVES spectrum of \feige\ (grey) and final model (red).}\label{fig:fspec_feige}
\end{figure*}
\begin{figure*}
\ContinuedFloat
\centering
\includegraphics[width=24.2cm,angle=90,page=4]{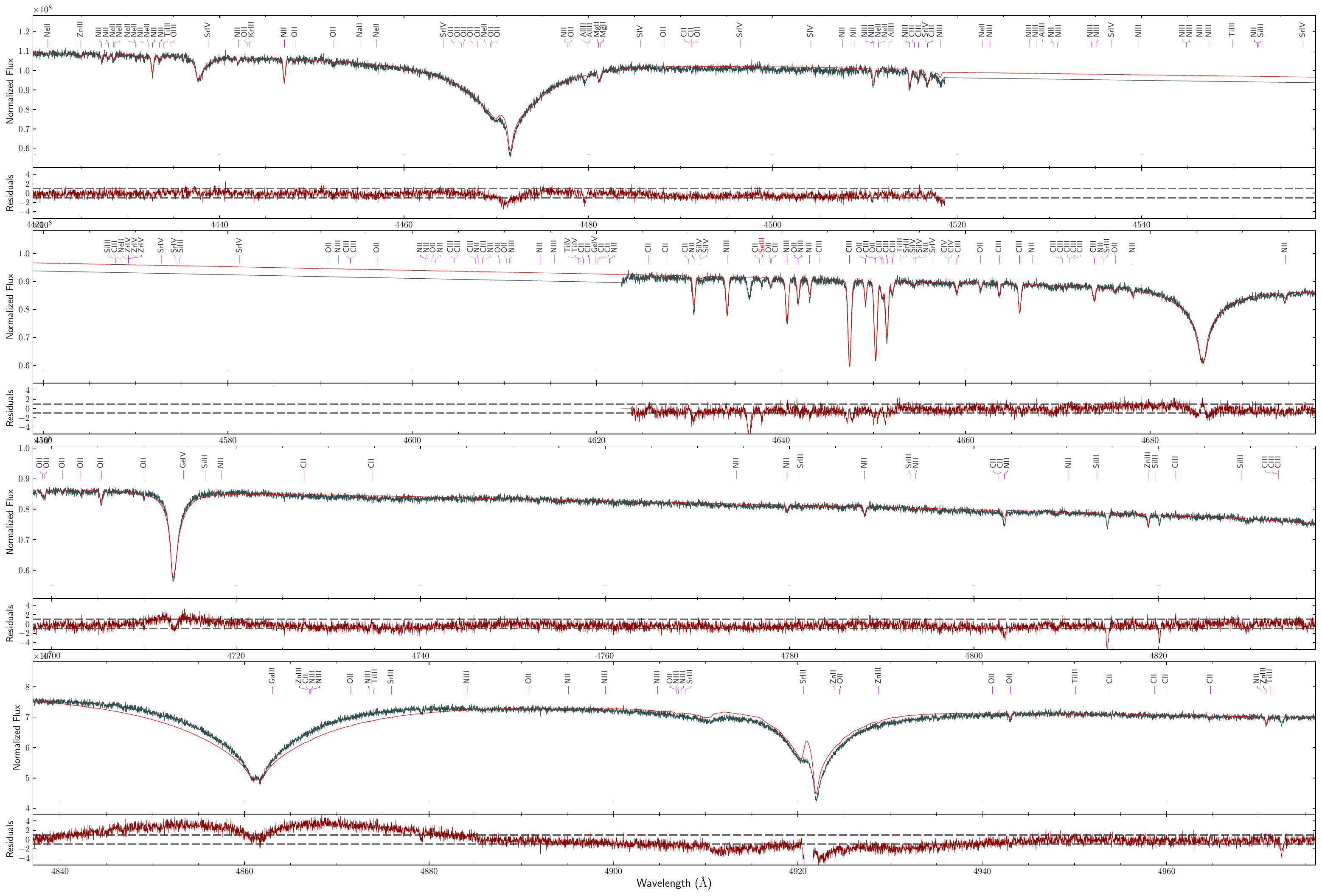}
\caption{UVES spectrum of \feige\ (grey) and final model (red).}
\end{figure*}
\begin{figure*}
\ContinuedFloat
\centering
\includegraphics[width=24.2cm,angle=90,page=5]{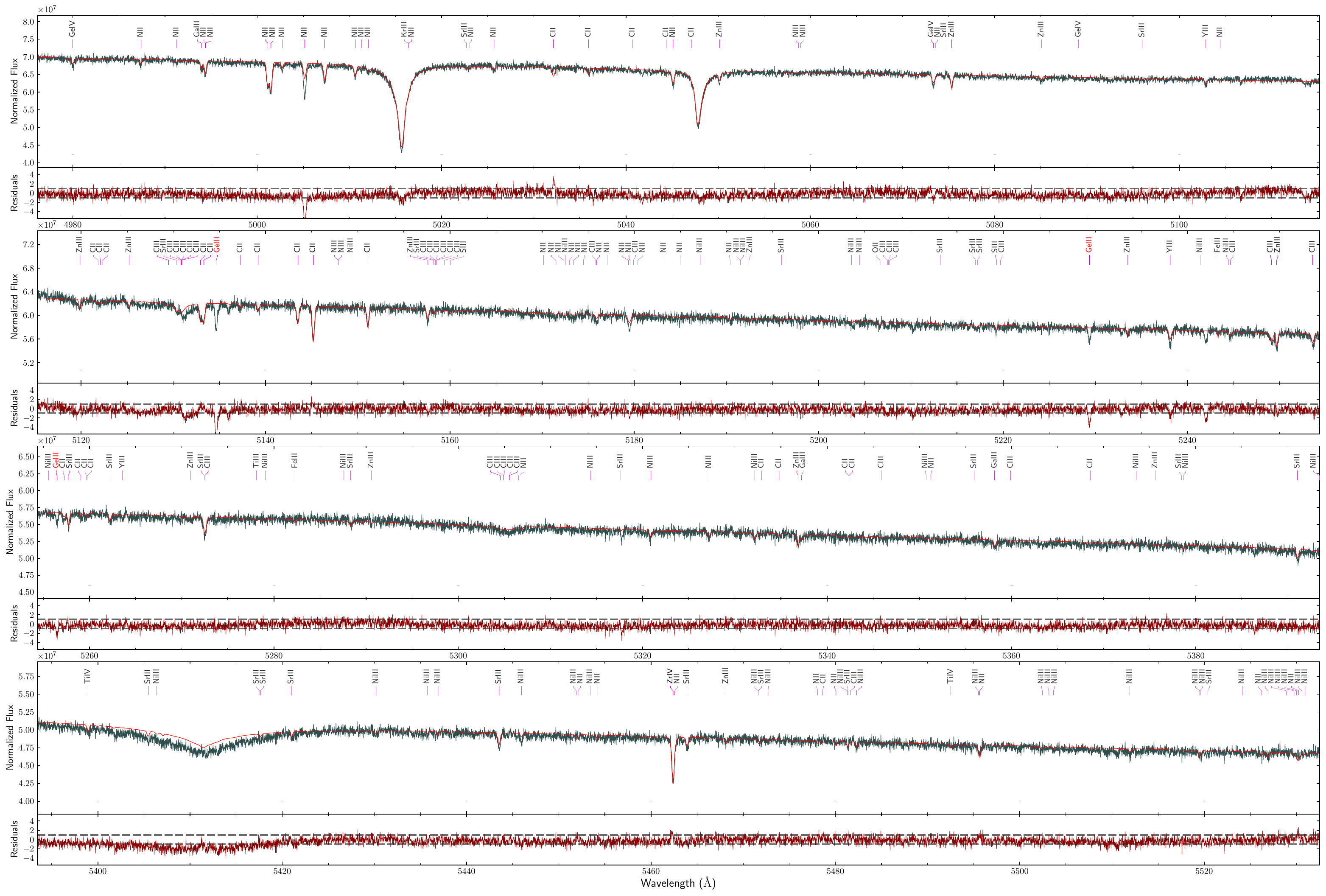}
\caption{UVES spectrum of \feige\ (grey) and final model (red).}
\end{figure*}
\begin{figure*}
\ContinuedFloat
\centering
\includegraphics[width=24.2cm,angle=90,page=6]{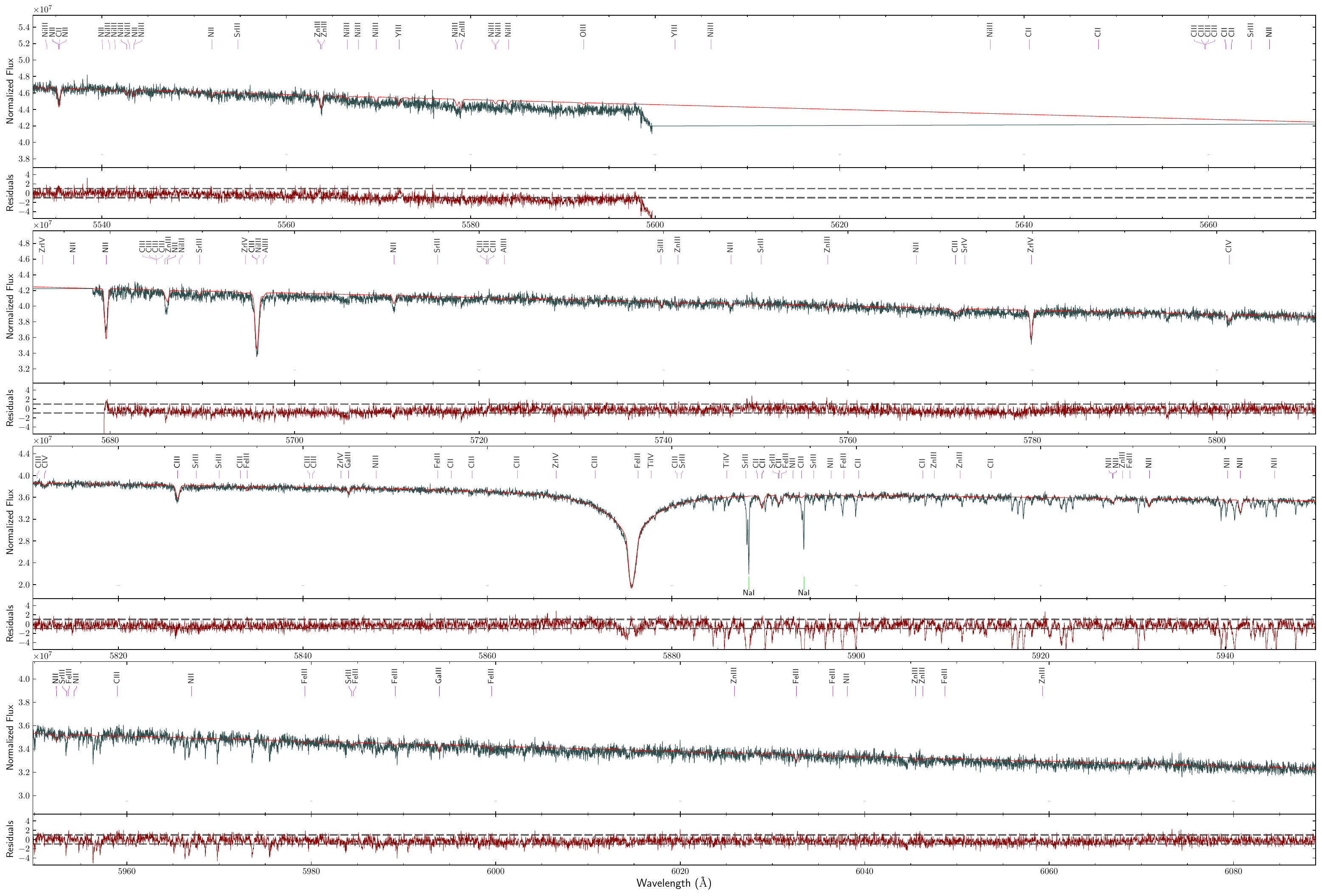}
\caption{UVES spectrum of \feige\ (grey) and final model (red).}
\end{figure*}
\begin{figure*}
\ContinuedFloat
\centering
\includegraphics[width=24.2cm,angle=90,page=7]{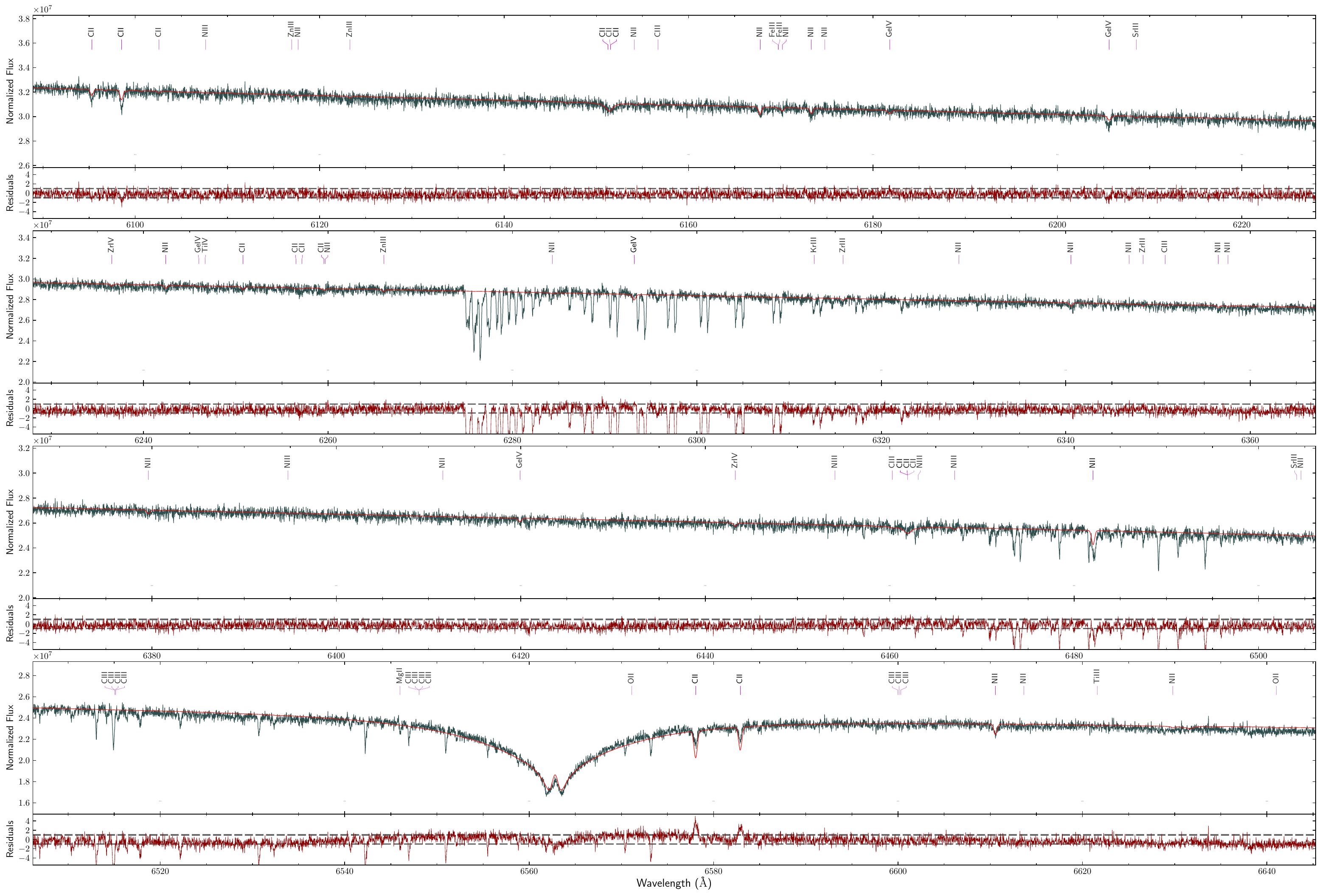}
\caption{UVES spectrum of \feige\ (grey) and final model (red).}
\end{figure*}

\captionsetup[ContinuedFloat]{labelformat=continued}
\begin{figure*}
\centering
\includegraphics[width=24.2cm,angle=90,page=2]{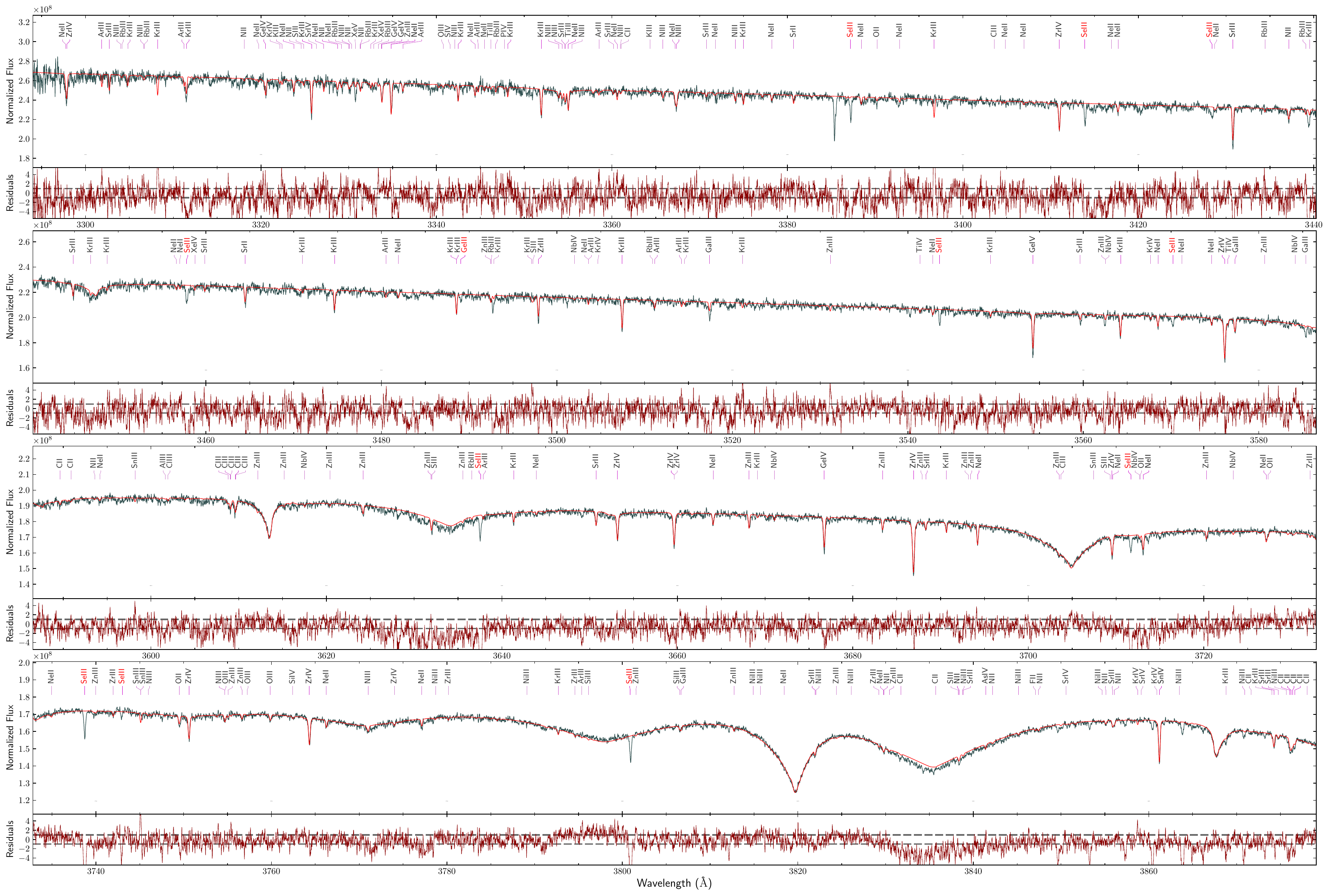}
\caption{UVES spectrum of \lsiv\ (grey) and final model (red).}\label{fig:fspec_lsiv}
\end{figure*}
\begin{figure*}
\ContinuedFloat
\centering
\includegraphics[width=24.2cm,angle=90,page=3]{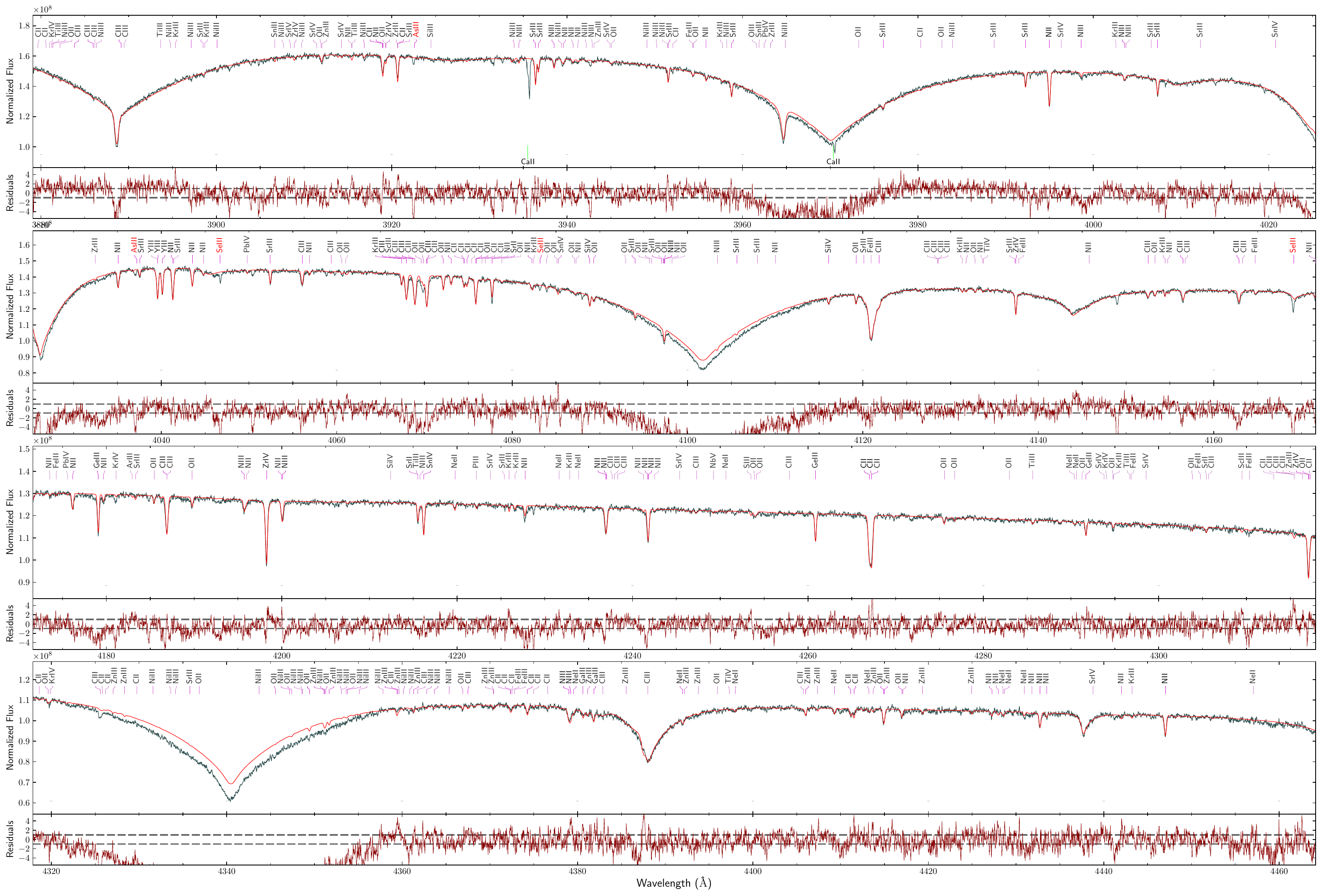}
\caption{UVES spectrum of \lsiv\ (grey) and final model (red).}
\end{figure*}
\begin{figure*}
\ContinuedFloat
\centering
\includegraphics[width=24.2cm,angle=90,page=4]{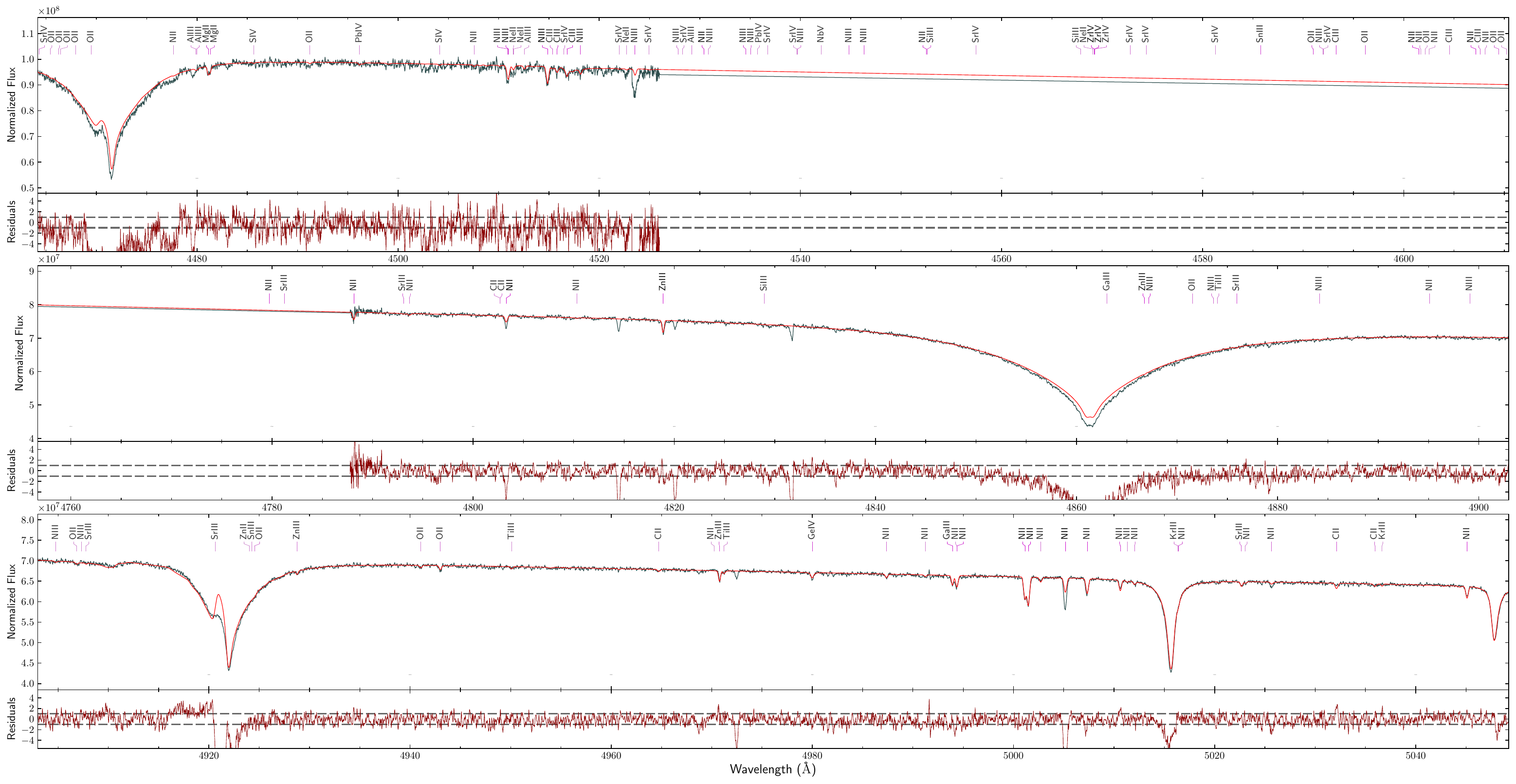}
\caption{UVES spectrum of \lsiv\ (grey) and final model (red).}
\end{figure*}
\begin{figure*}
\ContinuedFloat
\centering
\includegraphics[width=24.2cm,angle=90,page=5]{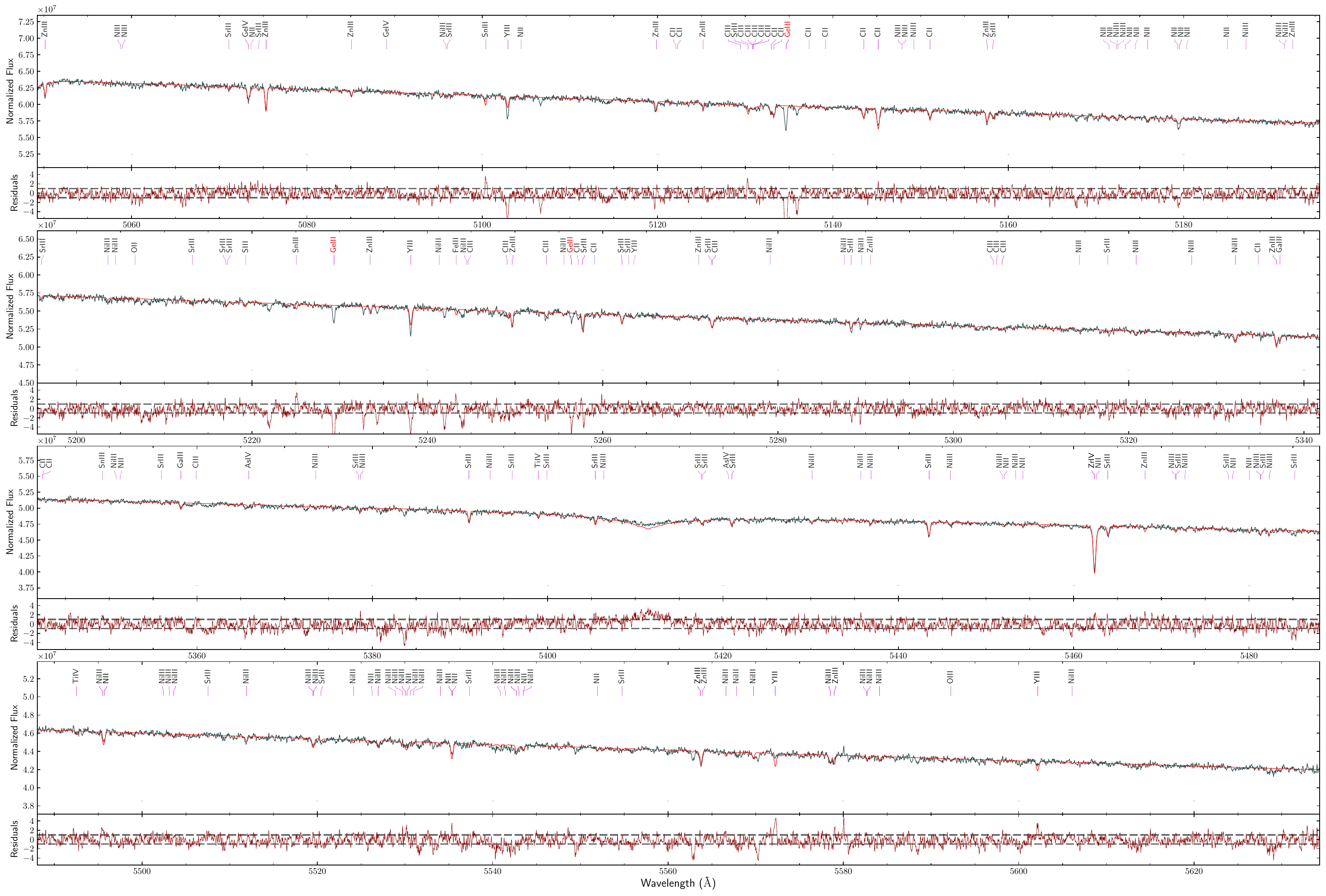}
\caption{UVES spectrum of \lsiv\ (grey) and final model (red).}
\end{figure*}
\begin{figure*}
\ContinuedFloat
\centering
\includegraphics[width=24.2cm,angle=90,page=6]{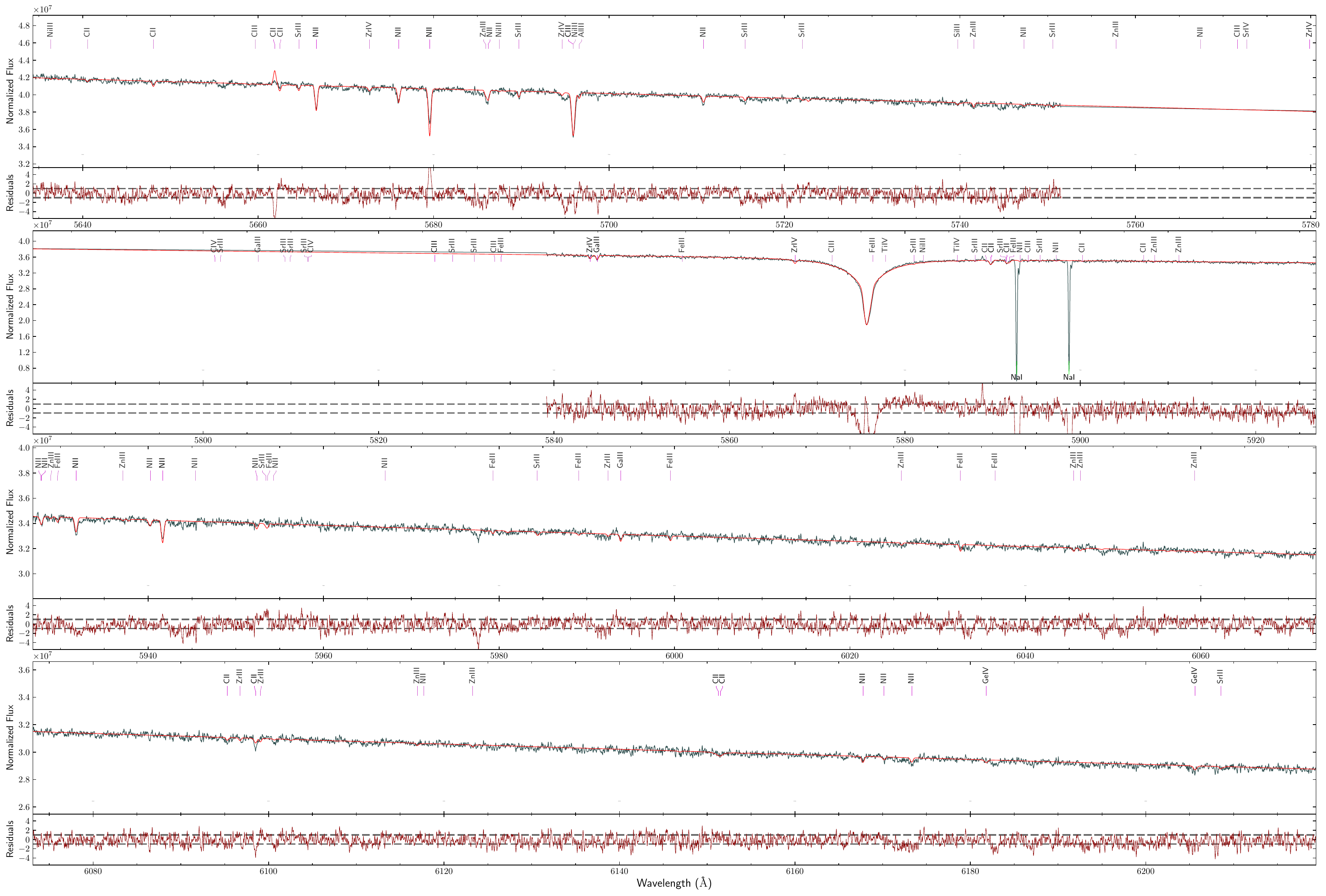}
\caption{UVES spectrum of \lsiv\ (grey) and final model (red).}
\end{figure*}
\begin{figure*}
\ContinuedFloat
\centering
\includegraphics[width=24.2cm,angle=90,page=7]{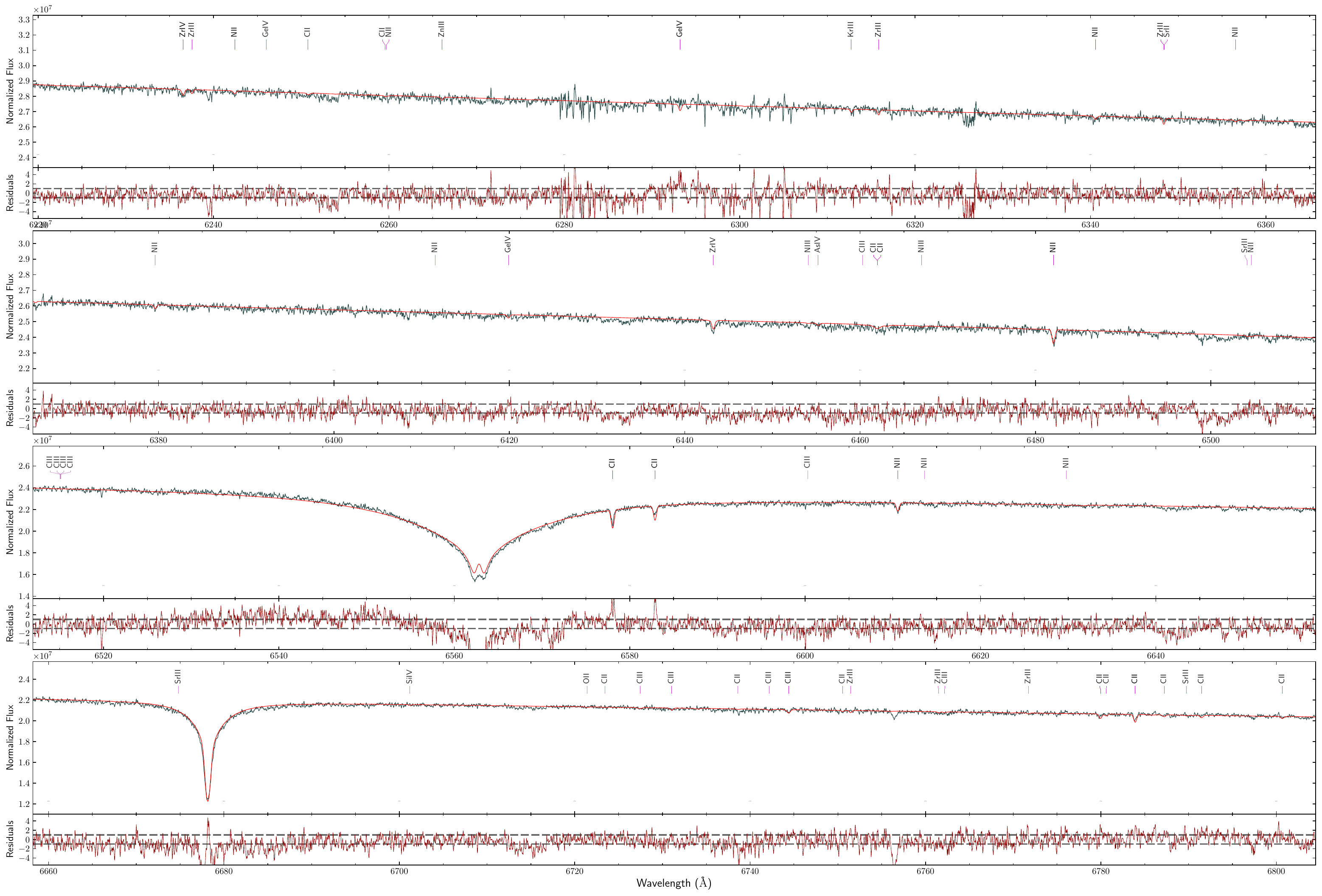}
\caption{UVES spectrum of \lsiv\ (grey) and final model (red).}
\end{figure*}

\section{SED of Feige\,46}\label{appendix_photometry}

\begin{figure*}
\sidecaption
\includegraphics[width=0.705\textwidth]{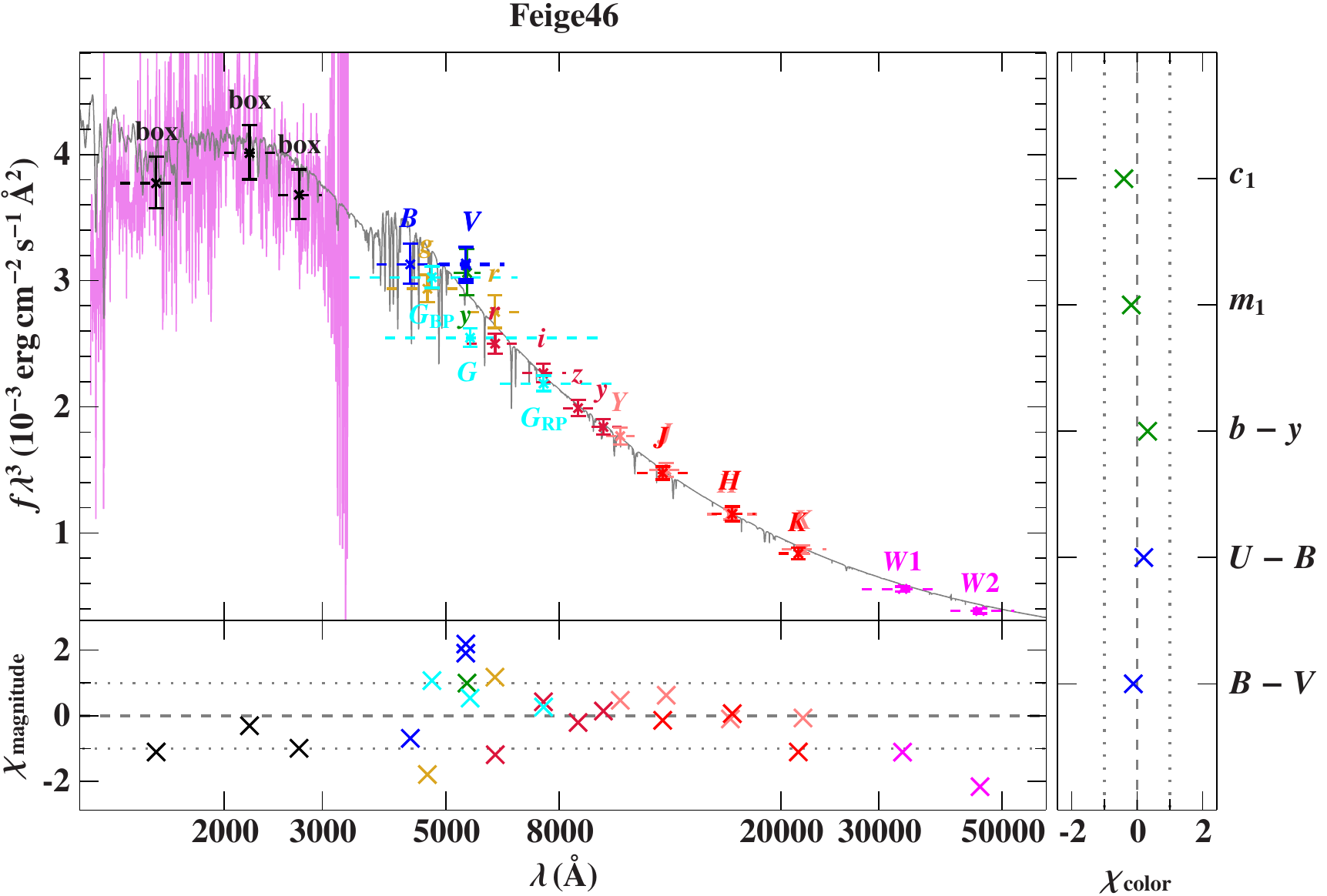}
\caption{Same as Fig.~\ref{fig:sed_fit} but for \feige. 
The following colour codes are used to identify the photometric systems: SDSS \cite[yellow,][]{Henden_2015}, Pan-STARRS1 \cite[red,][]{2016arXiv161205560C}, Johnson-Cousins \cite[blue,][]{Mermilliod_1994,Henden_2015}, Str\"omgren \cite[green,][]{Hauck_1998}, \textit{Gaia}  \cite[cyan,][]{Gaia2018_VizieR}, UKIDSS \cite[rose,][]{Lawrence_2007}, 2MASS \cite[bright red,][]{Cutri2003_2MASS}, and WISE \cite[magenta,][]{Schlafly_2019}. Three IUE spectra were used to construct the box filters (SWP17466RL, SWP20342L, LWR16264LL).}
\label{fig:sed_feige}
\end{figure*}

\end{appendix}

\end{document}